\documentclass[10pt]{article}
\usepackage{epsfig}
\usepackage[intlimits]{amsmath}
\usepackage{amssymb}
\usepackage{indentfirst} 
\usepackage{nonfloat}
\usepackage{caption}
\usepackage{graphicx}
\usepackage[affil-it]{authblk}

\makeatletter
\renewcommand*{\@fnsymbol}[1]{\ensuremath{\ifcase#1\or \rm{*}\or \rm{b}\or 
\ddagger\or \mathsection\or \mathparagraph\or \|\or **\or \dagger\dagger
    \or \ddagger\ddagger \else\@ctrerr\fi}}
\makeatother

\usepackage{abstract}
\begin{document}

\thispagestyle{myheadings}

\title{
\LARGE{\bf{Analysis of multichannel measurements of rare processes\\ 
with uncertain expected background and acceptance}}
}
\author{I. B. Smirnov\thanks{E-mail: Igor.Smirnov@cern.ch}} 
\date{26 May 2013}
\affil{\small{Petersburg Nuclear Physics Institute, Gatchina 188300, Russia}}
\date{}
\twocolumn[
\begin{@twocolumnfalse}
\maketitle
\begin{abstract}
A typical experiment in high energy physics is considered.
The result of the experiment is assumed to be a histogram
consisting of bins or channels with numbers of corresponding
registered events.
The expected background and expected signal shape or acceptance 
are measured in separate auxiliary experiments,
or calculated by the Monte Carlo method with finite sample size,
and hence with finite precision. 
An especially complex situation 
occurs when the expected background 
in some of the channels 
happens to be zero due to either a fluctuation 
of the auxiliary measurement (or simulation) or because it is truly zero.
Different statistical methods give different
confidence intervals
for the full signal rate and different significances 
of the signal$+$background hypothesis versus the pure background hypothesis.
Detailed analysis and numerical tests are presented.
\\
\\

\end{abstract}
\end{@twocolumnfalse}
]
\saythanks

\section{Introduction}

Rates of rare processes in high energy physics
have sometimes to be estimated
from a few observed events.
This can happen during the research of very rare processes or
at the beginning of any research.
Reconstruction of such rates
is a complex and ambiguous problem \cite{WhyNotBayesian}, 
especially in the presence of uncertain nuisance parameters.

\subsection{Typical experiment}

The result of an experiment is frequently represented
by a histogram
consisting of several, $k$, bins or channels, $k \ge 1$.
Each of these channels keeps the number of events $n_i$ 
registered in this channel, where $i$ is the channel number.
The number $n_i$ is sampled from the Poisson distribution with a
parameter, which is unique for each channel.
Events in each channel are expected to be produced by 
background processes, called background,  
and by a studied process called a signal,
all distributed according to the Poisson law.
The expected background $b_i$ and expected signal (shape) or acceptance $a_i$
(we prefer the term ``expected signal'' and 
omit the word ``shape'' for briefness and because we do not require 
$a_i$ to be normalized, c.f. Refs. \cite{Heinrich_04,Heinrich_05})
are either known exactly or measured in separate auxiliary experiments,
or calculated by the Monte Carlo method with finite sample sizes,
and hence with a finite precision.
In the general case these {\it nuisance parameters} can correspond 
to different exposures
or luminosities, so the expected full rate in the main
experiment is expressed by 
\begin{eqnarray}
f_i = t_a a_i  s + t_b b_i \, ,
\label{mu_i}
\end{eqnarray}
or in the vector notation
\begin{eqnarray}
\vec{f} = t_a \vec{a}\,  s + t_b \vec{b} \, ,
\label{mu_i_1}
\end{eqnarray}
where $t_a$ and $t_b$ are the ratios of exposures of the main and respective
auxiliary experiments and $s$ is the real signal rate, 
the absolute value, or the value relative to the expected signal rate.

We consider here only the stochastic uncertainties of $a_i$ and
$b_i$. The uncertainties that are 
presumably non-stochastic can usually be assumed stochastic 
in some more general sense and can be handled by similar methods.
Both $a_i$ and $b_i$ are assumed to be 
``measured'' in the respective ``auxiliary experiments'' 
as the numbers $n_{ai}$ and $n_{bi}$
sampled from the Poisson distributions with the corresponding parameters
$a_i$ or $b_i$. 

The purpose of the experimental research 
is to determine the most probable full signal rate, 
a confidence interval for it, and the significance of
the signal (plus background) hypothesis versus the pure background hypothesis.

\subsection{Zero channels of expected background}

An especially challenging situation
arises when the expected background
in some of the channels happens to be zero in the auxiliary 
background experiment
due to either a fluctuation of the auxiliary measurement or 
because it is truly zero. 
Such a situation can happen in the search of very rare
processes in experiments with very good background rejection.
For such experiments it is difficult to perform
the Monte Carlo simulation of background with large final statistics
(i.e. to obtain a large number of background events passed through all 
triggers, reconstructions, 
``cuts'', and selections),
because in order to do this one needs to run huge initial statistics. 
Hence it is not possible to distinguish
the cases of a downward background fluctuation and the true zero background
on the basis of existing information. 
It is then unclear both conceptually and numerically, how to interpret
a non-zero result of the main experiment in this channel.

Literature does not offer any recipe for dealing with 
zeros in the expected background.

One could simply ignore such channels.
Or 
zeros can be removed by
unification or smoothing of neighboring or presumably similar
channels, some of which are non-zero.
The similarity of the channels is typically indicated by neighboring values
of a response variable produced by multivariate analysis methods.
But any zero-removing procedure can lead, briefly speaking,
to an unexpected change of the precision and complicates its estimation.
The most doubtful case is
when a few zero channels appear in the distribution of the
expected background by the response variable
at the end of the distribution, where 
the expected background is minimal and 
the expected signal is maximal.
When the expected-background spectrum finishes with 
several zero bins and there is nothing to the right-hand side of them, 
any interpolation or smoothing is ambiguous.
In this case, the end of the spectra, which has to be the most important
region for the results, is subjected to effectively arbitrary treatment.

Therefore, 
it is interesting to investigate which 
statistical methods,
if any,
provide correct results in such problems, when 
the channels' content
is taken 
as is, without 
forbidding of zeros or smoothing.

\subsection{Tests of intervals}

The idea is to generate a sequence of pseudo-experiments
with some small enough true background,
to divide the experimental results into 
different numbers of channels, 
to treat them
by all available statistical methods, and to compare the results.
In particular, it is interesting to see whether an optimal number of channels
can be chosen on the basis of the data
available for the experimenter, for instance, by minimizing
the interval width reconstructed by the data of the single experiment
(not the average width).
We will study a rough optimization
with selection of the best division
from several divisions with different numbers of channels
incremented by a factor of 1.5--3: 1, 2, 3, 5, 10, 30.
Channels in each division have the same width by the response variable.

If the obtained interval for the searched parameter
includes (``covers'') its true value, 
whatever it is, with the stated
probability, or frequency in a long sequence of experiments, 
it is said that the coverage is provided.
Then, from the formal classical viewpoint the method is acceptable.
The coverage probability is usually allowed to be greater than stated.
In this case the coverage is called ``conservative''.
Nuisance parameters are assumed to have some true values 
in this sequence too.
During the analysis they are unknown, but they do not need to be reconstructed
or ``covered''.
This is effectively a ``frequentist'' or a ``classical'' 
viewpoint. Both terms are accepted, see, for example, 
Ref. \cite{WhyNotBayesian}, \S26.1 of Ref. \cite{Kendall_2A}, 
and Ref. \cite{DAgostini_1998}.
The intervals obtained by the minimization of width 
mentioned above or
by other methods of the binning selection
and calculated with different binning for each given experiment can
also be tested this way and can be accepted, 
if they provide the stated coverage probability.

Such an investigation is interesting not only in the context of the
problem with zeros, 
but in a much wider context.
Many statistical methods
provide accurate coverage for the problems with known
expected signal and background, but do not guarantee
it for the problems with nuisance parameters.
The Bayesian method and the profile likelihood (or likelihood ratio) method 
do not guarantee frequentist coverage for samples of finite sizes \cite{WhyNotBayesian}, but
it is interesting to see whether they provide it in practice.
If the parameter of interest is restricted, typically to be
non-negative, some classical methods can
produce empty or unphysically small
intervals in the case of a downward
fluctuation of background and small true signal
(c.f. Refs. \cite{F&C,Mandelkern_00}).
Some test statistics allow one to obtain finite
intervals 
by maximizing the upper limit and
minimizing the lower limit with respect to 
nuisance parameters
or by the projection of the confidence set
to the parameter of interest.
However, these procedures can require unrealistic computing resources
in the case of many nuisance parameters,
the optimal values of the latter can be incompatible with measurements, and
the resulting interval can reportedly ``badly over-cover''\cite{Cousins_05}.
Another issue which is discussed in the literature
is the coverage of the upper border of the classical intervals for
the experiments microscopically dependent on the signal \cite{Read_00}.

Frequentist tests 
reported in Refs. \cite{Heinrich_07,Junk_11,Rolke_2005,Tegenfeldt_2004}
do not include some interesting methods and cases, such as, in particular,
zeros in the expected background.

\subsection{Tests of significance}
\label{Sect_intro_sig}

The notion of coverage 
does not exist in the case of
significance. We can only calculate the significance itself in a different way,
for example, with different test statistic and (or) for different sample or
subset of experiments.
For example, the uniformity of a $p$-value distribution is sometimes
tested \cite{Cousins_08,Demortier_07,Robins_2000}.
Denoting the estimate of the $p$-value by $\rho$ (in order to distinguish it
from the probability density, which is denoted by $p$ herein)
and its probability density 
distribution by $p(\rho)$ one can check whether  
$\int_0^{\rho_t} p(\rho) \, \mathrm{d} \rho \leq \rho_t$
for any small enough threshold $\rho_t$.
If the sign is ``$<$'', this is the conservative case, better than assumed to be
necessary.
The opposite case ``$>$'' is considered as a signature 
of overestimated significance, see a discussion in Section 1.3 of 
Ref. \cite{Robins_2000}.
The equality means that the $p$-value is uniform.
If $\rho$ is discrete,
we imply here that the equality occurs for 
$\rho_t$ equal to  all possible $\rho$.
The integral in this inequality can be considered as 
a $p$-value for the test statistic $\rho_t$ or for the correspoding 
significance,
assuming that the small values of $\rho_t$ or large values of 
the correspoding significance
indicate disagreement with the null hypothesis.
This alternative $p$-value can differ from the regular $\rho_t$,
if nuisance parameters are involved.
The uniformity of the alternative $p$-value can be tested in the same way.
Alternative $p$-values are always uniform for the sample of experiments,
for which the corresponding regular $p$-values are calculated 
(the latter can be calculated internally with completely different
pseudo-experiments, depending on the method),
but the regular $p$-values are not necessarily uniform for this sample.
For some methods there exist samples for which
regular $p$-values are equal to alternative $p$-values and both are uniform
(see Section \ref{SSP-FMML_sec}).
Intuitively, the statistical analysis by such methods is more reliable.

The particular case of 
the one-channel ``on/off'' problem solved through the test of the ratio
of Poisson means \cite{Cousins_08}
yields the significance 
that appears to be independent of the background hypothesis,
though dependent on the total $n_{\mathrm{on}} + n_{\mathrm{off}}$ 
measurement. 
There is no similar simplification
known for the many-channel case. 

As with the confidence limits,
the significance calculated with certain test statistics
can have a non-zero minimum at some values of
nuisance parameters \cite{Demortier_07}.
But these values are sometimes incompatible with their 
measurements \cite{Berger_Boos_1994}, 
and it is very difficult to find this minimum in the case of many
nuisance parameters.
If the nuisance parameters are constrained by their
confidence limits with an arbitrary confidence level $\beta$ 
\cite{Berger_Boos_1994},
the sum of the obtained maximal $p$-value and $\beta$
can be used as the $p$-value for the test statistic which is this very sum.
This $p$-value is uniform or conservative, but
it depends on arbitrary $\beta$ and
can exaggerate the significance of a particular experiment,
if the true nuisance parameters are outside the confidence limits.
Finally, it is not easy to calculate the confidence limits 
for many correlated parameters.

\subsection{Content}

This paper is organized as follows.
In the next section the test problem is described in more detail.  
The optimization of division (binning) is briefly outlined.

The following main groups of methods are described and 
tested\footnote{The calculations on which this paper is based
were performed by  a self-made
C\texttt{++} software package independent of the external statistical software.}
in the following sections:
\begin{itemize}
\item[--] The Bayesian approach (Section \ref{Bayesian_approach_sec})
\cite{WhyNotBayesian,Heinrich_04,Cousins_09,Kendall_2B,Choudalakis_2011,Kovalenko_73,James}
with new ``safe'' priors and new modification of the central intervals. 
\item[--] The frequentist treatment of 
the maximum likelihood estimate
(Section \ref{sect_FML}) without nuisance parameters
\cite{Mandelkern_00,Ciampolillo_98} and  
many new methods of their inclusion.
\item[--] The profile likelihood 
or the likelihood ratio method (Section \ref{Likelihood_Ratio_sect}) --- 
currently both notations are
used 
in the literature, see Refs. 
\cite{WhyNotBayesian,Rolke_2005,Cousins_09}.
\item[--] The frequentist treatment of likelihood ratios or the 
$CL_s$ methods (Section \ref{The_CLs_method})
\cite{Read_00,Junk_99,Read_02,CMS_Combined,CMS_Paper_2012}
with different denominators and various old and new 
ways of nuisance parameters inclusion.
\end{itemize}

Features of confidence intervals 
are illustrated 
by many plots. 
Characteristics
of significance obtained by each appropriate method
despite the computational difficulties
are briefly described.
A comparison
of significance obtained 
by different methods for particular simple cases is 
given in Section \ref{on_off_Sec}.

Conclusions are presented in the last section.

\section{The test problem}
\label{TheTest_sect}
\subsection{Notations}

The conditional probability (density) of 
obtaining an experimental result $x$ at given parameter $y$
is denoted by $P(x|y)$ for the discrete case and
by $p(x|y)$ for the continuous 
case\footnote{The probability density 
is frequently denoted by $f(x|y)$ or $f(x; y)$
in the literature, but $p(x|y)$ is also used, 
see \cite{Heinrich_04,Grishin_1988,KassWasserman} 
and \S 2.3.5 of \cite{James}.}, except for 
the Bayesian prior distributions,
which are denoted by $\pi(x)$.
All $P$, $p$, and $\pi$ 
denote probabilities (densities for $p$ and $\pi$) of observing
corresponding values, 
but not functions with a fixed form.
We retain the same notations even when we consider them
as functions of  $y$ and call them``likelihoods''
(\S 8.22 of Ref. \cite{Kendall_1} and Ref. \cite{WhyNotBayesian}).

Then, for instance, the joint probability of obtaining $n_i$ events
in the $i$-th channel of the main experiment, 
 $n_{ai}$ in the auxiliary signal experiment, and 
 $n_{bi}$ in the auxiliary background experiment is denoted by
\begin{align}
P(n_i, & n_{ai}, n_{bi} | s, a_i, b_i) =  \nonumber \\ 
& P(n_i| s, a_i, b_i) P(n_{ai}| a_i) P(n_{bi}| b_i) = \nonumber \\ 
& P(n_i| t_a a_i s + t_b b_i) P(n_{ai}| a_i) P(n_{bi}| b_i).
\label{FullProb_i}
\end{align}
If all channels are involved, the corresponding multiplication
of probabilities is denoted, for short, by
  
\begin{align}
\prod_{i} P(n_i, n_{ai}, n_{bi} | & s, a_i, b_i) = 
P(\vec{n}, \vec{n}_{a}, \vec{n}_{b} | s, \vec{a}, \vec{b})
  = \nonumber \\ 
& P(\vec{n}| t_a \vec{a}\ s + t_b \vec{b}) 
P(\vec{n}_{a}| \vec{a}) P(\vec{n}_{b}| \vec{b}).
\label{FullProb}
\end{align}

The elementary probabilities $P(n|\mu)$ 
of observing $n$ events with the average expectation
$\mu$ for this work are assumed to follow the Poisson law:
\begin{eqnarray}
P(n|\mu) =  \mathrm{Poisson}(n,\mu) = \frac{\mu^{n} e^{-\mu}}{n!} \, .
\label{PoissonDist}
\end{eqnarray}

\subsection{Parameters and algorithm}
\label{Parameters_sect}

For most of the calculations 
in this paper it is assumed that
the multivariate response variable, denoted $x$, varies from 0 to 1.
This interval is equally divided into $k$ bins with step $1/k$
and end points $x_i$, $x_{i+1}$.
The true background and signal distributions are
\begin{eqnarray}
f_b(x) = Ce^{-qx},  &  \ \ \ \ \ \   & f_a(x) = Ce^{-q (1-x)}  \nonumber
\end{eqnarray}
respectively, where $q \geq 1$ and the normalization factor $C=q/(1 - e^{-q})$ is  needed
to make the integral equal to unity. 
The true parameters $a_i$ and $b_i$ are determined by equalities:
\begin{equation}
a_i = N_a \displaystyle \int\limits_{x_i}^{x_{i+1}}f_a(x) \, \mathrm{d}x \, ,
\ \ 
b_i = N_b \displaystyle \int\limits_{x_i}^{x_{i+1}} f_b(x) \, \mathrm{d}x \, ,
\label{a_i_b_i}
\end{equation}
where $N_a$ and $N_b$ are the 
mean total numbers of detected events
in the corresponding auxiliary experiments.

In order to generate the auxiliary pseudo-experiments one can either
generate the numbers of actual 
signal events by the Poisson law with the mean $N_a$
and similarly for the background events with $N_b$
and distribute these events randomly according to  Eqs. (\ref{a_i_b_i}),
or generate the numbers in each bin by the Poisson law
according to the means given by Eqs. 
(\ref{a_i_b_i}).
Two separate histograms, one with expected signal and the other with expected
background, are filled for each pair of auxiliary
pseudo-experiments. If the case of exactly known parameters is considered,
the corresponding histograms can be filled by $a_i$ or $b_i$ or both, 
depending on the case.
Similarly in the simulated main pseudo-experiment
the events are generated in either of two ways taking into
account  Eqs. (\ref{a_i_b_i}) and
Eq. (\ref{mu_i}) with $s = s_{\mathrm{true}}$,
the true signal rate,
which is unknown for the analysis program and has to be reconstructed by it.
To obtain identical events with a different number of channels
only the histograms with the largest number of channels are filled
by the method described above.
The other ones
are obtained by summing up the content
of the neighboring channels.
For tests of intervals 
we consider divisions with 1, 2, 3, 5, 10 and 30 channels.
This is all done in a separate ``main'' program.
For each division the main program calls the analysis program
and transmits to the latter three histograms: 
the ``main'' and two ``auxiliaries''.
The analysis program knows $t_a$ and $t_b$
and reconstructs the confidence interval for $s$, its most probable value,
and the significance.

After the end of the loop by divisions for each ``event''
(that is for each main and auxiliary experiments)
the main program
can compare the intervals obtained for each division and
choose the best according to any criteria.
After the end of the loop by experiments 
it can check the coverage of these intervals.
The significance is tested differently.

Most of the calculations of intervals for this work were made with
$t_a = 0.25$, $N_a  = 100$, 
$t_b = 5$, $N_b = 50$, 
$s_{\mathrm{true}} = 2$, and $q = 3$.
The behavior of many methods was also tested at
microscopic dependence on signal in the conditions of
$t_a = 0.25/20$, as well as for the zero true signal $s_{\mathrm{true}} = 0$
with normal dependence on it ($t_a = 0.25$).
The significance was also studied for many other configurations.  

With such parameters the probability of observing zeros in the expected
background is very high for many-channel cases.
For just the last channel
the probability of observing no events in it
is around 12\% for the 5-channel case, 40\% for the 10-channel case 
and 76\% for the 30-channel case.

Most of the tests of intervals
were performed in this paper for
the one-sided confidence level of 90\%.
In some cases, especially for the cases without uncertainties, which
are typically calculated very quickly,
the levels of 99\% and even 99.845\% ($3 \sigma$ level)
were tested.

\subsection{Optimization of division}
\label{Optimization}

The most basic method of optimization
is finding the division (binning) that 
provides the minimal interval width.
This should usually be combined with some additional conditions,
such as the absence of zeros in the expected-signal distribution, or 
another condition,
depending on the 
\linebreak
method. 
The mentioned condition
effectively excludes too many-channel divisions from consideration.
The optimization by separate limits 
usually results in the insufficient coverage probability for any method.
If the coverage for fixed divisions is  90\%,
usually something like 80\% is obtained by this optimization.

In the case with nuisance parameter uncertainties 
the minimization of the interval width 
usually reduces the coverage with respect to
the minimal coverage obtained at fixed divisions.
Thus, if there is no noticeable margin in the latter,
the coverage of the intervals optimized by widths
can be slightly less than required.

We will not consider here various other ways of optimization
that can eventually mix a channel with true zero expected background
with a neighboring non-zero channel, thus potentially reducing the sensitivity
of the experiment.

A simple way to provide the claimed
optimized coverage is to request
better fixed-division coverage.
Then it needs to know for how much it should be better.
In general this is an unclear issue.

A better way of improving the optimized coverage of any non-Bayesian method,
when the optimal divisions are obtained 
by the minimization of the interval width,
is the use of the Bayesian credible intervals
for finding the optimal division and presenting the interval obtained
by the non-Bayesian method for this division.
Optimization by a different interval-finding method,
in a simplistic explanation,
chooses the interval which is not always the shortest
for a given method, thus improving the optimized coverage of the given method.
This optimization
should not necessarily mix the zero-background channel
with others.
This method with modified (see Section \ref{Intervals_sect}) 
Bayesian credible intervals
is used throughout this work.

If for particular conditions
the lower limit is close to zero or almost always zero,
it can be non-informative for the purpose of division optimization. 
In particular, this can happen  if the confidence required
is very large, for example 99.9\%
for the test example studied 
in this work.
The optimization by width is then reduced to
the optimization by the upper limit and one can predict
the lack of coverage as if optimized by the single upper limit. 
If the problem is caused by the extremely 
high confidence level,
the optimization can be performed
by the intervals obtained with a smaller level.
Otherwise it is assumed, though not tested, that
another criterion that indirectly indicates
the distribution widths should be employed.

\section{The Bayesian approach}
\label{Bayesian_approach_sec}
\subsection{The Bayesian probability density}
\label{Bayesian_prob_dens}

There is a lot of discussions in the literature devoted to
an introduction to the Bayesian approach.
For the purpose of this paper let us formulate it
in the following way.
Let us assume that
the unknown parameter $s$ is a random value with the distribution $\pi( s )$,
and it is unknown which particular value occurs at the time and the place
of the experiment.
Let us assume that
the probability of the  observable $\vec{n}$ depends on $s$ and can be written
as $P( \vec{n} | s)$.
Then, the well known relations 
$p( s, \vec{n} ) = p( s | \vec{n} ) P(\vec{n}) = P( \vec{n} | s) \pi( s )$
and $P(\vec{n}) = \int P( \vec{n} | s ) \pi( s )  \, \mathrm{d}s $
indicate that
if there is a set of experiments with $s$ distributed according to $\pi(s)$,
the subset with $\vec{n}$ obtained ($P(\vec{n}) \neq 0$) has $s$ distributed
according to
\begin{eqnarray}
p( s | \vec{n} ) = \frac{ P( \vec{n} | s) \pi( s ) } 
{ \displaystyle \int P( \vec{n} | s ) \pi( s )  \, \mathrm{d}s  } \, .
\label{BayesForm}
\end{eqnarray}
This formula, and a similar formula in discrete notations, 
is traditionally referred to as ``Bayes' theorem'' 
(See, for example, \S 8.7 of Ref. \cite{Kendall_1} and Ref. \cite{Kendall_2B},
or, for example, \S 8 of Ref. \cite{Kovalenko_73} and 
\S 2.2.4 of Ref. \cite{James} 
with similar derivations of 
the discrete form of this theorem and discrete examples with a similar 
interpretation, which is often underestimated, especially for 
the continuous variables).

The limits of integration here and later can be chosen 
either from $-\infty$ to $+\infty$
with $\pi(s) = 0$ at $s < 0$, or from $0$ to $+\infty$.

In contrast,
in the frequentist approach the parameter of interest
is regarded as a constant from the viewpoint 
of the experimenter \cite{Neyman_37}.
Frequentists do not make any statements about the probability of
the unknown value 
\cite{Kendall_2B,Neyman_37}\footnote{
We do not consider here extreme cases of
the infinite intervals with 100\% confidence level or of
{\it a priori} empty intervals outside the working range.
Such intervals are useless.
Though, according to Ref. \cite{DAgostini_2010}, all
frequentist intervals are useless.}.

The Bayesian approach allows us to tell about the probability of the unknown.
The Bayesian probability for $s$ to fall within a particular interval
$[s_{\mathrm{L}}, s_{\mathrm{U}}]$ 
is given by 
$\int_{s_{\mathrm{L}}}^{s_{\mathrm{U}}} p( s | \vec{n} ) \, \mathrm{d}s$.

\subsection{The prior distribution for the parameter of interest}
\label{Prior_for_param_int_sec}

In the absence of prior information, we cannot give any preference
to any specific value of $s$. This idea is converted into 
the uniform or flat distribution $\pi( s )$ 
(\S 8.19--8.20 of Ref. \cite{Kendall_1})
with the exception that in our case it has to be zero at $s < 0$. 
This solution is known to be not unique, if Eq. (\ref{BayesForm}) can be 
rewritten
as a function of
some other variable $r$ with a non-linear 
(non-unique in the discrete case) relation between $s$ and $r$.
If $\pi(r)$ is taken as uniform too,
the probability $p(s) \, \mathrm{d}s$ expressed in terms of $s$
will not be identical in the general case to the same probability
$p(r(s)) |r(s + \mathrm{d}s) - r(s)| = p(r(s)) |r^{\prime}(s)| \, \mathrm{d} s$
obtained through $r$.

This ambiguity was a subject of long 
debate \cite{WhyNotBayesian,Kendall_2B,James,Kendall_1,KassWasserman}.
In order to obtain the identical result one has to use 
a non-uniform prior for $r$
which assures that $\pi(s)$ (which is constant in the given case) is
proportional, according to the known ``change-of-variables formula'', 
to  $\pi(r(s)) |r^{\prime}(s)|$ 
(\S 5.35 of Ref. \cite{Kendall_2B}, \S 8.25 of Ref. \cite{Kendall_1}).
For some classes of transformations $r(s)$, such as $r = s^g$, 
where $g$ is any non-zero power,
one can find ``invariant'' priors
that do not need to be changed
to assure the constant results for these specific transformations.
For the example of $r = s^g$ this is  $1/s$ \cite{WhyNotBayesian}.
This creates an illusion that there is no need to select
any specific parametrization.

However, there is no prior which is invariant in this sense 
for {\it any} possible transformation.
Furthermore, there is no strong argument why the prior used in the analysis
should be invariant at all.
If a particular form of prior provides the same physical results
for any parametrization,
this does not make it special in any other sense except this.
Indeed, 
the frequency interpretation
of the Bayes theorem described in the previous section 
implies the dependence on the prior anyway.
The credible intervals 
depend on the prior too.

There are many proposed alternative priors that
depend on the shape of likelihood and on auxiliary parameters or 
their measurements. For example, ``Jeffreys' general rule'' 
\cite{KassWasserman} leads to the so-called ``reference priors'', 
which vary according to
the change-of-variables formula and
depend on the shape of likelihood. 
This results in a strange dependence
of the ``prior knowledge'' 
on particular experimental features, such as resolution.
Some considerations are discussed in 
the introduction of Ref. \cite{Choudalakis_2011}
and in Ref. \cite{DiscussionWith_2011}.
 
The uniform prior is invariant
at the transformation $s = r+b$ with any $b$ \cite{KassWasserman}, 
but the condition
$s \geq 0$ is neglected here.
For a given problem there is neither need nor useful interpretation
of any transformation of $s$ to any other variable.
The uniform prior does not shift the most probable value of $s$
from the maximum likelihood value, thus preventing ambiguity, 
as to which of them
is more ``most probable''.
Given the probability density for the uniform prior
one can easily extract forecasts
for any non-uniform prior by simple multiplication and 
renormalization. 
 
Because of all these considerations the uniform prior was used for $s$.

In the case of known $\vec{a}$ and $\vec{b}$ 
the value of $P( \vec{n} | s)$, 
necessary for calculations by
Eq. (\ref{BayesForm}), is simply expressed through
$P(\vec{n}| t_a \vec{a} s + t_b \vec{b})$.

\subsection{The case of unknown $\vec{a}$ and $\vec{b}$}
\label{BayesUnknowmNuisance_sec}

If $\vec{a}$ and $\vec{b}$ are determined in an auxiliary
experiment with finite precision,
one can use their results $\vec{n}_a$ and $\vec{n}_b$ 
as the first approximation to $\vec{a}$ and $\vec{b}$.
Then, instead of $P(\vec{n}| s)$ in Eq. (\ref{BayesForm}) one should use 
$P(\vec{n}| t_a \vec{n}_a s + t_b \vec{n}_b)$.
Numerical checks have shown that for the considered example this approximation
does not work.

The probability $P(\vec{n} | s)$ can be more accurately expressed according to
the complete probability formula by the convolution
with probability densities 
of parameters:
\begin{eqnarray}
P( \vec{n} | s) = \iint
P(\vec{n} | s, \vec{a}, \vec{b}) \,
p(\vec{a} | \vec{n}_{a})  \,
p(\vec{b} | \vec{n}_{b}) \, \mathrm{d} \vec{a} \, \mathrm{d} \vec{b} \, .
\label{BayesForm_1}
\end{eqnarray}

The probability densities of  $\vec{a}$ and $\vec{b}$
can be reconstructed from the auxiliary measurements and
expressed by Bayes' formula too:
\begin{eqnarray}
 p(\theta_{i} | n_{\theta i}) = \frac{ P(n_{\theta i} | \theta_{i} ) \, 
   \pi( \theta_{i} ) } 
{ \displaystyle \int P(n_{\theta i} | \theta_{i} )  \, 
\pi( \theta_{i} )  
  \, \mathrm{d} \theta_{i}  } \, ,
\label{BayesForm_2}
\end{eqnarray}
where $\theta$ stands for $a$ or $b$.
The limits of integration are 
either from $-\infty$ to $+\infty$
with $\pi(\theta_i) = 0$ at $\theta_i < 0$, or from $0$ to $+\infty$.

After substitution of Eqs. (\ref{BayesForm_2}) 
into Eq. (\ref{BayesForm_1}), 
and 
the result into Eq. (\ref{BayesForm})
the denominators in Eqs. (\ref{BayesForm_2}) 
are canceled
and the result appears to be
\begin{eqnarray}
p( s | \vec{n}, \vec{n}_a, \vec{n}_b ) = \frac{ N(s) } 
{ \displaystyle \int N(s)  \, \mathrm{d} s  }  \, ,
\label{BayesForm_3}
\end{eqnarray}
where
\begin{align}
N(s) = \iint &
P(\vec{n}, | s, \vec{a}, \vec{b})  \, \pi( s )\,  \nonumber \\ 
\times & P(\vec{n}_{a} | \vec{a} ) \, \pi( \vec{a} )\, 
P(\vec{n}_{b} | \vec{b} ) \, \pi( \vec{b} ) \, \mathrm{d} \vec{a} 
\, \mathrm{d} \vec{b} \, .
\label{BayesForm_4}
\end{align}
A similar formula in different notation, with different nuisance parameters
and initially, as a rule, with a non-facto\-rized prior
appears in many sources,
see, for example, 
\S 3.5 of Ref. \cite{Kendall_2B},
Refs. \cite{Heinrich_04,Choudalakis_2011}, 
and \S 1(b) of Ref. \cite{Neyman_37}.
In our case the prior is automatically factorized.
Interestingly, in these resulting formulas
there is no evident difference between
the roles of the main and the auxiliary experiments,
in contrast with their roles in its derivation.
Therefore this approach can technically be used for more general problems,
for example, when the background auxiliary experiment contains 
a small admixture of signal.

Equations (\ref{BayesForm_3}) and (\ref{BayesForm_4}) mean that
if there is a set of experiments
with $s$ distributed according to $\pi(s)$,
$\vec{a}$ according to $\pi(\vec{a})$ and
$\vec{b}$ according to $\pi(\vec{b})$,
the subset with obtained $\vec{n}$, $\vec{n}_a$, and $\vec{n}_b$
has $s$ distributed
according to these equations.

These equations also show,
on careful investigation, that
in the assumed presence of the background 
any prior $\pi(s)$ 
equal to any negative power of $s$ (thus infinite at  $s = 0$) 
results in the infinite posterior probability density at  $s = 0$.
Forbidding these strange posteriors means forbidding
such priors, which gives an additional argument
for the use of the uniform main prior (c.f. \S 6.30 of Ref. \cite{Kendall_2B}).

\subsection{The prior distributions for the nuisance parameters}

Table 1 shows the parameters of resulting distributions obtained 
by 
Eq. (\ref{BayesForm_2}).
\begin{table}[t]
\centering
\caption{Parameters of Bayesian posterior probability density distributions
for the Poisson distribution of observations with different priors. 
}
\begin{tabular}{|c|c|c|c|} \hline
  prior  &  mean & $\sigma^2$ & maximum     \\   \hline
uniform 		& $n+1$	& $n+1$ & $n$ \\  
$1 / \sqrt \mu$		& $n+0.5$ & n + 0.5 & $\max(n-0.5, 0)$ \\   
$1 / \mu$		       & $n$	& $n$ & $\max(n-1, 0)$  \\  
\hline  
\end{tabular}
\label{Param}
\end{table}
It may seem surprising that the mean of the Poisson parameter $\mu$ 
($a_i$ or $b_i$ in the case of Eq. (\ref{BayesForm_2})),
if restored with the help of the uniform prior from a single measurement,
is not equal to this measurement (in contrast with $\overline{n} = \mu$
for a range of measurements with fixed $\mu$),
but exceeds it by unity.
The equality to the measurement
is obtained only with the inverse prior (that is $1 / \mu$, which is denoted 
by this term everywhere in this paper unless otherwise specified), 
but in this case
the most probable value is smaller than the measurement
by unity (with an exception of zero measurement).

After many repetitions of the experiment
the average of $n$, $\overline{n}$, 
will be equal to the corresponding true parameter.
But the effective reconstructed nuisance parameter will always
overestimate the true value if the analysis is more susceptible to
the average assumed nuisance parameter
and the uniform prior is used,
or will always underestimate it if 
the analysis is more susceptible to
the most probable nuisance parameter
and the inverse prior is used.
It is easy to understand that if the background is overestimated 
or underestimated,
lesser or greater $s$, respectively, is enough to describe the observed data.
The same is true for the acceptance.

In the data analysis either averages or the maxima appear to be more important,
depending on the conditions. Intermediate cases are of course possible too.
Therefore, 
in order to provide the 
coverage
of interval boundaries for all conditions,
in the case of the Poisson distributions
the inverse priors should be used for calculation of the upper limit
and the uniform priors for the lower limit.
In a more general form, presumably applicable for a broader class of problems,
this empirical rule demands that
the upper (lower) limit be obtained with such nuisance priors that guarantee
that both the mean and the maximum of the posterior 
nuisance-parameter distributions 
are as close as possible but
not greater (not less) than the corresponding auxiliary measurements.
Such priors will here be called ``safe priors'' or ``safe nuisance priors''.

The frequentist methods that depend on the priors (some of these methods are
sometimes called hybrid or mixed in the literature, 
but we call them frequentist if they treat the parameter-of-interest
in the frequentist way; the methods that treat all parameters in the frequentist way
can use priors for the calculation of the test statistic)
are more complex than the Bayesian
method,
and the priors
affect them in a more complex way, depending on the particular method.
But, surprisingly, the result of this influence is, as a rule, similar,
and they therefore need the same priors except for a few methods
that produce good upper boundaries
both with inverse and with uniform priors.

If allowed data distributions
are limited somehow, the difference between priors and the 
width of confidence intervals can perhaps be reduced
for Bayesian, as well as for some frequentist methods.
For instance, the allowed values of the parameter $q$ from Section 
\ref{Parameters_sect} can be limited.
If $q$ is not limited,
the only alternative choice of priors found so far
is the safe (uniform) prior for the lower limit and
the safe (inverse) prior at $n_a=0$ or $n_b=0$ for the upper limit.
At $n_a>0$ or $n_b>0$ for the upper limit
the prior should be equal to $1/\mu^{g(n)}$, where  
$g(n) \approx 0.66669 + 0.01957 / n$, and $n$ stands for $n_a$ or $n_b$.
This $g(n)$ ensures that
the median $\mu$ of the resulting posterior distribution is 
equal to the observed $n$.
Such ``hybrid median priors'' can be used, but they
depend on the measurements and
do not strongly improve the resolution.

Heinrich \cite{Heinrich_05} mentioned
earlier some pathologies 
caused by the uniform prior in the Bayesian analysis.
He found that an overestimate of $a$ and $b$ (in our notations)
led to an underestimate of $s$ for many channels.
He concluded that
the inverse priors ``are matched to this Poisson case''.
However, our calculations confirm this only for the upper interval boundary,
for which ``hybrid'' priors can probably be applied too.

Equation (\ref{BayesForm}) with substitution 
of Eqs. (\ref{BayesForm_1}) and (\ref{BayesForm_2}) 
or Eqs. (\ref{BayesForm_3}) and (\ref{BayesForm_4})
give the probability density of the parameter.
Obviously, the maximum of this function gives the most probable signal rate,
which at $\pi(s) = \mathrm{const}$ coincides with the maximum likelihood
$P(\vec{n}| s)$ integrated over nuisance parameter distributions. 
To calculate the most probable parameter of interest the researcher can
compute
the arithmetic average of the most probable signal rates
calculated with safe (or median) nuisance priors
for the lower and for the upper limits.
The average of this average over many pseudo-experiments
appears to be very close to the true parameter.

\subsection{Numerical calculations}
When the prior distribution is
$1/\mu^g$, $0<g<1$,
the multiplication of this prior by
the Poisson distribution for zero observation diverges at $\mu \rightarrow 0$.
To get useful result from Eq. (\ref{BayesForm_2})
such a prior is considered starting not from zero,
but from a very small threshold, actually from $10^{-100}$ to $10^{-300}$ for this paper, 
and is considered equal to zero below this limit.
Most of the results
do not vary noticeably when varying the order of this power
and hence do not depend on this specific choice.
The typical exclusion is the significance 
obtained by some methods for the one-channel problem with measured $n_b=0$,
for which the limit at zero threshold (infinity) should be given.
Similarly, all priors, inverse as well as uniform,
are cut at some large enough value of $s$, above which all probability 
densities are effectively zero anyway.

Fast and accurate calculations of integrals 
in Eq. (\ref{BayesForm_4}) pose a complex problem,
but this is outside the scope of this paper.

\subsection{The credible intervals}
\label{Intervals_sect}

\begin{figure}[t]
\centering
\includegraphics[width=1.0\linewidth]
{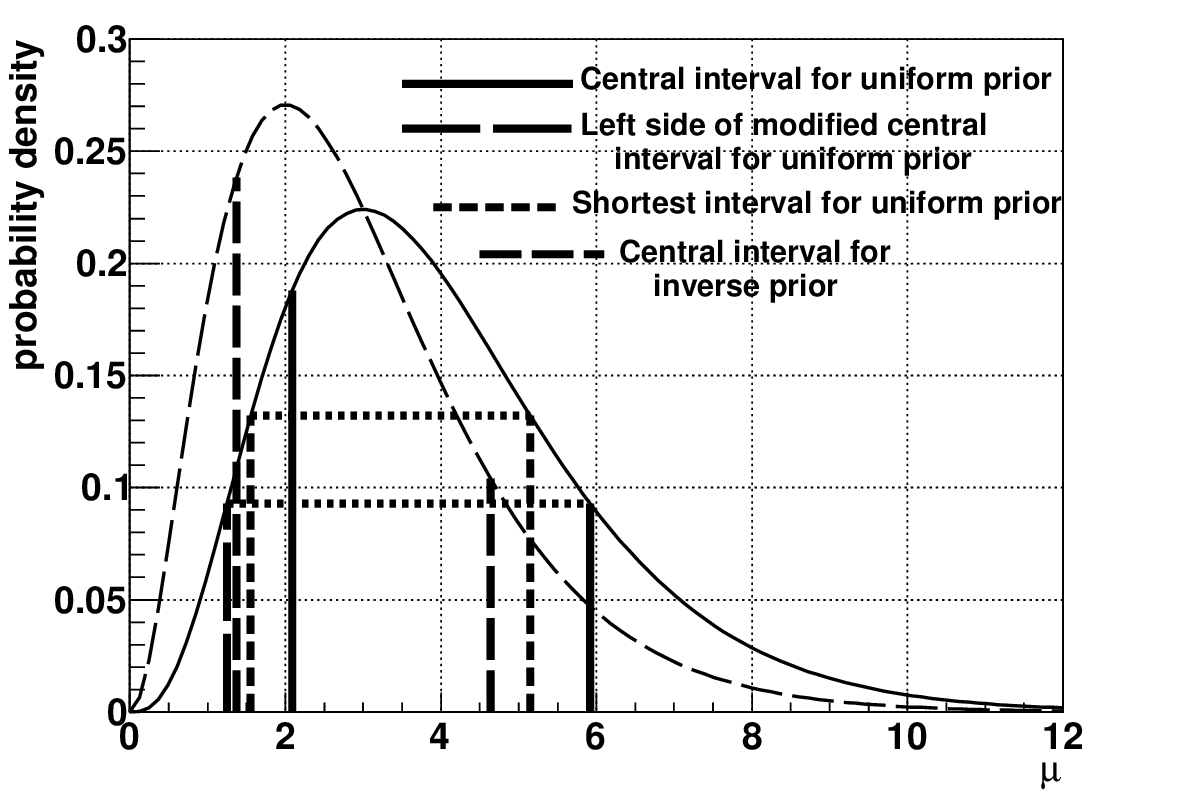}
\caption{
Probability densities for $\mu$ of the Poisson distribution 
(the one-channel problem without background),
if 3 events are observed. 
The solid (dashed) smooth line depicts the probability density distribution
for the uniform (inverse) prior.
The intervals calculated with $\alpha = 0.158655$
are shown by solid and dashed thick lines with different length of dashes.
Dotted lines show horizontal levels.
}
  
\label{CheckModHup_np_3}
\end{figure}

The most frequent and obvious choice of intervals are the so-called
central intervals \cite{WhyNotBayesian}, 
which are defined by cutting off left and right
tails with equal areas, see Fig. \ref{CheckModHup_np_3}. 

If the area cut from each side of the distribution is denoted by $\alpha$ 
(and restricted by $\alpha<0.5$),
the lower and upper boundaries 
$s_{\mathrm{L}}$ and $s_{\mathrm{U}}$ are defined by
\begin{equation}
\displaystyle \int_{s_{\mathrm{L}}}^{\infty} p( s | \vec{n} ) 
\, \mathrm{d} s  = 
1 - \alpha,  \ \  
\displaystyle \int_{0}^{s_{\mathrm{U}}} p( s | \vec{n} ) 
\, \mathrm{d} s  = 1 - \alpha.
\label{BayesForm_6}
\end{equation}
Cousins \cite{WhyNotBayesian} showed 
that for the one-channel Poisson measurement with known nuisance parameters 
the use of the uniform prior for the main parameter
results in an upper limit that covers the true value
exactly with the stated probability and in a lower limit
that covers with lower probability.
Conversely, with the use of the inverse prior
the lower limit covers correctly and 
the upper limit insufficiently.
Another problem is that if the most probable $s$ is zero or close to zero,
its value can be excluded from the credible interval,
which raises doubts about the consistency of the whole approach.

\begin{figure*}[t]
\centering
\includegraphics[width=0.7\linewidth]
{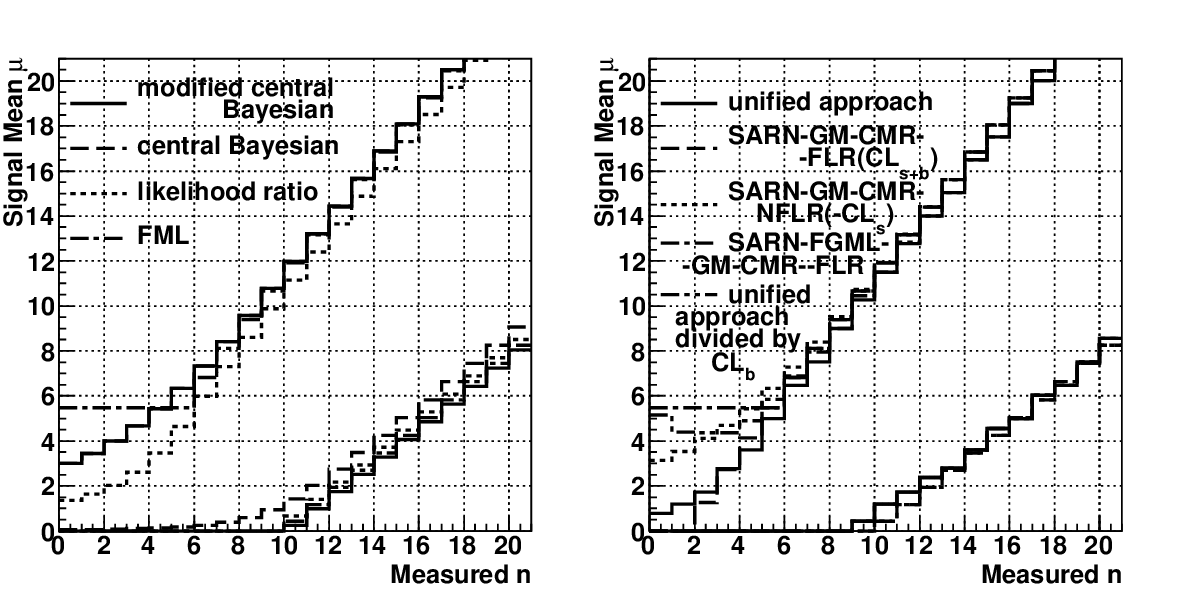}
\caption{
The confidence intervals for the one-channel problem
with known auxiliary parameters $a=1$ and $b=5$, $t_a = t_b = 1$,
for different observed $n$ by different methods.
The upper limit of the central Bayesian interval is not seen because
it coincides by definition with
the upper limit of the modified central Bayesian interval.
The lower limit of the central Bayesian interval
is the highest among the other lower limits.
The upper limit for SARN--GM--CMR--FLR ($CL_{s+b}$) is very
close to zero for $n$ equal to $0$ and $1$, but is not zero.
The lower limit for SARN--GM--CMR--NFLR (LHC $CL_{s}$) 
coincides by definition with that of  SARN--GM--CMR--FLR ($CL_{s+b}$) 
and both are lower than the lower limit for the unified approach.
The lower limit of the unified approach coincides by definition
with that of the unified approach divided by $CL_{b}$.
The upper limit for the latter is defective because
it increases for reducing $n \leq 3$.
The lower limit of SARN--FGML--GM--CMR--FLR turns out to be numerically equal
to that of SARN--GM--CMR--NFLR in this problem.
}
\label{compar_lim}
\end{figure*}

For example, 
the confidence intervals
calculated by different methods for a one-channel problem
with known auxiliary parameters $a=1$ and $b=5$, $t_a = t_b = 1$,
for different observed $n$ are shown in Fig. \ref{compar_lim}.
One can see that the lower limit of the Bayesian central credible interval
does not include $s=0$ even at $n < 5$, although it is very low.  

The shortest interval includes the maximum due to the method of 
its construction,
but does not provide good coverage, as was shown in \cite{WhyNotBayesian} and
obtained also for the examples considered herein.

One can also construct a level-based interval
as it is done in the likelihood
ratio method, that is using a
level found with the Gaussian approximation.
Writing the integral of the Gaussian in the form
\begin{eqnarray}
F(z) = \frac{1}{\sqrt{2\pi}} \displaystyle \int_{-\infty}^{z}e^{-t^2/2}
\, \mathrm{d}t
\label{IntGaussian}
\end{eqnarray}
one obtains $z$ for given p-value, which is the same as $\alpha$
in our notations, by $z = F^{-1}(1-\alpha)$.
If  $\hat{s}$ is the most probable value that maximizes
$p( s | \vec{n} )$, see Eq. (\ref{BayesForm_3}) or (\ref{BayesForm_4}),
the interval boundaries are set at the probability density
smaller by the factor of $e^{-z^2/2}$.
One has to find the lowest and the uppermost boundaries 
$s_{\mathrm{L}}$ and $s_{\mathrm{U}}$ such that
$p( s_{\mathrm{L}} | \vec{n}) = p( s_{\mathrm{U}} | \vec{n}) = 
p( \hat{s} | \vec{n})  e^{-z^2/2}$.
If non-negative $s_{\mathrm{L}}$ does not exist, it is set to zero.
These intervals
appear to be close to the shortest intervals 
shown in Fig. \ref{CheckModHup_np_3}
and have the same benefits and drawbacks.

The use of the right boundary of the central interval with the uniform prior
and the left boundary of the central interval with the inverse prior
(see Fig. \ref{CheckModHup_np_3}) 
provides coverage, but does not always provide the inclusion of
the most probable value. The left boundary can never be exactly zero.

But all mentioned problems are solved
if one takes the right boundary from the central 
interval Eq. (\ref{BayesForm_6})
computed with the uniform prior,
and sets the left boundary at the same level of probability density
as that for the right boundary, see
Fig. \ref{CheckModHup_np_3}.
Graphically, one should draw a horizontal line from the upper edge of
the right boundary 
to
the left till its crossing with the distribution.
If the non-negative 
$s_{\mathrm{L}}$ 
satisfying this condition does not exist, 
it is equated to zero.
If the left boundary of the classical central interval 
is lower 
for any reason, it has to be used instead of this modified boundary. 
Calculations of the simple one-channel problem 
without background for various $n$
show that 
this modified boundary is always lower and usually almost equal to 
the lower boundary of 
the central interval for the inverse prior,
which allows one to conclude that it should not undercover.
Hence these ``modified central intervals'' cover by both
ends for this simple problem. 

In this method the probability of violating the lower limit 
can be smaller than $\alpha$.
In the case of small signal it can even be zero.

\subsection{The coverage and width
of modified central intervals}
\label{sec_Bayes_cover}

\begin{figure}[t]
\centering
\includegraphics[width=1.0\linewidth]
{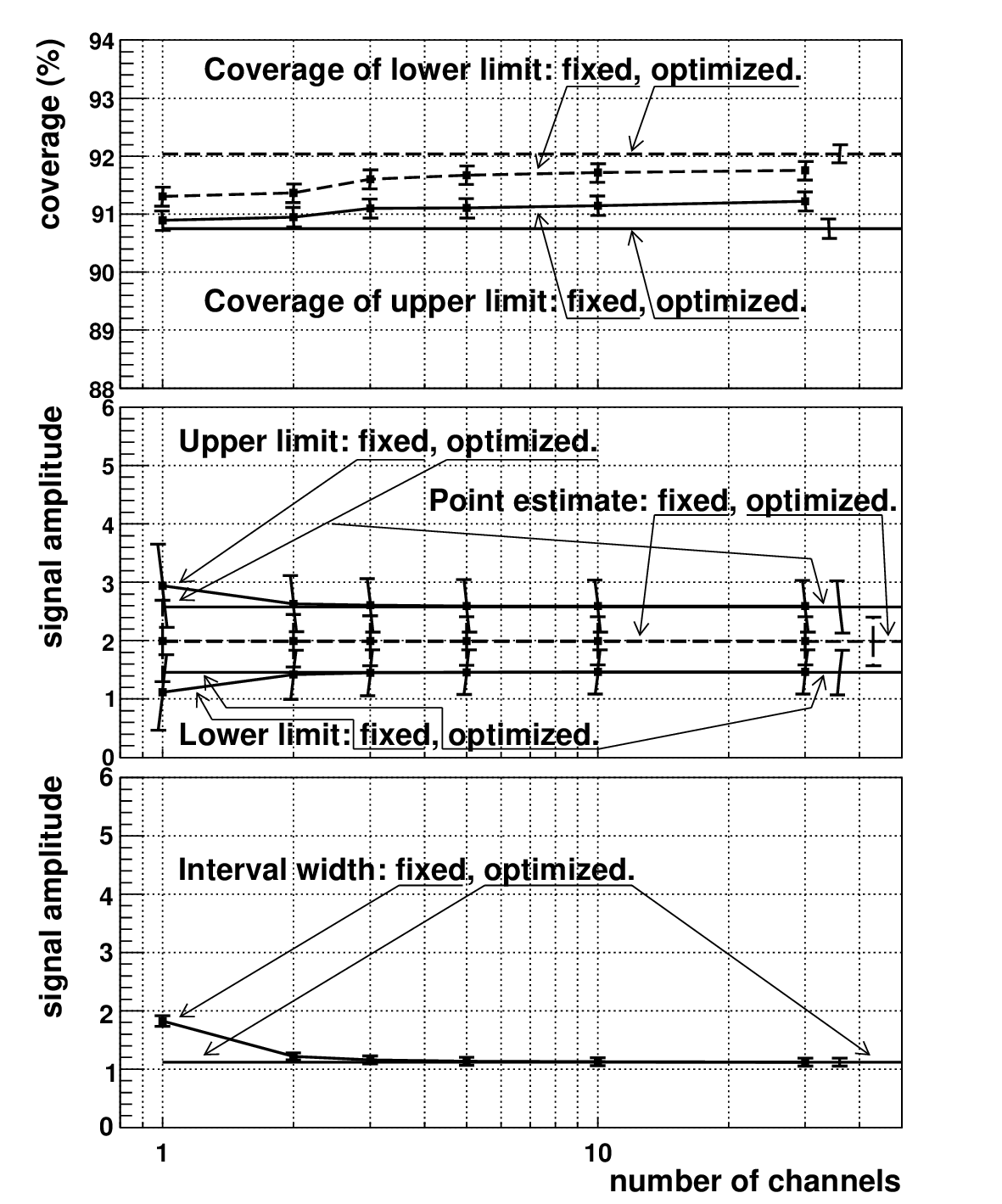}
\caption{
The Bayesian method, the modified central intervals for 90\% ($\alpha = 0.1$),
no nuisance parameter uncertainties.
The horizontal straight lines are results of optimization by interval widths.
The optimization by separate interval boundaries results in
coverage of about 83\%.
}
\label{Bayes_center_mod_no_unc_0}
\end{figure}

If nuisance parameters are known,
the coverage of the Bayesian modified central intervals
is provided for all fixed divisions, 
see Fig. \ref{Bayes_center_mod_no_unc_0}.

In this figure and in the following ones
the $x$-axis is identical in all three plots
and represents the number of channels. 
The $y$-axes are different.
In the top plot the $y$-axis shows the probability of coverage in percent.
In the two lower plots $y$-axes are measured in the units of
the total signal rate (or ``signal amplitude'') $s$, 
but have different meanings.
In the middle plot the $y$-axis means the position of the
interval boundaries (i.e. the confidence limits) or of the point estimates.
Since two different values are plotted in a single plot, 
the axis is labeled by the unit of their measurement ``signal amplitude''.
In the lower plot it means the interval widths and labeled similarly.

The points located at 1, 2, 3, 5, 10 and 30 channels and connected by lines 
display the results for
fixed divisions. 
To show that the optimized-division results are not linked to
a certain fixed number of channels, the optimized results 
are plotted as horizontal 
straight lines going beyond the used 30-channel limit with single error bars
positioned somewhere at a larger number of channels (this position does not have 
any other meaning except showing that this is not the real number of channels).
Recall that a different division can be chosen for each experiment,
when the division is ``optimized'' using the information available in
the particular experiment.

The points connected by the 
solid and dashed 
lines in the upper plot
represent the coverage of the upper and lower interval boundary, respectively.

In the middle plot
both the upper and lower limits are shown by the points connected by
solid 
lines.
The point estimates (the maxima of the Bayesian posterior)
are the points connected by the 
dashed 
line. Obviously, the latter reside between the former.

The error bars in the uppermost coverage plots indicate the 
uncertainty of calculations for the standard 68\% confidence level.
These are frequentist uncertainties for the binomial distribution
at the given number of experiments \cite{ROOT}.
These uncertainties appear owing to the limited statistics 
of Monte Carlo simulations performed for this paper. 
In this particular plot they are very small
due to very large simulated statistics.
The Bayesian analysis without nuisance uncertainties is very quick.

The points and horizontal lines in the two lower plots
show the arithmetic averages of the respective values over many experiments.
All errors drawn in the two lower plots
express the fluctuations of the respective values
occurring 
experiment by experiment. 
To make the image more clear, the bars corresponding to
upper limits are slightly inclined to the left, and
the bars corresponding to lower limits to the right.
The same inclination (not seen clearly 
in Fig. \ref{Bayes_center_mod_no_unc_0} because the bars are too short)
is present also in the coverage plots.
The errors in the two lower plots are calculated as
the root-mean-square deviations and hence correspond to the standard 68\% 
confidence level too.

Interestingly enough, the coverage for fixed numbers of channels 
presented in the uppermost plot,
is almost constant and stays near 91\%
for all divisions for the studied example. 
But the interval widths and boundaries reach the plateau
starting from 2 channels.
For this case without nuisance uncertainties
all the other reasonable methods behave similarly.
According to similar calculations
the lower boundary of the Bayesian central intervals (not modified) does not
provide the stated coverage, as expected.

Thus, for the case with known nuisance parameters
the modified Bayesian central intervals provide frequentist coverage.

The mean point estimates in the middle plot almost coincide
with the true value of the parameter of interest, $s_{\mathrm{true}}=2$, both
for fixed and for optimized divisions, so one cannot
distinguish visually two dashed lines in this plot.

\begin{figure}[t]
\centering
\includegraphics[width=1.0\linewidth]
{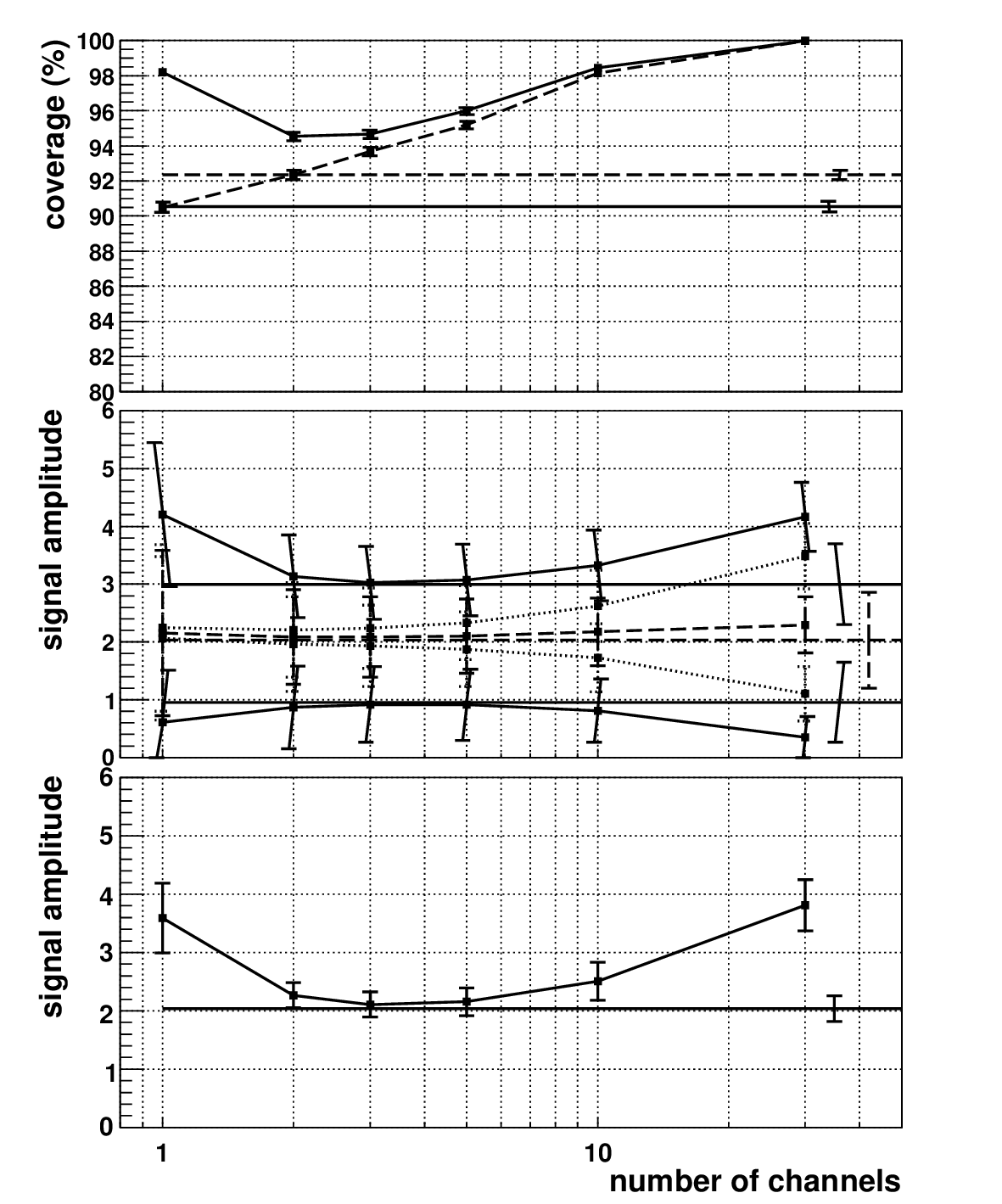}
\caption{
The Bayesian method, the modified central intervals for 90\% ($\alpha = 0.1$),
the uncertainty of the expected background, the known expected signal, 
safe nuisance priors.
The horizontal lines are results of optimization by interval widths.
Other details are described in Section \ref{sec_Bayes_cover} 
and in Fig. \ref{Bayes_center_mod_no_unc_0}.
}
\label{Bayes_center_mod_geny_anay_5_0}
\end{figure}

If the background uncertainties are switched on,
the modified Bayesian method with safe priors
behaves as shown in
Fig. \ref{Bayes_center_mod_geny_anay_5_0}. 
The same case with hybrid priors gives almost an identical
picture.
In such figures
the upper and lower limits form a valley with narrowing in the middle.
In the middle plot one can also see additional 
dotted lines, which display the mean point estimates
(maxima of the posteriors) calculated with the priors 
appropriate for the upper and lower limit. It is seen that they deviate
from the optimal position simultaneously with the
corresponding limits with the increase in the number of channels.
Obviously this divergence is entirely due to the priors.
Their arithmetic averages drawn by the dashed line for
fixed divisions and by the straight dashed line for
optimized divisions are very close to the true $s$.

\begin{figure}[t]
\centering
\includegraphics[width=1.0\linewidth]
{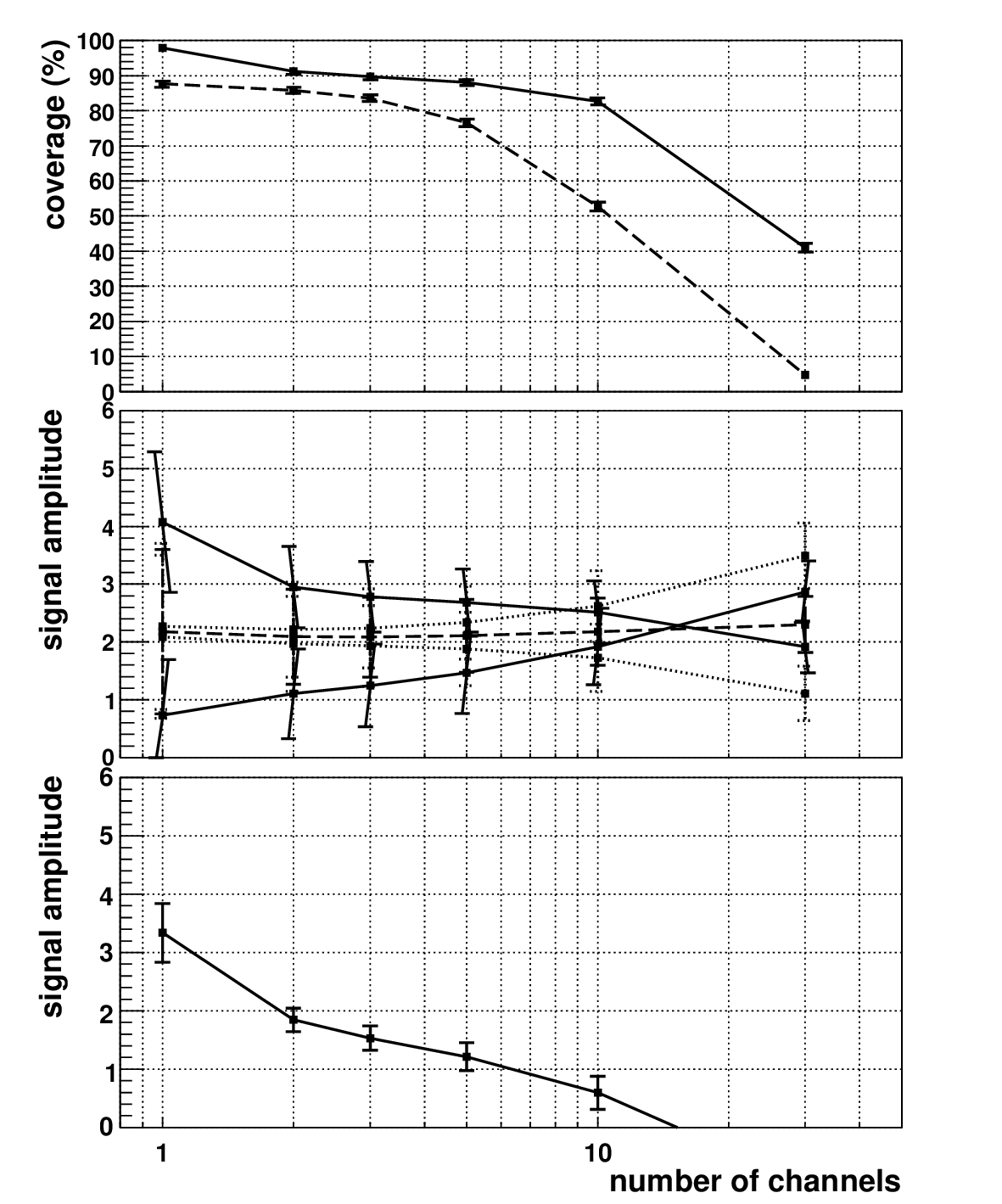}
\caption{
The Bayesian method, the modified central intervals for 90\% ($\alpha = 0.1$),
the uncertainty of the expected background, the known expected signal, 
{\it exchanged} safe nuisance priors.
Other details are described in Section \ref{sec_Bayes_cover} 
and in Fig. \ref{Bayes_center_mod_no_unc_0}.
}
\label{Bayes_center_mod_geny_anay_5_exch_pr_n4i}
\end{figure}

The same calculations with exchanged safe priors,
where the uniform prior is used for the upper limit and
the inverse prior for the lower one,
give catastrophic results shown in Fig. 
\ref{Bayes_center_mod_geny_anay_5_exch_pr_n4i}.
At more than 15 channels the lower limit becomes greater than the upper limit!

The priors $1/ \sqrt{\mu}$ produce almost exact point estimates
and not diverging limits for this example,
but at the other parameters 
they lead to deviations anyway.

As shown in Fig. \ref{Bayes_center_mod_geny_anay_5_0},
the optimization of the division
by the interval width provides almost perfect 90\% coverage
for the Bayesian case.
Obviously, the algorithm usually takes one of the
medium divisions, which provides the shortest interval for
given main and auxiliary experiments.

Thus, for the case of unknown expected background
the modified Bayesian central intervals with safe nuisance priors
provide frequentist coverage,
which is sometimes conservative.
The same intervals with hybrid nuisance priors
are almost identical.

\begin{figure}[t]
\centering
\includegraphics[width=1.0\linewidth]
{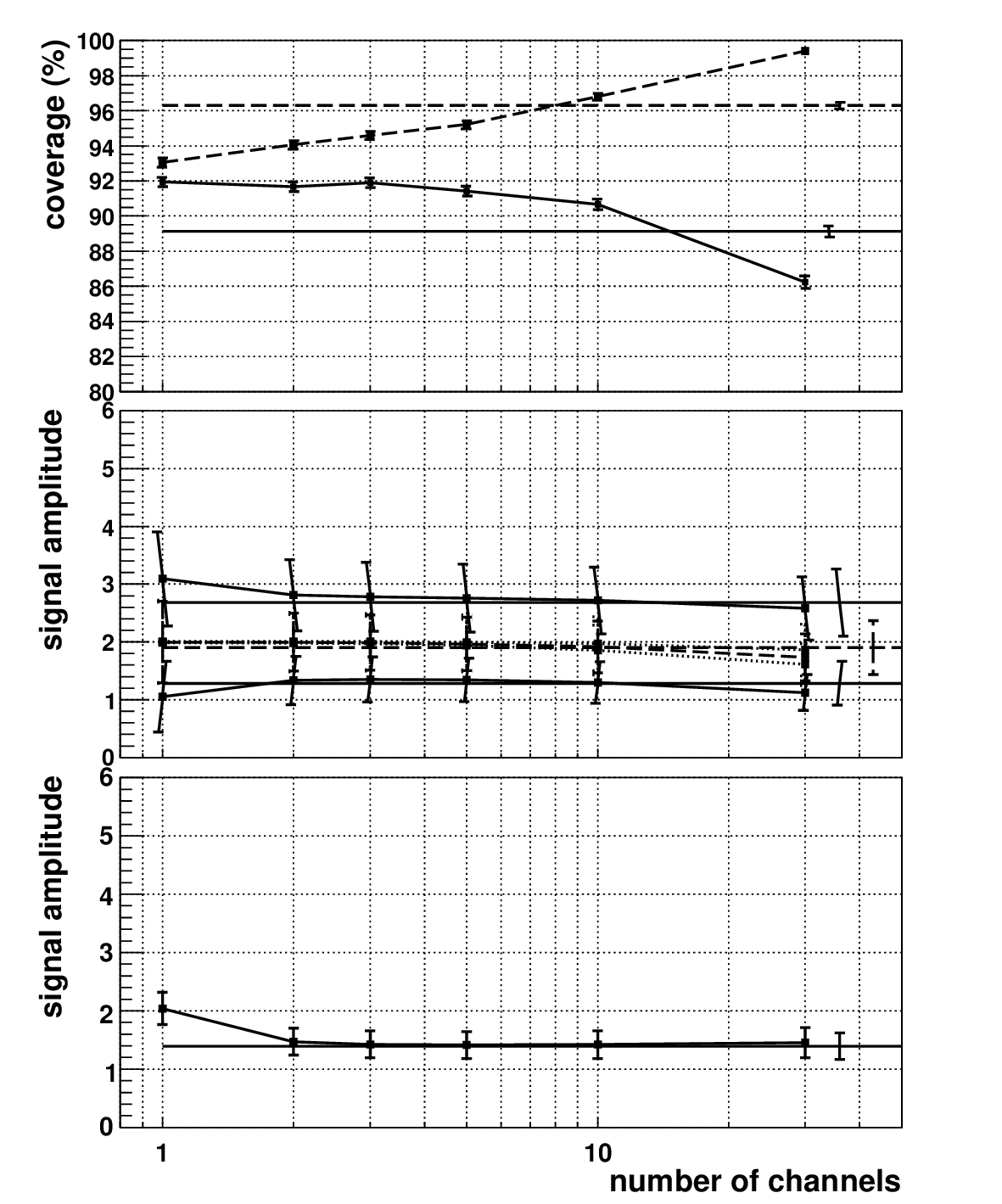}
\caption{
The Bayesian method, the modified central intervals for 90\% ($\alpha = 0.1$),
the uncertainty of the expected signal,
the known expected background, safe nuisance priors.
The horizontal lines are results of optimization by interval widths
without zeros in the expected signal.
Other details are described in Section \ref{sec_Bayes_cover} 
and in Fig. \ref{Bayes_center_mod_no_unc_0}.
}
\label{Bayes_center_mod_geny_anay_11}
\end{figure}

When only the uncertainty of the expected signal is present,
only the lower limit was found to cover the true parameter
with no less than stated probability for all fixed divisions
for this method,
see Fig. \ref{Bayes_center_mod_geny_anay_11}.

The coverage of the upper limit falls from 92\% for 1 channel to 
86\% for 30 channels. This effect is even stronger with hybrid priors.
The same effect appears in all other methods that use the safe or hybrid priors.
The exchanged priors provide even worse
coverage for the upper limit.

The reason for this pathology is simpler to illustrate for the Bayesian case.
It has similar reasons for the other cases. 
The channels having the downward fluctuation of the expected signal
and $n_{a i} = 0$
do not influence the result because in such channels
$a_i$ is distributed very close to zero due to the use of the inverse prior
and
$f_i = t_a a_i s + t_b b_i$ does not depend on $s$.
Only the rest of the channels, where $n_{a i}$ could fluctuate upward,
influence the result.
Since $a_i$ seems to be greater for such channels than it is in the average,
a smaller signal is enough to describe the observed result.
Calculations indicate that not only zero  $n_{a i}$ but also
low non-zero values of $n_{a i}$ affect the result too.
Apparently, the channels with downward fluctuations of
$n_{a i}$ are more strongly masked by the background and participate less 
in the result, than the channels with larger $n_{a i}$.

Choosing the division with the least width
without zeros in the distribution of the expected signal
allows one to obtain the upper limit
with coverage slightly smaller than requested,
as indicated by 
the horizontal solid lines in
the upper plots in Fig. 
\ref{Bayes_center_mod_geny_anay_11}.
Some small lack of coverage 
is deemed to be tolerable.

It seems unlikely that
one will ever have zero or close to zero
content in a channel of the expected-signal distribution.
This problem is more probable for the expected background.

The exchange of the priors affect less on the lower limit
in the case of the uncertain expected signal,
but it still affects strongly on the upper limit.

When both uncertainties of expected background and signal are switched on,
the behavior of all characteristics for the test example studied is
qualitatively the same as for the case only with the background uncertainty.

\section{Frequentist Treatment of\\ Maximum Likelihood\\ Estimate }
\label{sect_FML}

\subsection{Introduction, the case without uncertainties.}
\label{FML_introduction}

Ciampolillo \cite{Ciampolillo_98} and, independently,
Mandelkern and Schultz \cite{Mandelkern_00} 
recently pointed out 
that the maximum likelihood estimate of 
the parameter of interest is a good test statistic
for constructing 
frequentist confidence intervals
for Poisson measurements with known expected signal and background.
As they found, this test statistic allows one to avoid
unphysical empty or nearly empty intervals
in the case of downward background fluctuations,
from which the frequentist analyses with other test statistics 
suffer\footnote{Ref. \cite{Mandelkern_00} does not 
consider nuisance parameters at all.
In Ref. \cite{Ciampolillo_98} only one sentence about them is found.
It recommends maximizing the total likelihood over the nuisance background.}.
It can be added that obtaining limits for the parameter of interest
by testing this very parameter is more straightforward, as well as convenient,
than doing this by testing another variable, such as a likelihood ratio,
whose behavior is difficult to predict in practical situations.

Here we call this method ``FML'', 
which means ``Frequency of Maximum Likelihood''.

\begin{figure}[t]
\centering
\includegraphics[width=1.0\linewidth]
{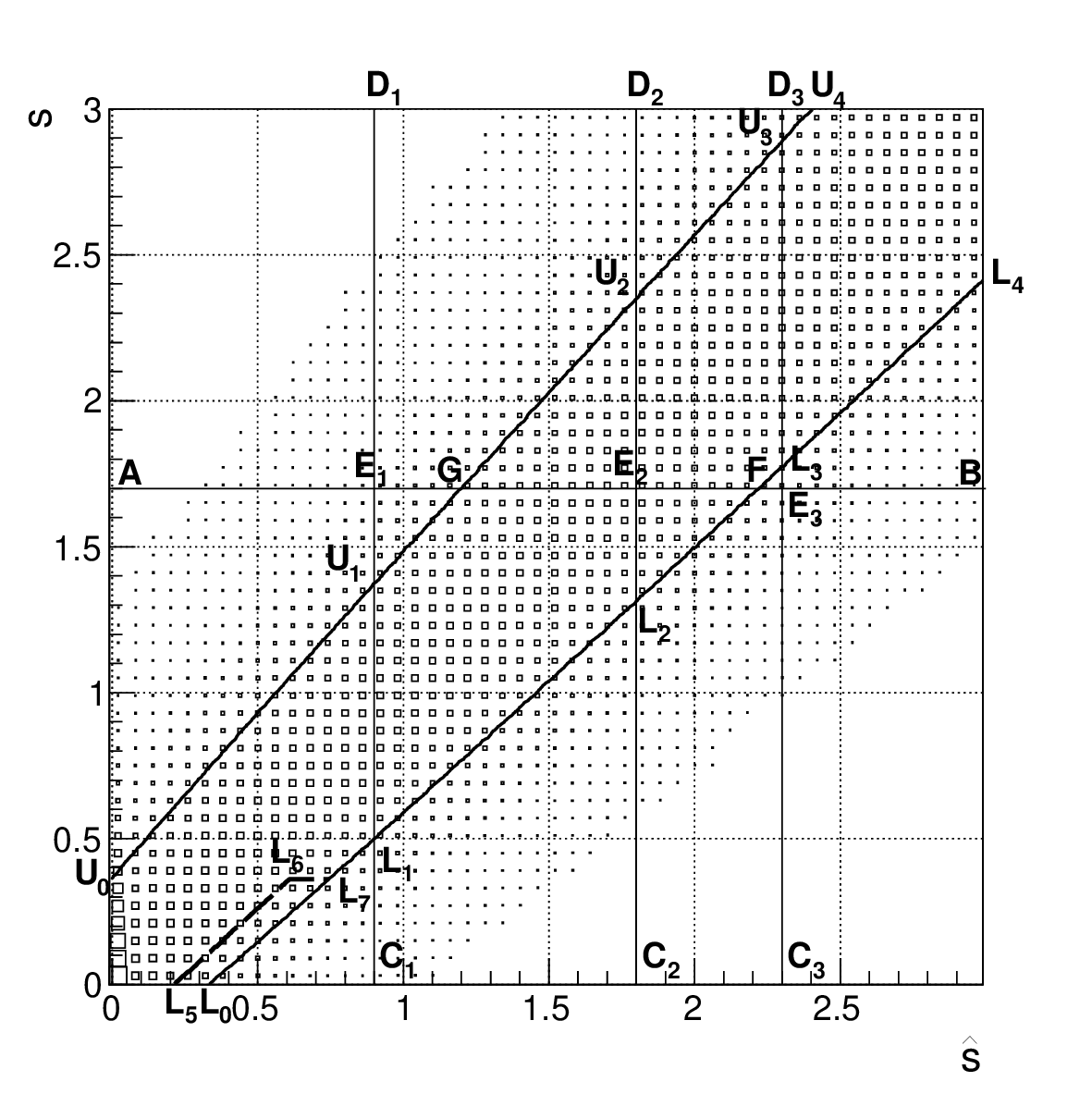}
\caption{
Distribution of the observed most probable $s$ (denoted as $\hat{s}$)
as a function of the true $s$
for 5 channels 
with known expected signal and background.
See text for other notations.
}
\label{FMLConfBelt}
\end{figure}

The typical confidence belt for FML is shown in Fig. \ref{FMLConfBelt}.
This figure depicts the case of 5 channels with standard parameters
for 10\% one-sided confidence level with known 
expected signal and background.
The notation $\hat{s}$ means the value of $s$ that maximizes 
$P(\vec{n} | s)$, which is here equal to 
$P(\vec{n}| t_a \vec{a} s + t_b \vec{b}) = \prod_{i=1}^{k} \mathrm{Poisson}(n_i, t_a a_i s + t_b b_i)$.
It is assumed that $\hat{s}$ is searched for in the non-negative interval
$[0, \infty[$.
For each assumed or possible $s$ we can 
simulate a set of pseudo-experiments and obtain the distribution of
$\hat{s}$. These distributions are shown in this figure by the
horizontal rows of 
boxes with variable size.

After choosing a specific value of $s$ for the current trial 
one generates a set of pseudo-experiments with it.
This process will be called subgeneration, 
in order to distinguish it from
the generation of ``real'' experiments. In this work
the latter are simulated by the Monte Carlo method too,
but this is done in the separate main program
with the {\it true} parameters (which is not the case for
subgeneration, see Section \ref{Parameters_sect}).

The 
probability density distribution
of 
$\hat{s}$ at given $s$ in the ``subgenerated'' experiment
is denoted by $p( \hat{s}_{\gamma} | s )$.
The index $\gamma$ is included in order to
indicate that the result is obtained by subgeneration. 
The integrals 
\begin{eqnarray}
\displaystyle \int_{\hat{s}_{\mathrm{right}}}^{\infty} 
p( \hat{s}_{\gamma} | s ) 
\, \mathrm{d} \hat{s}_{\gamma}  
= \alpha
\label{FML_form_0.1}
\end{eqnarray}
and
\begin{eqnarray}
\displaystyle \int_{0}^{\hat{s}_{\mathrm{left}}} 
p( \hat{s}_{\gamma} | s ) 
\, \mathrm{d} \hat{s}_{\gamma}  = 
\alpha
\label{FML_form_0.2}
\end{eqnarray}
allow us to plot the boundaries of the confidence region,
which are depicted in Fig. \ref{FMLConfBelt} by thick inclined solid 
trajectories
$[\mathrm{U}_0, \mathrm{U}_4]$ and $[\mathrm{L}_0, \mathrm{L}_4]$.

For the measurement C$_2$
the confidence interval
is given by  $[\mathrm{L}_2, \mathrm{U}_2]$, which includes the true
$s$ if it is depicted, for example, by $[\mathrm{A}, \mathrm{B}]$.
For the measurement C$_3$
the confidence interval $[\mathrm{L}_3, \mathrm{U}_3]$
does not include this $s$.
The proof of the one-sided coverage of the lower limit 
is based on the idea that
the probability for $\mathrm{L}_3$ to be higher than $s$
is equal to the probability for $\mathrm{E}_3$ to be 
to the right of $\mathrm{F}$,
and the latter is equal to $\alpha$ according to Eq. (\ref{FML_form_0.1})
or possibly smaller than $\alpha$
in the discrete case. 
The coverage of the upper limit
is proved similarly by the points 
$\mathrm{U}_1$, $\mathrm{E}_1$ and $\mathrm{G}$.

These trajectories look like lines and this was assumed in
the paragraph above, but 
in reality $\hat{s}$ can have discrete values only.
The values of the limits in these points are only important.
In the conditions of Fig. \ref{FMLConfBelt} these points are very
close to each other and are merged in lines. 
Let us search the solution of Eq. (\ref{FML_form_0.1})
by replacing 
$\hat{s}_{\mathrm{right}}$ by $\hat{s}$ obtained for observed $\vec{n}$,
which we denote here by $\hat{s}_{\mathrm{obs}}$.
Then we have to fit $s$ to obtain 
the equality 
in this expression.
Let us assume that the statistics of subgeneration is (nearly) infinite.
Then the corresponding equation  
in the discrete form is
\begin{eqnarray}
\sum_{\substack{ \vec{n}_{\gamma}: \\ 
 \scriptsize{ \makebox[0.9cm][l]{\( \hat{s}_{\gamma}(\vec{n}_{\gamma}, \vec{a}, \vec{b}) 
\geq \hat{s}_{\mathrm{obs}} \)} } }}
P(\vec{n}_{\gamma} | s, \vec{a}, \vec{b}) = \alpha ,
\label{FML_form_0.1.5}
\end{eqnarray}
where 
$P(\vec{n}_{\gamma} | s, \vec{a}, \vec{b})
= \prod_{i=1}^k \mathrm{Poisson}(n_{\gamma i} , t_a a_i s + t_b b_i)$.
It is meant here that only those $\vec{n}_{\gamma}$
are included for which the condition under the sign of sum is satisfied.
If this sum is greater than $\alpha$ at $s=0$,  zero $s$ is taken as the solution.  
This solution is the lower confidence limit
$s_{\mathrm{L}}$.
Note that this summing up should start from 
$\hat{s}_{\mathrm{obs}}$,
not from the next allowed $\hat{s}_{\gamma}$ as it might seem 
at first glance. 
It is important because 
the value of this very sum calculated for $s = 0$ 
can be used as the significance
of signal$+$background hypothesis versus the simple background hypothesis. 
Obviously, the greater is $\hat{s}_{\mathrm{obs}}$,
the more the event is signal-like.
Significance is 
estimated 
as the probability for the test statistic
to exceed the observed value or to {\it coincide} with it.

If we neglect discrete effects and some other mathematical details,
which can make the coverage conservative,
we can find out that  
the accurate coverage of intervals for any $\alpha$ 
is guaranteed by construction
if and only if 
the subgenerated $\hat{s}_{\gamma}$
calculated at the true $s$
is distributed exactly as
the experimental $\hat{s}$ after the imaginary repetitions
of the experiment.
If $\vec{a}$ and $\vec{b}$ are not known exactly,
the coverage is not guaranteed by construction.

Since $\hat{s}$ is restricted to be non-negative,
in the experiments where the formal $\hat{s}$ found in the interval
$]-\infty, +\infty[$ is negative,
$\hat{s}$ found in
$[0, +\infty[$ is usually zero.
This results in the appearance of a spike at zero
in the distribution of $\hat{s}$, which is
described by the $\delta$-function
with a certain weight.
This weight is negligible at high $s$.
As $s$ decreases,
this weight increases and at some point it becomes greater than $\alpha$.
In Fig. \ref{FMLConfBelt}
this is crossing of the upper limit $[\mathrm{U}_0, \mathrm{U}_4]$
with the $y$-axis, that is the point $\mathrm{U}_0$.
This spike should be excluded entirely
from the integral in Eq. (\ref{FML_form_0.2})
and included in the confidence region.
The sign ``$=$'' 
in Eq. (\ref{FML_form_0.2}) has to be replaced by
``$\leq$''.

If we want to keep constant the area inside the 
two-sided interval,  we have to shift the right boundary
in order to cut off the right tail with 
the $2\alpha$ area instead of single $\alpha$.
Thus, the full lower boundary will pass through the points 
$\mathrm{L}_5, \mathrm{L}_6, \mathrm{L}_7, \mathrm{L}_1, \mathrm{L}_2, \mathrm{L}_3, \mathrm{L}_4$. 
This very case was considered in Refs. \cite{Mandelkern_00,Ciampolillo_98}.
In this case the probability for the obtained lower limit 
to be higher than
the true $s$ is unknown. It varies as a function of the true $s$
and can be either $\alpha$ or $2\alpha$, depending on
the position of the point $\mathrm{U}_0$.
If the latter is not determined and reported, it will also be unknown.

For comparison, boundaries of the shortest intervals and the low boundary
of the modified central
intervals in the Bayesian analysis 
cut the variable probability,
but it can be easily calculated. 
Here the coverage cannot be directly calculated, 
and cannot be calculated at all,
if one strictly follows the frequentist approach and does not consider
the probability distribution of the true parameter of interest.

On the other hand,
if the researcher does not shift the lower boundary
when $s$ is below $\mathrm{U}_0$,
the coverage by the lower limit will be constant, but the
simultaneous ``two-sided'' coverage of the true $s$ by both limits
will be either $1-\alpha$ or 
$1-2\alpha$.
However, this two-sided coverage is less important
in practice.
There are exceptions, but usually this probability
does not have any useful meaning.
The violation of the lower and upper border
usually leads to different physical conclusions
and their separate confidence levels are the only values which are important.
Therefore, the confidence belt restricted by
$[\mathrm{U}_0, \mathrm{U}_4]$ from above and
$[\mathrm{L}_0, \mathrm{L}_4]$ from below is tested in this research.

Calculations indicate that such a technique provides 
plots almost identical to the plots obtained by
the modified central Bayesian intervals
for the case of known nuisance parameters, 
see Fig. \ref{Bayes_center_mod_no_unc_0}.
The one-sided coverage of both upper and lower boundaries for fixed divisions,
as well as for the divisions optimized by the interval width, stays near 90\%
in all cases.
Differences in the lower two plots are negligible.  
Fig. \ref{compar_lim} indicates that at the low observed signal
the upper limit by FML can be higher than that for the Bayesian method.
For the case with nuisance parameter uncertainties the
method is split into many modifications, which
will be described in the next section.

\subsection{The case of unknown $\vec{a}$ and  $\vec{b}$ }
\label{FML_Unknown_sec}

If the values $\vec{a}$ and $\vec{b}$ are unknown,
we have to use some approximations in the form of their assumed point values
or probability densities.
As in the Bay\-esian case,
the naive ignoring of these uncertainties and the use
of $\vec{n}_a$ instead of $\vec{a}$ and $\vec{n}_b$ instead of $\vec{b}$,
as well as many other simple approaches, 
do not work well enough for FML in the example studied.
More advanced assumptions 
are needed 
for the maximum likelihood finding with the data of the real experiment 
($\hat{s}$),
for the subgeneration of the experiment and for 
the maximum finding with the ``subgenerated'' data ($\hat{s}_{\gamma}$). 
We consider only the methods in which
$\hat{s}$ and $\hat{s}_{\gamma}$
are found by an identical method.
If $\hat{s}$ and $\hat{s}_{\gamma}$ are found
differently, this subgeneration (with analysis) could never be realized as generation,
that is we could not imagine such a sequence of experiments,
for which our coverage and  significance would be
``true by construction''. 
When this feature is present, we call it
the ``modeling interpretation'' or just the ``interpretation''.
Arguably, this modeling interpretation
can be sufficient,
if the frequentist coverage is unknown, but the model for 
nuisance parameters and the method of analysis are reasonable.

\subsubsection{The SSP--FMML, SHP--FMML, and SEP--FMML methods}
\label{SSP-FMML_sec}

A simple method based on the assumption that
the nuisance parameters are distributed randomly
according to  their posterior Bayesian probability
density distributions (Eq. (\ref{BayesForm_2})) with safe (or hybrid) priors
works well.
In the following we assume that both $\vec{a}$ and $\vec{b}$
are unknown. The following expressions are simplified in an obvious way
if one of them is known.
The random values $a_{\gamma i}$ and $b_{\gamma i}$
are inserted in Eq. (\ref{mu_i}) or (\ref{mu_i_1})
and the result $f_{\gamma i} = t_a a_{\gamma i} s + t_b b_{\gamma i}$
is used to obtain $n_{\gamma i}$ of the subgenerated
main experiment
according to the Poisson distribution with mean $f_{\gamma i}$.  
One has to find the  maximum 
$\hat{s}_{\mathrm{obs}}(\vec{n}, \vec{n}_a, \vec{n}_b)$
of $p( s \, | \vec{n}, \vec{n}_a, \vec{n}_b )$ and
the maximum
$\hat{s}_{\gamma}(\vec{n}_{\gamma}, \vec{n}_a, \vec{n}_b)$
of $p( s \, | \vec{n}_{\gamma}, \vec{n}_a, \vec{n}_b )$
for the observed and subgenerated data, respectively.
In both cases
the probability density distributions are
given by Eq. (\ref{BayesForm}) with
substitution of Eqs. (\ref{BayesForm_1}) and (\ref{BayesForm_2}) 
or by Eq. (\ref{BayesForm_3}) with substitution of Eq. (\ref{BayesForm_4})
with the uniform prior for $s$
and with safe (or hybrid) priors for auxiliary  $\vec{a}$ and $\vec{b}$.
Because of the uniform prior for $s$,
it is enough to find the maximum of Eq. (\ref{BayesForm_1}) with substitution of
Eq. (\ref{BayesForm_2}) or the maximum of Eq. (\ref{BayesForm_4}).
Equation (\ref{FML_form_0.1.5})
can be rewritten as
\begin{align}
\displaystyle \int 
\, p(\vec{a}_{\gamma}|\vec{n}_a)
\displaystyle \int  
\, p(\vec{b}_{\gamma}|\vec{n}_b) 
\sum_{\substack{ n_{\gamma}:  \\
\scriptsize{ \makebox[0.9cm][l]{\( \hat{s}_{\gamma}(\vec{n}_{\gamma}, \vec{n}_a, \vec{n}_b)  \geq
\hat{s}_{\mathrm{obs}}(\vec{n}, \vec{n}_a, \vec{n}_b) \)} } } }
P( \vec{n}_{\gamma}  | 
s , \vec{a}_{\gamma}, \vec{b}_{\gamma} ) 
\, \mathrm{d} \vec{b}_{\gamma} \, \mathrm{d} \vec{a}_{\gamma}
= \alpha \, .
\label{FML_form_0.3}
\end{align}
Here 
$P(\vec{n}_{\gamma} | s, \vec{a}_{\gamma}, \vec{b}_{\gamma})
= \prod_{i=1}^k \mathrm{Poisson}(n_{\gamma i} , t_a a_{\gamma i} s + 
t_b b_{\gamma i})$, which is equivalent to saying that
$\vec{n}_{\gamma}$ is generated with current 
$s$, $\vec{a}_{\gamma}$ and $\vec{b}_{\gamma}$.
The densities $p(\vec{a}_{\gamma}|\vec{n}_a)$ and 
$p(\vec{b}_{\gamma}|\vec{n}_b)$ are calculated by Eq. (\ref{BayesForm_2}).
Since Eq. (\ref{FML_form_0.3}) gives the lower limit,
the uniform nuisance priors are used for calculations of
 $p(\vec{a}_{\gamma}|\vec{n}_a)$, $p(\vec{b}_{\gamma}|\vec{n}_b)$,
$\hat{s}_{\gamma}(\vec{n}_{\gamma}, \vec{n}_a, \vec{n}_b)$ and
$\hat{s}_{\mathrm{obs}}(\vec{n}, \vec{n}_a, \vec{n}_b)$.
For the upper limit the inverse nuisance priors should be used 
(the hybrid median prior is allowed too).
Some mathematical and numerical subtleties can be present in Eq. 
(\ref{FML_form_0.3}) and in other similar equations for different 
methods discussed here, 
because of the limited
statistics and limited number of trials, as well as complex features 
of the methods.
In particular, {\it the least} $s$ that satisfies the equation,
should always be searched for.
An analogous reversed approach is used for the upper limits.

Obviously, in the case of zero $s$ the expression at the left-hand side
of Eq. (\ref{FML_form_0.3}) can be used
as an estimate of $p$-value similarly to Eq. (\ref{FML_form_0.1.5}).

These variants of FML can be briefly denoted by 
FMML, ``Frequency of Marginalized Maximum Likelihood'',
or more explicitly by
SSP--FMML or SHP--FMML, 
where the prefixes mean the Subgeneration with Safe Priors or 
Hybrid Priors, respectively.
Other priors do not work satisfactorily.
Note that the safe (or hybrid) priors are used not only for subgeneration, 
but for
marginalization too, 
that is for the calculation of
$\hat{s}_{\gamma}(\vec{n}_{\gamma}, \vec{n}_a, \vec{n}_b)$ and
$\hat{s}_{\mathrm{obs}}(\vec{n}, \vec{n}_a, \vec{n}_b)$. 
It is implied unless otherwise specified.

According to Ref. \cite{Demortier_07},
the $p$-value obtained by Eq. (\ref{FML_form_0.3}) at $s=0$
belongs to the category of ``prior predictive $p$-values''.
This notation can be confusing because 
$p(\vec{a}_{\gamma}|\vec{n}_a)$ and $p(\vec{b}_{\gamma}|\vec{n}_b)$
are posteriors for $\vec{n}_a$ and $\vec{n}_b$.
But the ``posterior predictive $p$-values'' assume that 
the posteriors should also depend on
$n$, which is not the case here.

These methods ensure the coverage by construction provided that
the assumption at the beginning of this section is true.
This is easily realized in practice if one does not repeat
the auxiliary experiments and treats the sequence of
the main experiments with the initially observed $\vec{n}_a$ and $\vec{n}_b$. 
So the reasonable modeling interpretation exists for this method.
Similarly, this method provides an interesting feature of
self-consistency of the $p$-value.
For given $\vec{n}_a$ and $\vec{n}_b$ 
the probabilities of $\vec{n}_{\gamma}$ used for calculation
of $p$-value by the left-hand side of Eq. (\ref{FML_form_0.3}) 
do not depend on $\vec{n}$.
Let us denote 
$p$-values calculated for any $\vec{n}_1$ and $\vec{n}_2$ and for the same
$\vec{n}_a$ and $\vec{n}_b$ 
by $\rho(\vec{n}_1, \vec{n}_a, \vec{n}_b)$ and 
$\rho(\vec{n}_2, \vec{n}_a, \vec{n}_b)$, respectively.
Then for any such $\vec{n}_1$ and $\vec{n}_2$,
if $\hat{s}_{\mathrm{obs}}(\vec{n}_1, \vec{n}_a, \vec{n}_b) > \hat{s}_{\mathrm{obs}}(\vec{n}_2, \vec{n}_a, \vec{n}_b)$,
all $\vec{n}_{\gamma}$ that are taken into account for 
$\rho(\vec{n}_1, \vec{n}_a, \vec{n}_b)$
should also be taken into account for 
$\rho(\vec{n}_2, \vec{n}_a, \vec{n}_b)$,
but at least one $\vec{n}_{\gamma} = \vec{n}_2$ that is taken into account for
$\rho(\vec{n}_2, \vec{n}_a, \vec{n}_b)$ should not be taken into
account for $\rho(\vec{n}_1, \vec{n}_a, \vec{n}_b)$. 
Therefore 
$\rho(\vec{n}_2, \vec{n}_a, \vec{n}_b) > \rho(\vec{n}_1, \vec{n}_a, \vec{n}_b)$.
This means that if one uses the $p$-value $\rho$ as the test statistic
for calculation of another $p$-value, one obtains an alternative $p$-value
(see Section \ref{Sect_intro_sig}), which should be equal to 
the regular $p$-value. 
The both $p$-values are also uniformly (taking into account discreteness) 
distributed
in $[0, 1]$ for fixed $\vec{n}_a$ and $\vec{n}_b$.
This equality and uniformity is not guaranteed
for many other methods, for which
the probabilities of $\vec{n}_{\gamma}$ used for calculation
of $p$-values are different for different $\vec{n}$.

To test different priors SSP--FMML was also run with
exchanged priors, so that
the uniform prior was used for the upper limit
and the inverse prior was used for the lower limit.
This method is denoted here by prefix SEP (Subgeneration with Exchanged Priors)
with full notation SEP--FMML.
The exchanged priors are used for marginalization too.

\begin{figure}[t]
\centering
\includegraphics[width=1.0\linewidth]
{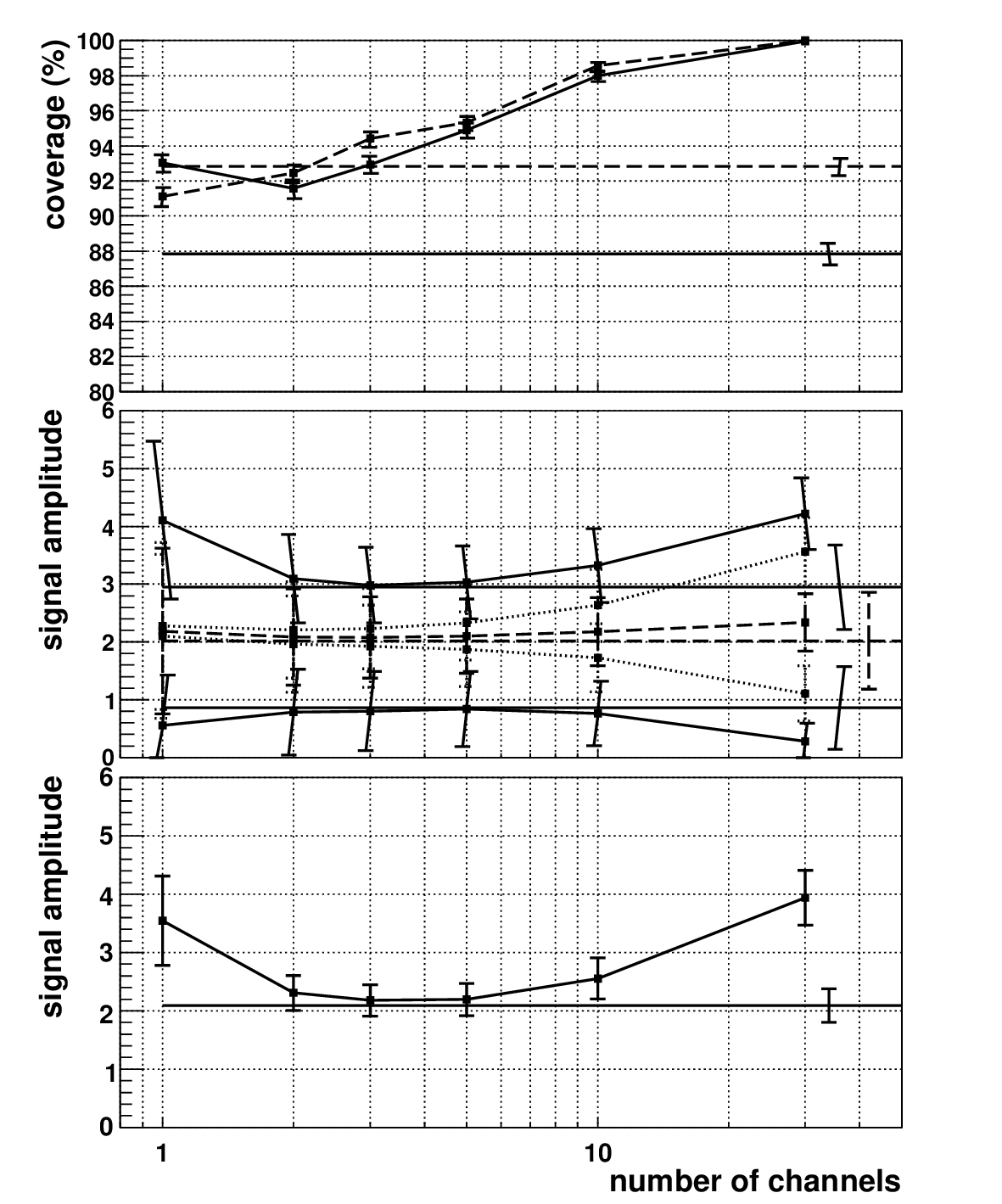}
\caption{
The SSP--FMML method, confidence intervals for 90\% one-sided confidence
($\alpha=0.1$),
the uncertainty of the expected background, the known expected signal.
The horizontal lines are results of optimization by the widths
of modified central Bayesian intervals.
Other details are described in Sections  \ref{SSP-FMML_sec},  
\ref{sec_Bayes_cover}, and in Fig. \ref{Bayes_center_mod_no_unc_0}.
}
\label{FMML_geny_anay_5_1}
\end{figure}

The calculations have shown that the SSP--FMML and SHP--FMML 
methods 
have characteristics that are very similar to those of the Bayesian methods
with respective priors and with modified
central intervals, described earlier.
For example, the case with uncertainty of the expected background is shown in 
Fig. \ref{FMML_geny_anay_5_1}.
The upper limit for SHP--FMML is lower apporximately by 0.1.
The coverage of SHP--FMML is similar.
The optimization of the division with the Bayesian modified
central intervals provides reasonably good coverage of the
optimized limits obtained by both methods.
The optimization with the intervals obtained by these methods themselves
provides slightly worse coverage of the upper limit.
Each of these methods, as well as the Bayesian one, produces 
two point estimates,
which have to be averaged.

The analysis by SEP--FMML
behaves similarly to the Bayesian analysis with modified central intervals
and with exchanged priors, which is described in Section \ref{sec_Bayes_cover}.
This method can yield completely wrong results.

Since the lower limit and the $p$-value are calculated
effectively by the same equation (\ref{FML_form_0.3}),
the lower limits by SSP--FMML are reliable, 
its modeling interpretation is convincing and 
the significance is self-consistent for fixed $\vec{n}_{a}$ and $\vec{n}_{b}$,
one might expect that the significance by this method is reliable too.
It is however difficult to find any exclusive numerical feature
of significance by SSP--FMML besides
self-consistency, which is inherent to many other methods as well.
One can compare the significance calculated by Eq. (\ref{FML_form_0.3}) 
with the exact significance $z_{\mathrm{ef}}$, 
whose $p$-value 
is defined by
\begin{align}
\rho_{\mathrm{ef}}(\vec{n}, \vec{n}_{a}, \vec{n}_{b})  = 
\sum_{\substack{ \vec{n}_{\gamma p}: \\ 
\scriptsize{  \makebox[1.5cm][l]{\( \hat{s}_{\gamma}(\vec{n}_{\gamma}, \vec{n}_{a}, \vec{n}_{b}) 
\geq \hat{s}_{\mathrm{obs}}(\vec{n}, \vec{n}_{a}, \vec{n}_{b}) \)} } }}
P(\vec{n}_{\gamma} | s=0, \vec{a}, \vec{b}),
\label{precise_p_form}
\end{align}
where 
the subscript ``e'' means ``exact'', ``f'' means ``fixed'', 
that is the fixed nuisance parameter
measurements $\vec{n}_{a}$ and $\vec{n}_{b}$,
and for the fitting of $\hat{s}$-values
one should take into account that the nuisance parameters
are unknown. The corresponding significance will be called $z_{\mathrm{ef}}$.
If only the expected signal is unknown,
the approximate significance (that is the estimate of 
significance by SSP--FMML, whose $p$-value
is calculated according to Eq. (\ref{FML_form_0.3}))
turns out to be
identical to $z_{\mathrm{ef}}$, but this holds
also for SEP--FMML, which gives slightly greater
significance for many-channel problems.
This holds also for any other methods
with prefixes SSP or SEP, described later.
Note also that the exact $p$-value can be defined
with random subgenerated nuisance parameter measurements 
and obtained from Eq. (\ref{precise_p_form}) by replacement
of $\hat{s}_{\gamma}(\vec{n}_{\gamma}, \vec{n}_{a}, \vec{n}_{b})$
by 
$\hat{s}_{\gamma}(\vec{n}_{\gamma}, \vec{n}_{a \gamma}, \vec{n}_{b \gamma})$.
Let us denote this by the subscript ``r'', random.
Comparing the approximate significance 
with $z_{\mathrm{ef}}$ or $z_{\mathrm{er}}$
we simply assume different 
frequentist interpretations of our approximate significance.
When only the expected background is unknown,
the approximate significance by SSP--FMML is usually less than $z_{\mathrm{er}}$
(which is acceptable), but not always. 
It is not usually less than $z_{\mathrm{ef}}$.
Moreover,
both exact significances, 
{\it minimized} with
respect to $\vec{a}$ and $\vec{b}$, are usually zero, except the case of $z_{\mathrm{ef}}$ and $\vec{a}$, see above.
However, there is another test statistic based on likelihood ratios
with marginalization, described in Section \ref{Sect_SARN_CLs},
and providing nontrivial minima of $z_{\mathrm{er}}$.
Calculations indicate that
the approximate significance by SSP--FMML is usually less than
this minimal significance, but there are better approximate methods.
So the significance by this method can be used, but
there are more reliable methods.
The confidence intervals by this method are very reliable.

\subsubsection{The SSP--FGML, SHP--FGML, and SEP--FGML  methods}
\label{SSP-FGML_sec}

Another approach alternative to (SSP--)FMML 
\linebreak
consists in 
finding the global maximum of 
the common 
\linebreak
likelihood given by Eq. (\ref{FullProb}) and expressed by
\linebreak
$P(\vec{n}, \vec{n}_a, \vec{n}_b | 
\hat{s}_{\mathrm{obs}}, \hat{\vec{a}}_{\mathrm{obs}}, \hat{\vec{b}}_{\mathrm{obs}})$ for the case of the observed data, 
instead of the maximum of the Bayesian posterior 
as required for FMML.
The subgeneration can be performed exactly as for SSP--FMML.
For the analysis of the subgenerated experiments 
it needs to find the global maximum
$P(\vec{n}_{\gamma}, \vec{n}_{a}, \vec{n}_{b} | 
\hat{s}_{\gamma}, \hat{\vec{a}}_{\gamma}, \hat{\vec{b}}_{\gamma})$
with respect to $\hat{s}_{\gamma}$,
$\hat{\vec{a}}_{\gamma}$, and $\hat{\vec{b}}_{\gamma}$.
Equation (\ref{FML_form_0.3}) is not changed, except that
the values $\hat{s}_{\gamma}(\vec{n}_{\gamma}, \vec{n}_a, \vec{n}_b)$
and $\hat{s}_{\mathrm{obs}}(\vec{n}, \vec{n}_a, \vec{n}_b)$ 
have a different sense, which is described above.
The modeling interpretation of this method is similar to
that of SSP--FMML. The $p$-values are self-consistent.

In this method the Bayesian priors are used only
for subgeneration. For maximization the priors are not used.
Hence the observed $\hat{s}_{\mathrm{obs}}(\vec{n}, \vec{n}_a, \vec{n}_b)$ 
is single and should not be averaged to obtain the final point 
estimate as necessary
for the Bayesian and FMML cases.
It has some systematic shift, but the latter is not large.

This variant of FML can be called SSP--FGML or SHP--FGML,
Subgeneration with Safe (or Hybrid, respectively) Priors, 
Frequency of Global Maximum Likelihood.
As with SEP--FMML, one can consider
FGML with Subgeneration with Exchanged Priors, SEP--FGML,
but it does not provide satisfactory results.

Calculations indicate that all performance characteristics of 
SSP--FMML and SSP--FGML (or SHP--FMML and SHP--FGML, respectively)
are almost the same with four exceptions which 
are worth mentioning.
First, the upper limit for one channel has almost 100\% coverage.
Second, the upper and the lower limit diverge less at 30 channels
for FGML than they do for FMML in Fig. \ref{FMML_geny_anay_5_1}.
Instead of the average interval width equal 
to approximately 3.9 units for SSP--FMML (about 3.8 for SHP--FMML) 
the SSP--FGML method gives about 3.5 units (about 3.3 for SHP--FGML). 
Third, the coverage of the upper optimized limit ($90.0\pm0.6\%$) 
is slightly higher than that for FMML ($87.9\pm0.6\%$, see Fig. \ref{FMML_geny_anay_5_1}).
Fourth, the calculations by FGML are faster with the existing program
than that by FMML.
However, FGML finds the most probable values of the main parameter  
taking into account
{\it the most probable} nuisance parameters and 
{\it ignoring the other possible values} of them,
whereas FMML takes into account all of them. 
The latter is more 
appealing conceptually and also technically,
if the nuisance parameter is predicted from general
theoretical considerations
as an interval of allowed values, with unknown and hence
equal probabilities inside this range.
Another example of failure to determine $\hat{s}$ by FGML is
the one-channel problem with expected-signal uncertainty
at $n_a = 0$ and $b \geq n$, where the likelihood does not
depend on $s$ (in this work it is assumed that $\hat{s}=0$ 
for this case). 
Advantages of the ``integrated likelihood'' are also
discussed in Ref. \cite{Berger_1999}.
Faster calculations by global maximization by our software 
and taking into account all possible values of nuisance
parameters with possibility to apply plain distributions
in the case of marginalization
are inherent to all the other discussed methods
that use these approaches (we will not repeat this each time).

\subsubsection{The SSPRN--FMML and SSPRN--FGML\\ methods}
\label{The_SSPRN--FMML_and_SSPRN--FGML_methods_sect}

In both SSP--FMML and SSP--FGML
the auxiliary measurements are not generated at the subgeneration stage.
The question is whether one could obtain
a method with better characteristics which
uses the ``subgenerated'' auxiliary measurements.
First of all, we can simply add the generation of the auxiliary measurements
at the subgeneration stage into SSP--FMML and SSP--FGML
and keep everything else the same.
Then, Equation (\ref{FML_form_0.3}) is converted into
\begin{align}
\displaystyle \int 
\, p(\vec{a}_{\gamma}|\vec{n}_a)
\displaystyle \int 
\, p(\vec{b}_{\gamma}|\vec{n}_b) & \times  \nonumber \\
\sum_{\substack{ \vec{n}_{\gamma}, \vec{n}_{a \gamma}, \vec{n}_{b \gamma}:  \\
\scriptsize{ \makebox[1.5cm][l]{\( \hat{s}_{\gamma}(\vec{n}_{\gamma}, \vec{n}_{a \gamma}, \vec{n}_{b \gamma}) \geq
\hat{s}_{\mathrm{obs}}(\vec{n}, \vec{n}_a, \vec{n}_b)  \)} } } }
 P( & \vec{n}_{\gamma}, \vec{n}_{a \gamma} , \vec{n}_{b \gamma}  | 
s , \vec{a}_{\gamma}, \vec{b}_{\gamma} ) 
\, \mathrm{d} \vec{b}_{\gamma} 
\, \mathrm{d} \vec{a}_{\gamma}
= \alpha \, .
\label{FML_form_0.3.1}
\end{align}
Here 
$n_{a \gamma i}$ and $n_{b \gamma i}$ are generated according to
the Poisson distributions with parameters $a_{\gamma i}$ and $b_{\gamma i}$,
respectively. The value 
$\hat{s}_{\gamma}(\vec{n}_{\gamma}, \vec{n}_{a \gamma}, \vec{n}_{b \gamma})$
is calculated as usually for SSP--FMML or SSP--FGML.
Inserting the suffix ``RN'' (Random Nuisance) into the old notations we obtain
the notations
SSPRN--FMML and SSPRN--FGML.
In the case of SSPRN--FMML the use of Eq. (\ref{BayesForm}) with
substitution of Eqs. (\ref{BayesForm_1}) and (\ref{BayesForm_2})
for the fitting of 
$\hat{s}_{\gamma}(\vec{n}_{\gamma}, \vec{n}_{a \gamma}, \vec{n}_{b \gamma})$
implies that $\vec{a}_{\gamma}$ and $\vec{b}_{\gamma}$
are distributed according to 
$\vec{n}_{a \gamma}$ and $\vec{n}_{b \gamma}$, 
while they are really distributed
according to $\vec{n}_{a}$ and $\vec{n}_{b}$ 
during the subgeneration.

In general, non-``RN'' methods effectively 
(here the term ``effectively'' means that we ignore for the moment
technical details like the type of the test statistic and many dimensions)
compare $\vec{n}$ with  $\vec{n}_{\gamma}$,
but the ``RN'' methods compare some effective generalized relation 
of $\vec{n}$ and  $\vec{n}_{b}$ together with $\vec{n}_{a}$
with a relation of $\vec{n}_{\gamma}$ and  
$\vec{n}_{b \gamma}$ together with $\vec{n}_{a \gamma}$.
This can lead to strange situations when
an experiment with $t_b \vec{n}_{b \gamma}$ effectively greater than
$\vec{n}$ is not included in the $p$-value,
if $\vec{n}_{\gamma}$ is yet greater.
When $\vec{b}$ is known, the non-``RN'' methods give the ``exact'' $p$-values
in the sense that this or greater test statistic
should be observed with exactly this probability independently of
the unknown $\vec{a}$ after many repetitions of this experiment (see Section \ref{SSP-FMML_sec}), 
but these $p$-values are different for different methods
in the general case.
It is easier to speed up the calculations
by  memorizing and recovering
$\hat{s}_{\gamma}$
from some tables for non-``RN'' methods, than for ``RN'' methods
because of greater dimensionality of these tables in the last case.

Both SSPRN--FMML and SSPRN--FGML do not provide 
a realistic modeling interpretation of intervals and
the corresponding modeled coverage by construction. 
The model from SSP--FMML 
does not work here because after the repetition of auxiliary experiments
one would restore varying distributions of 
$\vec{a}_{\gamma}$ and $\vec{b}_{\gamma}$
and varying confidence regions.
This is not a problem for the calculation of the $p$-value,
which is given by the left-hand side of Eq. (\ref{FML_form_0.3.1}) at $s = 0$.
As with the FMML and FGML methods without the suffix ``RN'',
the $p$-value has to be reproduced
in a long range of main and auxiliary experiments 
provided that $\vec{a}$ and $\vec{b}$
are distributed according to Eq. (\ref{BayesForm_2}) calculated with
the initially measured $\vec{n}_{a}$ and $\vec{n}_{b}$.
However, the self-consistency of $p$-values is not guaranteed.
For any two measurements denoted by subscripts ``1'' and ``2'',
if $\hat{s}_{\mathrm{obs}}(\vec{n}_1, \vec{n}_{a1}, \vec{n}_{b1}) > \hat{s}_{\mathrm{obs}}(\vec{n}_2, \vec{n}_{a2}, \vec{n}_{b2})$, the value
$\rho(\vec{n}_2, \vec{n}_{a2}, \vec{n}_{b2})$ should not necessarily be
greater than $\rho(\vec{n}_1, \vec{n}_{a1}, \vec{n}_{b1})$.
For example, if $n=67$, $t_a=1$, $a=1$, $t_b=2$, and $n_b=15$
(an example from table 1 of Ref. \cite{Cousins_08},
discussed also in Section \ref{on_off_Sec}
herein), then $\hat{s}_{\mathrm{obs}} = 36.4179$ and $\rho = 0.000105$.
If $n=65$ and $n_b=14$ with the same other parameters,
then $\hat{s}_{\mathrm{obs}} = 36.3893$ and $\rho = 0.000075$.
Therefore the self-consistency cannot be proved.

\begin{figure}[t]
\centering
\includegraphics[width=1.0\linewidth]
{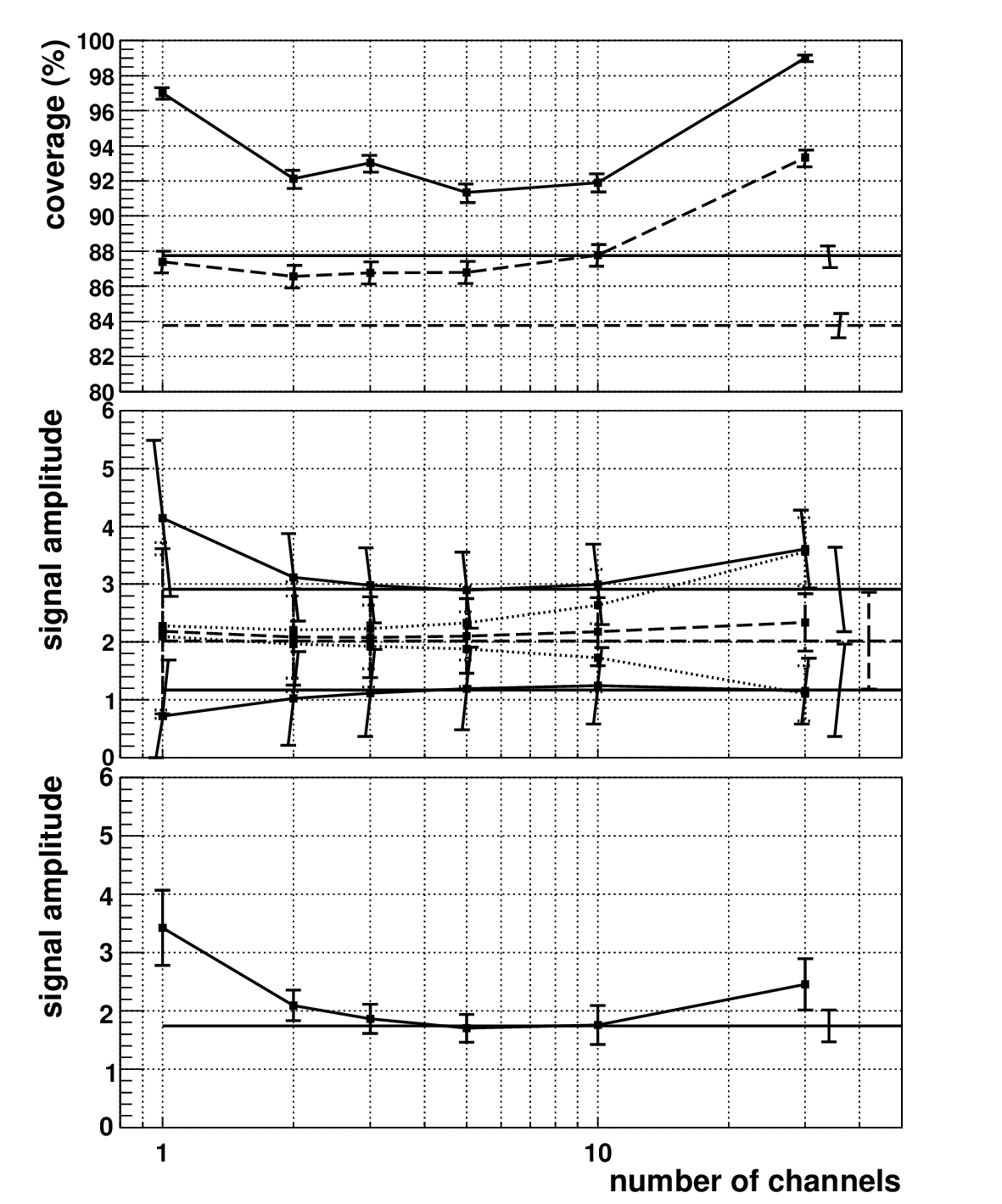}
\caption{
The SSPRN--FMML method, confidence intervals for 90\% one-sided confidence
($\alpha=0.1$),
the uncertainty of the expected background, the known expected signal.
The horizontal lines are results of optimization by the widths
of modified central Bayesian intervals.
Other details are described in 
Sections \ref{The_SSPRN--FMML_and_SSPRN--FGML_methods_sect}, 
\ref{sec_Bayes_cover}, 
and in Fig. \ref{Bayes_center_mod_no_unc_0}.
}
\label{SSPRM_FMML_unc_5}
\end{figure}

Numerical tests show that both SSPRN--FMML
(see Fig. \ref{SSPRM_FMML_unc_5})
and SSPRN--FGML behave similarly and 
do not provide frequentist coverage
for the lower limit. 
Interestingly, 
the coverage is minimal for intermediate numbers of channels, 
2--5 channels.
When only the uncertainty of the expected signal is present,
both methods behave similarly to the Bayesian approach,
including the loss of the coverage of the upper limit
for a large number of channels owing to zeros in the expected
signal distributions. 
The significance calculated by them is usually greater and less reliable
than that for SSP--FMML and SSP--FGML for unknown expected background 
and slightly less at unknown expected signal.
If the priors are exchanged (these methods can be denoted by 
SEPRN--FMML and SEPRN--FGML) at unknown background, 
the coverage of the lower limits gets even worse, 
while the upper limits are not strongly 
changed. None of these methods can be recommended.

\subsubsection{The SMRN--FGML and  SARN--FGML methods}
\label{SMRN-FGML_sec}

In the SMRN--FGML method the subgereration is done with 
the most probable nuisance parameters 
$\hat{\vec{a}}_{\mathrm{obs}}$ and $\hat{\vec{b}}_{\mathrm{obs}}$,
which are determined by the global
maximization of 
\linebreak
$P(\vec{n}, \vec{n}_{a}, \vec{n}_{b} 
| \hat{s}_{\mathrm{obs}}, \hat{\vec{a}}_{\mathrm{obs}}, \hat{\vec{b}}_{\mathrm{obs}})$.
The prefix SMRN means 
the Subgeneration with the Most probable observed nuisance parameters
and Random Nuisance parameter measurements. 
The global maximization 
is proposed in
Ref. \cite{Ciampolillo_98} (p. 1421) and more recently in Ref. \cite{Muller_10}.
In the SARN--FGML method the subgeneration is done with 
$\hat{\hat{\vec{a}}}_{\mathrm{obs}}$ and 
$\hat{\hat{\vec{b}}}_{\mathrm{obs}}$ that maximize 
$P(\vec{n}, \vec{n}_{a}, \vec{n}_{b} 
| s, \hat{\hat{\vec{a}}}_{\mathrm{obs}}, \hat{\hat{\vec{b}}}_{\mathrm{obs}})$
for each
{\it given}
$s$. 
This idea is borrowed from 
the LHC-style $CL_s$ method, 
which is described in Section \ref{The_CLs_method}.
On the other hand, this method can be considered as a modification of
SMRN--FGML.
The values 
$\hat{\hat{\vec{a}}}_{\mathrm{obs}}$ and
$\hat{\hat{\vec{b}}}_{\mathrm{obs}}$ 
can be seen as adjusted for given $s$, which changes the abbreviation
from SMRN to SARN: Subgeneration with Adjusted nuisance parameters
and Random Nuisance parameter measurements.

The value $\hat{s}_{\mathrm{obs}}$ has to be used as 
the observed test statistic value.
The ordinary FGML is applied to the subgenerated data.
For the subgenerated experiments it needs to find the global maximum of
$P(\vec{n}_{\gamma}, \vec{n}_{a \gamma}, \vec{n}_{b \gamma} | 
\hat{s}_{\gamma}, \hat{\vec{a}}_{\gamma}, \hat{\vec{b}}_{\gamma})$
with respect to $\hat{s}_{\gamma}$,
$\hat{\vec{a}}_{\gamma}$, $\hat{\vec{b}}_{\gamma}$.
The result $\hat{s}_{\gamma}$ is compared with $\hat{s}_{\mathrm{obs}}$.
Equation (\ref{FML_form_0.1.5})
can be rewritten by
\begin{align}
\sum_{\substack{ n_{\gamma}, \vec{n}_{a \gamma}, \vec{n}_{b \gamma}:  \\
\scriptsize{ \makebox[1.5cm][l]{\(  \hat{s}_{\gamma}(\vec{n}_{\gamma}, \vec{n}_{a \gamma}, \vec{n}_{b \gamma}) \geq
\hat{s}_{\mathrm{obs}}(\vec{n}, \vec{n}_{a}, \vec{n}_{b}) \)} } } }
& P( \vec{n}_{\gamma}, \vec{n}_{a \gamma}, \vec{n}_{b \gamma}| 
s , \hat{ \vec{a} }_{\mathrm{obs}}, \hat{ \vec{b} }_{\mathrm{obs}} ) 
 = \alpha .
\label{FML_form_0.3.5}
\end{align}
for SMRN--FGML and the same 
with replacement of 
$\hat{\vec{a}}_{\mathrm{obs}}$ and $\hat{\vec{b}}_{\mathrm{obs}}$
by $\hat{\hat{\vec{a}}}_{\mathrm{obs}}$ and 
$\hat{\hat{\vec{b}}}_{\mathrm{obs}}$, respectively, for SARN--FGML.

These methods do not have a reasonable interpretation of intervals. 
Indeed, if the main experiment is repeated,
$\hat{\vec{a}}_{\mathrm{obs}}$ and  $\hat{\vec{b}}_{\mathrm{obs}}$ for SMRN 
and $\hat{\hat{\vec{a}}}_{\mathrm{obs}}$ and 
$\hat{\hat{\vec{b}}}_{\mathrm{obs}}$ for SARN,
which are used for subgeneration after each repetition,
would be different each next time whether one repeats the auxiliary experiments
or not, because they depend on $\vec{n}$.
Even if the true 
$\vec{a}$ and $\vec{b}$ coincide with initially observed
$\hat{\vec{a}}_{\mathrm{obs}}$ and  $\hat{\vec{b}}_{\mathrm{obs}}$,
the confidence belt would be different at each next repetition
and the procedure used for the initial subgeneration could not be reproduced.
The self-consistency of $p$-values is not guaranteed.

\begin{figure}[t]
\centering
\includegraphics[width=1.0\linewidth]
{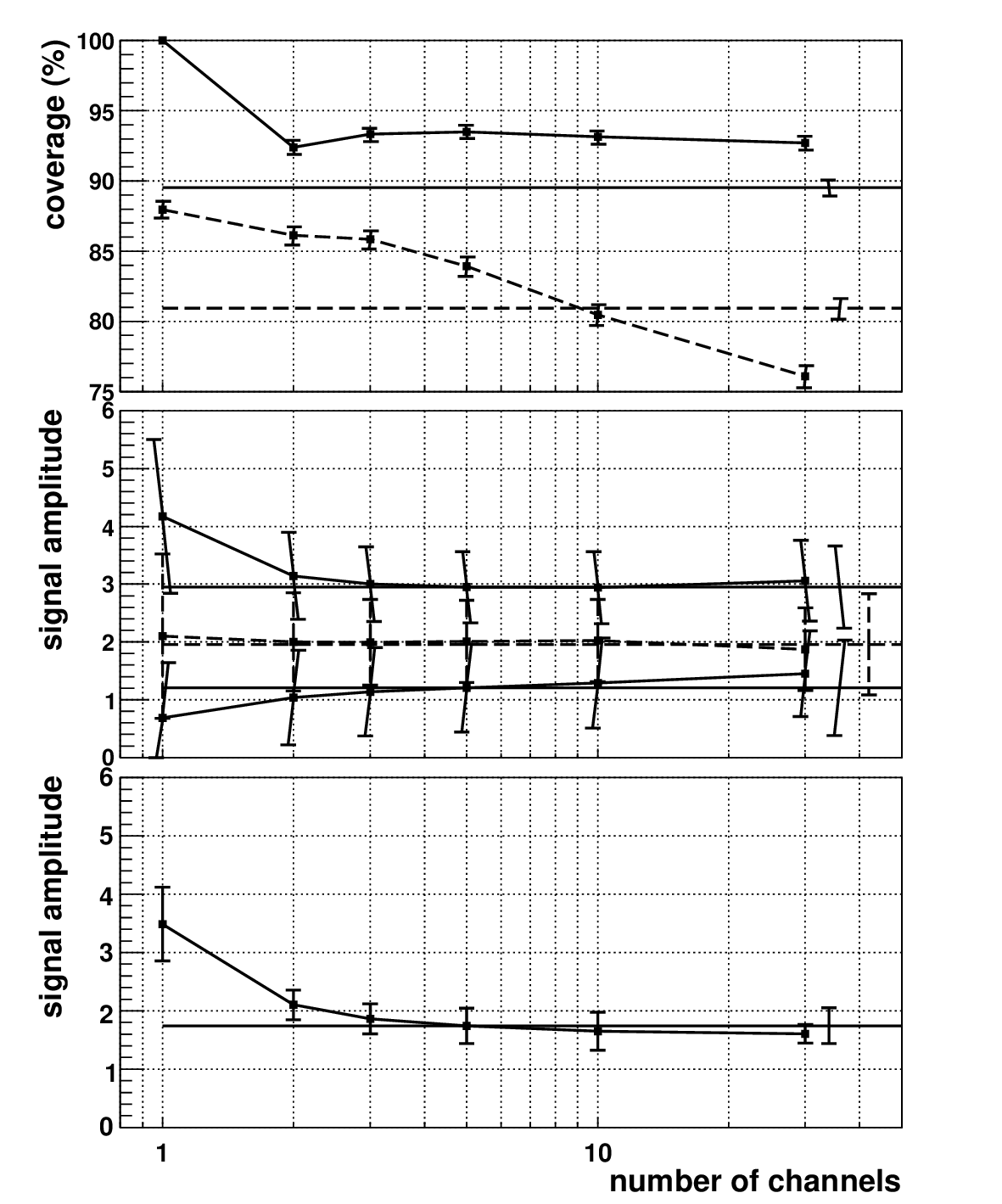}
\caption{
The SMRN--FGML method, confidence intervals for 90\% one-sided confidence
($\alpha=0.1$),
the uncertainty of the expected background, the known expected signal.
The horizontal lines are results of optimization by the widths
of modified central Bayesian intervals.
Other details are described in Sections \ref{SMRN-FGML_sec},
\ref{sec_Bayes_cover}, 
and in Fig. \ref{Bayes_center_mod_no_unc_0}.
}
\label{SMRN_FGML_unc_5}
\end{figure}
\begin{figure}[t]
\centering
\includegraphics[width=1.0\linewidth]
{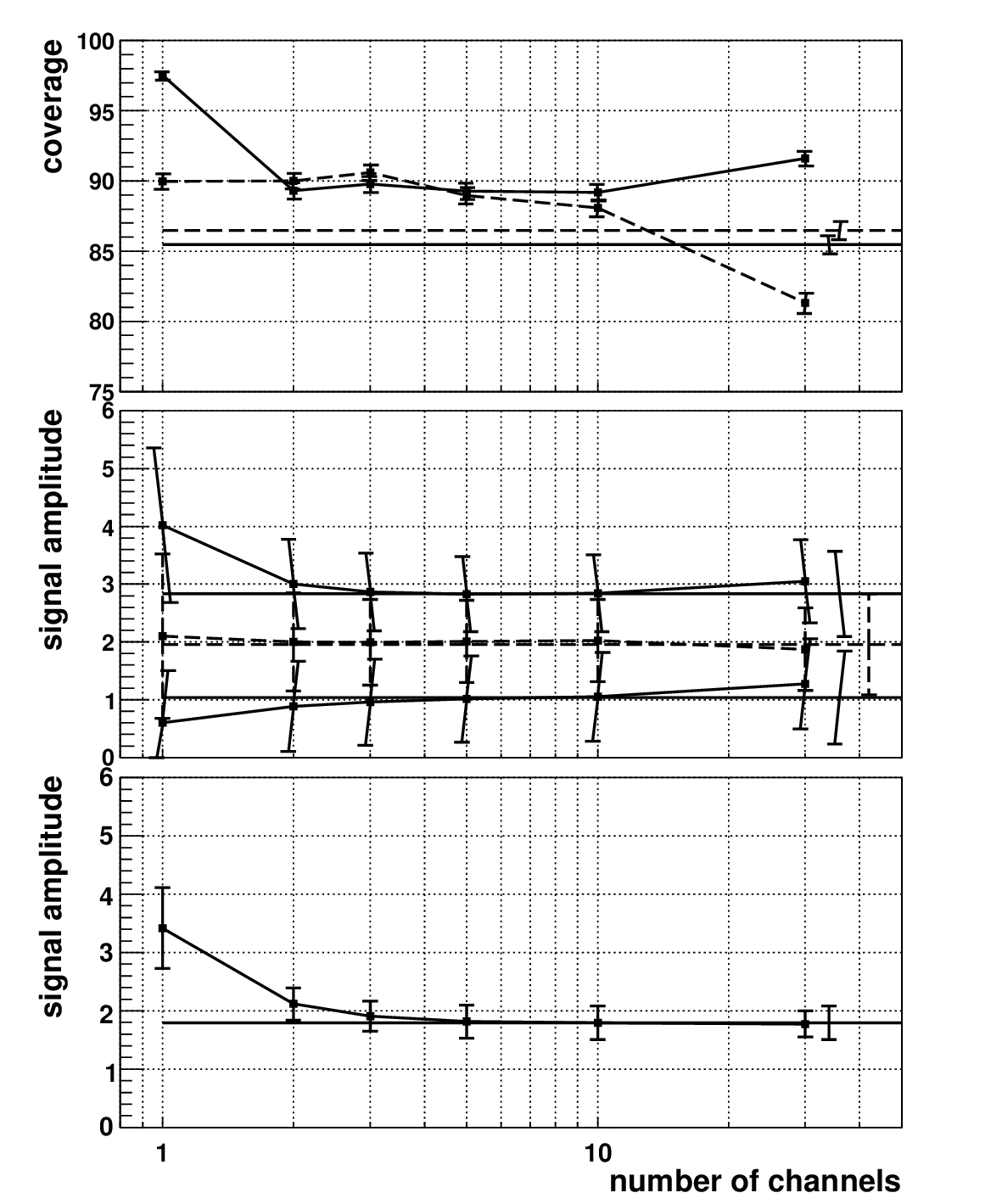}
\caption{
The SARN--FGML method, confidence intervals for 90\% one-sided confidence
($\alpha=0.1$),
the uncertainty of the expected background, the known expected signal.
The horizontal lines are results of optimization by the widths
of modified central Bayesian intervals.
Other details are described in Sections
 \ref{SMRN-FGML_sec}, \ref{sec_Bayes_cover},
and in Fig. \ref{Bayes_center_mod_no_unc_0}.
}
\label{SARN_FGML_unc_5}
\end{figure}

According to calculations
the frequentist coverage is not provided for the lower limit,
see Figs. \ref{SMRN_FGML_unc_5} and \ref{SARN_FGML_unc_5}.
The ``standard'' optimization of divisions (see Section \ref{Optimization}) 
shown in these figures
results in undecoverage of the lower limit for SMRN--FGML and moderate
undecoverage of both limits for SARN--FGML.
If the optimization is necessary and the coverage of the upper limit
is important, one can choose to use SMRN--FGML for the upper limit and
SARN--FGML for the lower one.
A good optimized coverage of all these methods is obtained by
the rejection of divisions with zeros
in the expected-background distribution and choosing the most detailed
division without zeros.
But this optimization cannot be recommended because of its assumed poor performance
in the presence of
the truly zero channels in the expected background distribution (see Section \ref{Optimization}) and
because of the possibility of splitting into 
too many channels at large statistics.
When only the uncertainty of the expected signal is present,
the coverage of the upper limit by both methods is not reduced with the increase in the
number of channels, but the optimized coverage is lower than necessary,
about 86--88\%.

The interpretation of the $p$-value
is based on single and arguable values of $\vec{a}$ and $\vec{b}$
without considering alternatives,
which can be unconvincing.
The calculations show that significance is usually much greater
than that for SSP--FMML and SSP--FGML for the case of uncertain background.
Since the exact significances for SARN--FGML are the same as for
SSP--FGML, the former should be, as a rule, less reliable.

\section{Likelihood Ratio}
\label{Likelihood_Ratio_sect}

As mentioned earlier,
the probability density, given, for instance, by Eq. (\ref{FullProb}),
can be considered as the likelihood of the parameters.
We will not use here an additional notation for it (usually $L$). 
The likelihood ratio is denoted in some references by $\lambda$
and defined by
\begin{eqnarray}
\lambda(s) = 
\frac
{ 
{P( \vec{n}, \vec{n}_{a}, \vec{n}_{b} | s, 
\hat{\hat{\vec{a}}}, \hat{\hat{\vec{b}}})} }
{ 
P( \vec{n}, \vec{n}_{a}, \vec{n}_{b} 
| \hat{s}, \hat{\vec{a}}, \hat{\vec{b}}) }\, .
\label{LikRat_0}
\end{eqnarray}
Here the denominator is maximized with respect to all parameters,
and the nominator is maximized only with respect to the nuisance parameters
for the specified $s$.
Here, as well as everywhere in this paper, 
all parameters are limited to their physical values,
so they cannot be smaller than zero.
In principle, this method can be applied also 
without this restriction, as in Refs. \cite{Heinrich_07,Rolke_2005},
but here this option is not considered as having unclear physical sense.

Given $\alpha$ one should obtain $z$ as described after Eq. (\ref{IntGaussian}).
Then the lowest $s_{\mathrm{L}}$ and the uppermost $s_{\mathrm{U}}$
that satisfy
$\lambda(s_{\mathrm{L}})) =  \lambda(s_{\mathrm{U}})) = e^{-z^2/2}$
are taken as limits.
Alternatively, 
one can obtain the same limits from the fractile of the $\chi^2$-distribution
with one degree of freedom, as recommended in Ref. \cite{Rolke_2005}.
Given $2\alpha$ one obtains $z^2$ and proceeds in the same way.
If non-negative $s_{\mathrm{L}}$ does not exist, it is equated to zero.

Attractive features of this method are
the absence of priors, simplicity and applicability
for more generic problems, as well as its past success \cite{James}.
Its intervals should asymptotically converge to 
frequentist intervals for large statistics (see Ref. \cite{James}
and references in Ref. \cite{WhyNotBayesian}),
and they do not have meaning for non-Gaussian cases
with small statistics.

\begin{figure}[t]
\centering
\includegraphics[width=1.0\linewidth]
{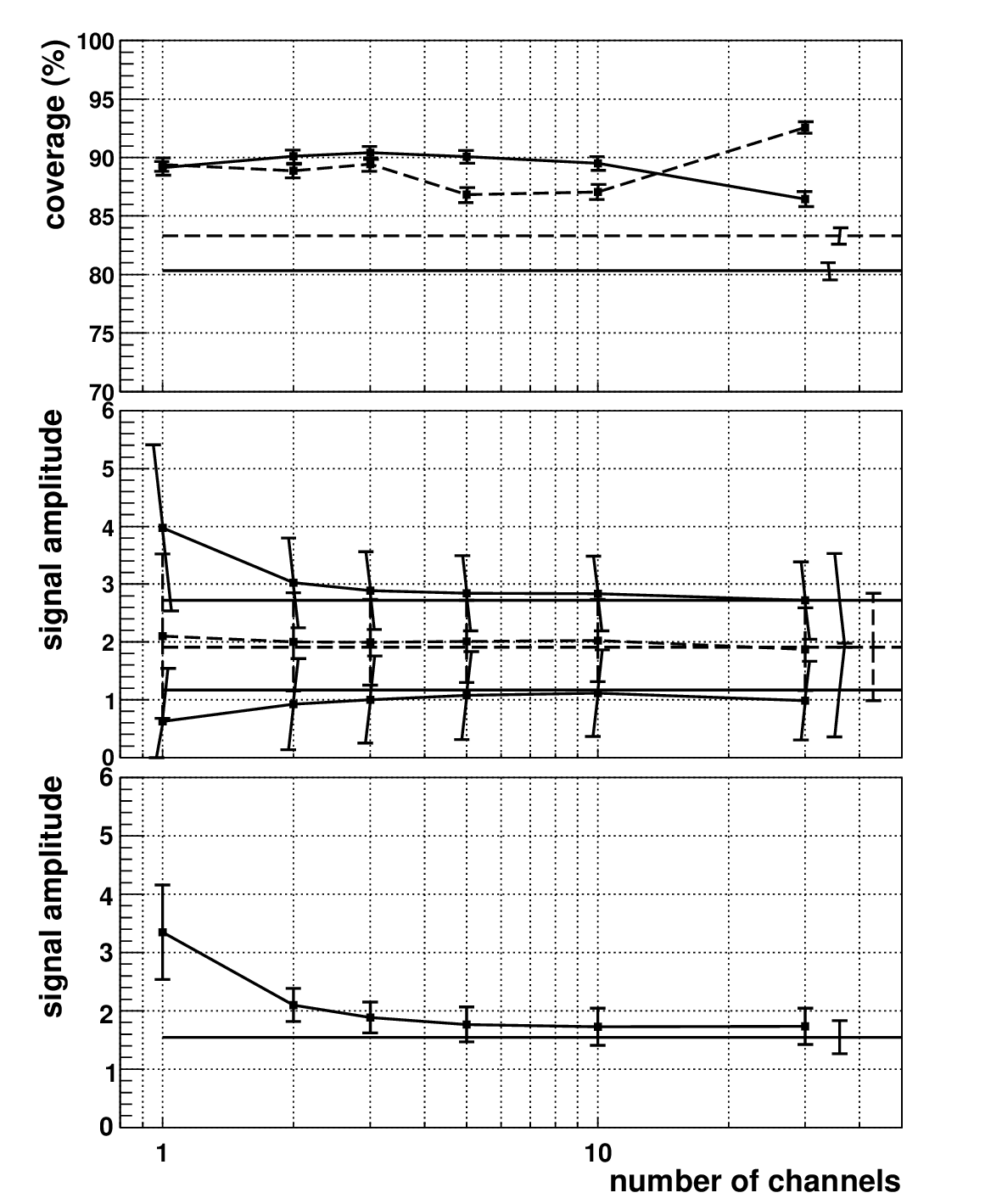}
\caption{
The likelihood ratio method, intervals for 90\% one-sided ``confidence''
($\alpha=0.1$), 
the uncertainty of the expected background,
the known expected signal.
The horizontal lines depict optimization by the width of these intervals.
Other details are described in Sections
\ref{Likelihood_Ratio_sect}, \ref{sec_Bayes_cover},
and in Fig. \ref{Bayes_center_mod_no_unc_0}.
}
\label{LR_geny_anay_5}
\end{figure}

The tests with the example studied here
show that the coverage is slightly unstable
and sometimes slightly insufficient, see Fig. \ref{LR_geny_anay_5}.
Optimization makes it worse.
The ``confidence'' interval is shorter than that for 
 the Bayesian method, SSP--FMML, and SSP--FGML, 
but it remains of the same order of magnitude, see Fig. \ref{compar_lim}.
So this method can be used for
fast estimates of intervals,
but these intervals may be inaccurate and usually
too short.
Obviously, 
just like the Bayesian method, 
it cannot provide significance
in a direct way, but an asymptotic approximation
to $CL_s$ methods described later is closely related to it.

\section{Frequentist treatment of the likelihood ratio}
\label{The_CLs_method}
\subsection{Introduction, the case without uncertainties}
\label{The_CLs_method_intro}

Following notations of Read \cite{Read_00} we now denote the
likelihood ratio by $Q$ and write
\begin{eqnarray}
Q = 
\frac{ 
P( \vec{n} | s)}
{ P( \vec{n} | s_{\mathrm{ref}})}\, .
\label{CLs_form-1}
\end{eqnarray}
For the present we will ignore the issues of nuisance parameters. 
The parameter $s_{\mathrm{ref}}$ is some reference value of $s$.
According to the approach from Ref. \cite{Read_00} 
$s_{\mathrm{ref}} = 0$.
Here this approach will be called  
the ``Background-Related'' method and denoted by the abbreviation BR.
This approach 
can be used for estimation of significance at known or assumed $s$
or for estimation of the upper limit of $s$ for predefined
$\alpha$.
According to a newer method, formulated in CERN for 
standard model Higgs boson
search at the LHC \cite{CMS_Combined,CMS_Paper_2012},
$s_{\mathrm{ref}}$ maximizes\footnote{The report \cite{Read_00}
in p. 85 also proposes the use of $s_{\mathrm{ref}}$ that maximizes 
$P( \vec{n} | s_{\mathrm{ref}})$ 
``in the complete absence of background'' and ``observation of
one or more candidates''.} $P( \vec{n} | s_{\mathrm{ref}})$,
but it is constrained by $s_{\mathrm{ref}} \leq s$.
We interpret this constraint 
in such a way that $s_{\mathrm{ref}}$ 
is initially found in the $[0, \infty[$ interval
and, if it is greater than $s$, it is equated 
to $s$.
This can be used for
the upper limit. 
By analogy for the lower limit\footnote{I have not found so far any mentions
about the lower limits calculated by this method, so this recipe
is my extension of this method.}  
$s_{\mathrm{ref}}$ can be chosen such that it
maximizes $P( \vec{n} | s_{\mathrm{ref}})$ and is constrained by 
$s_{\mathrm{ref}} \geq s$.
The constraint allows us to set limits that exclude the fixed and
usually equal
portion of more signal-like events by the lower limit
and the same portion of less signal-like events by the upper limit.
This method will be denoted here by the suffix CMR,
which means
the ``Constrained-Maximum-Related'' method.
$s_{\mathrm{ref}}$ can also be found in the $[0, \infty[$ interval
without the additional constraints.
In this case we mix the experiments that are outside both limits.
One limit can exclude more experiments than another.
This method is known as ``the unified approach''  \cite{F&C}.
We will denote it here by the suffix UMR,
which means the ``Unconstrained(or Unified)-Maximum-Related'' method.

These methods can also be used for estimation of significance, 
if calculated with $s = 0$. In this case the constraint duplicates
the constraint assumed in the calculation of the global maximum 
$\hat{s} \geq 0$, 
so $s_{\mathrm{ref}}$ is simply equal to  $\hat{s}$
and there is no difference between the  CMR and UMR methods.
The first character can be omitted in this case.

For instance, for the background-related method in the fully binned Poisson case
the value of $Q$ is expressed  \cite{Read_00} by
\begin{align}
Q =  \frac{\prod_{i=1}^{k} 
\frac{e^{-(t_a a_i s + t_b b_i)} (t_a a_i s + t_b b_i)^{n_i}}{n_i !}}
{\prod_{i=1}^{k}
\frac{e^{-{t_b b_i}} (t_b b_i)^{n_i}}{n_i !}} = & \nonumber \\ 
=  e^{-\sum_{i=1}^{k}t_a a_i s } 
 \prod_{i=1}^{k}(1 & + \frac{t_a a_i s}{ t_b b_i}) \, . 
\label{CLs_form0}
\end{align}

The work \cite{Read_00} also claims that 
``the confidence in the signal$+$background hypothesis
is given by
the probability
that the test statistic is less than or equal
to the value observed in the experiment, $Q_{obs}$:
\begin{eqnarray}
CL_{s+b} = P_{s+b}(Q \leq Q_{obs}) ",
\label{CLs_form-10}
\end{eqnarray}
provided that $\vec{n}$ which is used for calculating $Q$
(during subgeneration, according to our terminology) 
is distributed according to
the signal$+$background hypothesis,
which is indicated by the subscript $s+b$.
``Small values of $CL_{s+b}$ indicate poor compatibility with
the signal+background hypothesis and favor the background 
hypothesis'' \cite{Read_00}.
According to the earlier work of Junk \cite{Junk_99}
$CL_{s+b}$ is ``the confidence level for excluding the possibility
of simultaneous presence of new particle production and background
(the $s+b$ hypothesis)''.
So this is the usual exclusion of the impossible, expressed, for instance, by
Eq. (\ref{FML_form_0.2}) for the case of FML 
and used for setting the upper frequentist limit:
\begin{align}
CL_{s+b} =  P_{s+b}(Q \leq Q_{obs}) =
\sum_{\substack{ \vec{n}_{\gamma}: \\ 
\scriptsize{ \makebox[1.0cm][l]{\( Q(\vec{n}_{\gamma}, \vec{a}, \vec{b}, s) 
\leq Q(\vec{n}, \vec{a}, \vec{b}, s) \) } } }}
P(\vec{n}_{\gamma} | s, \vec{a}, \vec{b}) = \alpha\, .
\label{CLs_form-11}
\end{align}
The upper limit is obtained 
by finding the maximal $s$ that satisfies this equation.
For the constrained-maximum-related method 
we will have the same equation with $Q$
evaluated with $s_{\mathrm{ref}}$ equal to the minimum of $\hat{s}$ and $s$.
The background-related method is not used
directly for the lower limit setting, see comments in
Ref. \cite{Read_02}.
In the constrained-maximum-related method
the lower limit can be obtained as the lowest $s$ that satisfies this equation
with $Q$ evaluated with
$s_{\mathrm{ref}}$ equal to the maximum of $\hat{s}$ and $s$,
or zero, if the sum in this equation is greater than $\alpha$
at $s=0$.

Although the notation $CL_{s+b}$ is not used in the unified approach,
Eq. (\ref{CLs_form-11}) is valid for it too, provided that $\alpha$ denotes
the total excluded probability. Since we are studying the one-sided coverage
in this work, we assume that $\alpha$ is replaced by $2\alpha$ 
in Eq. (\ref{CLs_form-11}),
when it is applied for the unified approach.
At $s = \hat{s}_{\mathrm{obs}}$ the sum reaches its maximum,
the unity, and it falls at lower and higher $s$. 
The lower limit is
the lowest
$s < \hat{s}_{\mathrm{obs}}$ at which the sum is equal to $2\alpha$ and is decreasing,
or zero, if the sum is greater than $2\alpha$ at $s=0$.
The upper limit is found similarly.

The use of $Q$ for interval setting 
differs significantly from the ordinary ``Neyman construction'',
since here the observable test statistic $Q$ depends on the hypothesis about
the searched parameter $s$.
Therefore instead of vertical lines in the plots $s$ versus $\hat{s}$ like
$[\mathrm{C}_i, \mathrm{D}_i]$
plotted in Fig. \ref{FMLConfBelt} one has to consider curved inclined 
trajectories in the plots $s$ versus $\ln{Q(s)}$. 
These trajectories can even cross each other.
The picture can be weird enough, but in the absence of nuisance parameter 
uncertainties the coverage (possibly conservative) can be proved in a way 
similar to that for FML, see Section \ref{FML_introduction}.
Note, that one cannot use $P( \vec{n} | s)$ instead of the full ratio $Q$
by Eq. (\ref{CLs_form-1}) for calculations by Eq. (\ref{CLs_form-11}), 
because in the general case such an approach
does not satisfy 
the condition (ii) of the Proposition VII from Ref. \cite{Neyman_37}.

According to Ref. \cite{Read_00}
the significance (in the units of $p$-value) in the back\-ground-related method 
is estimated
by $1-CL_b$, where $CL_b$  
is calculated analogously to Eqs. (\ref{CLs_form-10}) or (\ref{CLs_form-11})
for $\vec{n}_{\gamma}$ distributed according to
the background hypothesis:
\begin{align}
CL_{b} =  P_{b}(Q \leq Q_{obs}) = 
\sum_{\substack{ \vec{n}_{\gamma}: \\ 
\scriptsize{ \makebox[1.0cm][l]{\(  Q(\vec{n}_{\gamma}, \vec{a}, \vec{b}, s) 
\leq Q(\vec{n}, \vec{a}, \vec{b}, s) \) } } }}
P(\vec{n}_{\gamma} | s_u = 0, \vec{a}, \vec{b}) \, .
\label{CLs_form-12}
\end{align}
Here $s$ used for the evaluation of $Q$ differs from
$s_u$ used for subgeneration, the latter is zero.
This $CL_b$ is expected to be close
to unity for good signal-like experiments.
Note that despite of using the background  $\vec{n}_{\gamma}$,
the value of $s$ in Eqs. (\ref{CLs_form-1}), (\ref{CLs_form0}) and
(\ref{CLs_form-12}) represents
the assumed signal during the calculation of $CL_b$.
It can be the maximum likelihood signal or the signal predicted by theory.
The literature describing this background-related method
does not offer concrete prescriptions about this.
The estimation of significance via $1-CL_b$
is approximate anyway because this excludes from the $p$-value
the probability of obtaining
the observed data.
 
In the maximum-related method
the corresponding $p$-value is correctly calculated by 
the sum in Eq.
(\ref{CLs_form-11}) with $s = 0$ and
with condition $s_{\mathrm{ref}} \geq 0$.
The value $CL_b$ is used in the constrained-maximum-related 
method only to correct the upper limit
in the case of microscopic signal dependence as described 
below. It is not used in the unconstrained-maximum-related 
method at all.
In the constrained method $CL_b$ is calculated by Eq. (\ref{CLs_form-12})
with $s_u = 0$ and with the given value of $s$, the same as used for the
calculation of $CL_{s+b}$.

If there are no nuisance parameter uncertainties, 
the upper limit obtained with the constrained $CL_{s+b}$ 
excludes the
true $s$ with probability $\alpha$
even for
the experiments microscopically susceptible to the signal.
During the application of the constrained  $CL_{s+b}$ to such experiments
there will be strong experiment-by-experiment fluctuations
of the reconstructed limit, due to which
this limit will be lower 
than the true $s$ with the probability $\alpha$.
This is mathematically correct, but is considered inappropriate in practice
\cite{Read_00}.
In the case of the unknown nuisance parameters this effect should occur too,
although it can have different size.
In the opinion of some physicists \cite{Read_00}, 
if an experiment is not susceptible 
to the signal,
there should not be a way for it to exclude the signal.
Moreover, another problem is that
the upper limit reconstructed by $CL_{s+b}$ 
can sometimes 
be ridiculously low, as in Fig. \ref{compar_lim}
(although it can never be exactly zero, 
which is mathematically forbidden in this case).
All of this is in contrast with
the upper boundaries of the Bayesian method,
of the frequentist treatment of maximum likelihood, 
of the likelihood ratio method and of the unconstrained-maximum-related method.
All of them (with slight exception for the two last methods) provide zero 
(almost zero for the two last methods) probability of non-coverage 
by the upper limit in such experiments. These facts are not obvious, 
but are obtained in calculations. 

In such experimental conditions
the upper limit lower than the true $s$ will occur
in the cases when the observed test statistic is lower than
some average or median test statistic expected from the background. 
Provided that the background is calculated correctly,
such cases can be interpreted as
the downward fluctuations of background.
Such experiments are characterized not only by the low
$CL_{s+b}$, but also by the low $CL_{b}$.
Dividing $CL_{s+b}$ by $CL_{b}$
the researcher takes into account how well
the experiment is described by the background.
The use of $CL_{s+b}/CL_{b} = \alpha$ instead of 
$CL_{s+b} = \alpha$
for the search of the upper limit
allows one to obtain more conservative limit with
 100\% coverage for
the experiments 
weakly susceptible to the signal.
See more detailed argumentation in 
Refs. \cite{Read_00,Junk_99,Read_02,CMS_Combined}.
The confidence-like value $CL_{s+b} / CL_{b}$ is
called $CL_s$ and
is traditionally used also as the common name of these methods.
The division of the unconstrained-maximum-related $CL_{s+b}$ by $CL_{b}$
is not necessary and even not reasonable because of the undesired
increase of the upper limit
with the decrease in the observation, see Fig. \ref{compar_lim}.

A similar correction applied to significance has been proposed \cite{Read_00},
but it does not seem to be used in practice.

We will not correct the lower limit 
obtained by the maximum-related method for $CL_{b}$ either.

Thus, the correction for $CL_{b}$ is used
only for finding the upper limit by the background-related method
and by the constrained maximum-related method.
When the correction for $CL_{b}$ is used
for  the upper limit,
the methods can be denoted by 
the abbreviation
NFLR, which means the Normalized (that is with $CL_{s+b}$ divided by 
$CL_b$) Frequency of Likelihood Ratio.
The notation ``$CL_s$'' is sometimes added
in parentheses for additional clarification.
The non-normalized case ($CL_{s+b}$) is denoted by simple FLR.
The lower limit and significance are always
calculated without the normalization even if the whole method is denoted
by NFLR ($CL_s$).

The meaning and interpretation of the FLR and NFLR ($CL_s$) methods
are very nontrivial.
It is much more convenient
to deal with the simple $\hat{s}$-values from FMML or FGML
than with nontrivial quantities like $Q$, $CL_{s+b}$, $CL_{b}$ and $CL_s$.

In the studied example in the absence of nuisance parameter 
uncertainties
the average confidence limits and their coverage calculated by 
BR--NFRL, CMR--NFRL
appeared
to be almost identical to the respective features of
all other correct methods.
The coverage of the lower limit for 
the UMR--FRL method at a small signal is lower than $\alpha$, because
at the small signal the lower limit cuts more than $\alpha$ 
fraction of experiments,
see, for example, Fig. \ref{compar_lim} and Ref. \cite{F&C}.
Although the two-sided coverage is correct, smaller and unknown coverage of
the lower limit is an unwanted feature for most of the applications. 

In the presence of uncertainties the frequentist coverage 
of intervals obtained by the frequentist likelihood ratio methods
is not guaranteed  ``by construction''
for any methods of taking this uncertainty into account and 
it has to be tested numerically. 
As usually, ignoring uncertainties and using $a_i = n_{ai}$,
$b_i = n_{bi}$ does not work well.

\subsection{The case of unknown $\vec{a}$ and  $\vec{b}$ }
\label{Sect_unknown_a_b}

Combining all the approaches described in the previous sections
we can
obtain a very large number of methods, which are difficult to test.
Since the maximum-related methods seem to be better justified,
they were tested with a much greater number of combinations 
of the other ingredients,
than that for the background-related methods.

\subsubsection{The background-related methods}

Among all tested background-related methods only
an approach similar to SSP--FGML (see Section \ref{SSP-FGML_sec})
can be accepted for the calculation of upper limits.
The lower limit was not calculated in BR--NFLR (see previous section for this notation).
The significance was not tested as well.
The subgeneration is carried out exactly in the same way as that for
SSP--FGML (and SSP--FMML).
From all checked priors only the case with the safe inverse prior
for the upper limit was found to work.
This method, 
as with SSP--FGML, 
has to provide the modeled coverage
if the nuisance parameters are distributed
in the assumed way.
Thus, the interpretation exists, but the test statistic used
is much more complex than the test statistic used in SSP--FGML, 
as was already mentioned in  Section \ref{The_CLs_method_intro}.
The calculations of the studied example show that 
the upper limit produced by this method for fixed divisions
has features similar to those of SSP--FGML.

In the used system of notations this method can be called 
SSP--GM--BR--NFLR($CL_s)$:
Subgeneration with Safe Priors, Global Maximization, Background-Related
 Normalized Frequency of Likelihood Ratio ($CL_s$).

The same method taken with subgenerated quantities 
$\vec{n}_{a\gamma}$ and $\vec{n}_{b\gamma}$
does not work well.

All forms of marginalization were found to be inappropriate for the
background-related $CL_s$ because of some numerical effects.

\subsubsection{The maximum-related methods}
\label{Sect_SARN_CLs}

According to the method described in Refs. \cite{CMS_Combined,CMS_Paper_2012}
the test statistic described by Eq. (\ref{CLs_form-1})
is replaced by an extended form, which in our notations is
\begin{eqnarray}
Q(\vec{n}, \vec{n}_a, \vec{n}_b, s) =  
\frac{ P(\vec{n}, \vec{n}_{a}, \vec{n}_{b} | 
s, \hat{\hat{\vec{a}}}, \hat{\hat{\vec{b}}})
}
{  P(\vec{n}, \vec{n}_{a}, \vec{n}_{b} | 
s_{\mathrm{ref}}, \vec{a}_{\mathrm{ref}}, \vec{b}_{\mathrm{ref}})
} \, ,
\label{label_CLs_11}
\end{eqnarray}
where, as usually, 
$\hat{\hat{\vec{a}}}$ and $\hat{\hat{\vec{b}}}$
maximize the likelihood for given $s$.
The value $s_{\mathrm{ref}}$ is calculated as described in Section
\ref{The_CLs_method_intro} for the maximum-related method.
The values $\vec{a}_{\mathrm{ref}}$ and $\vec{b}_{\mathrm{ref}}$
maximize the likelihood for this $s_{\mathrm{ref}}$.
For example, if the upper limit is calculated, 
$s_{\mathrm{ref}}$ is the minimum of $\hat{s}$ and $s$.
If $s_{\mathrm{ref}} = s$, then $Q(\vec{n}, \vec{n}_a, \vec{n}_b, s) = 1$.
The values 
$\hat{\hat{\vec{a}}}$ and $\hat{\hat{\vec{b}}}$,
obtained by the maximization of 
$P(\vec{n}, \vec{n}_{a}, \vec{n}_{b} | s, \hat{\hat{\vec{a}}}, \hat{\hat{\vec{b}}})$ with the observed data for given $s$,
including the case $s=0$ for $CL_b$,
are used for the subgeneration in the same way as 
$\hat{\hat{\vec{a}}}_{\mathrm{obs}}$ and $\hat{\hat{\vec{b}}}_{\mathrm{obs}}$ 
in the SARN--FGML method, see Section \ref{SMRN-FGML_sec} 
(here we omit the subscript ``obs''for briefness).
This method of subgeneration with random nuisance parameter measurements
will be labeled by the same prefix SARN.

The prefix SMRN will denote the subgeneration with
$\hat{\vec{a}}_{\mathrm{obs}}$ and $\hat{\vec{b}}_{\mathrm{obs}}$
like it is done in SMRN--FGML.
We can also apply the subgeneration by SSP, SEP and SSPRN approaches,
which are described in Sections \ref{SSP-FMML_sec} and 
\ref{The_SSPRN--FMML_and_SSPRN--FGML_methods_sect}.

The marginalization over nuisance parameters
was tested with the following alternative expression for $Q$:
\begin{eqnarray}
Q(\vec{n}, \vec{n}_a, \vec{n}_b, s) =  
\frac{ P(\vec{n} | s )
}
{  P(\vec{n} | s_{\mathrm{ref}} )
} \, ,
\label{label_CLs_17}
\end{eqnarray}
where $P(\vec{n} | s )$ is calculated according to Eqs. (\ref{BayesForm_1})
and (\ref{BayesForm_2}).
Of course, if the random nuisance parameter measurements are used 
(as in the SARN-methods),
$\vec{n}_{a\gamma}$ and $\vec{n}_{b\gamma}$ are substituted to these equations
to obtain the test statistic for subgenerated data.
$s_{\mathrm{ref}}$ is calculated according to Section \ref{The_CLs_method_intro}.
Maximization and marginalization over nuisance parameters are 
denoted by suffixes GM and MM, respectively. 
Safe priors are used, unless otherwise specified. 

\label{label_combined_page}
One can compare the subgenerated and observed $Q$-values and $\hat{s}$-values
simultaneously, with exclusion from the confidence set by the logical ``or'',
which should make the result more reliable.
The test of $Q$-values is susceptible to the consistency
of channels at many-channel measurements,
which is not tested directly when comparing $\hat{s}$ values, and vice versa.
Thus two tests applied simultaneously
allow one to test data from two different perspectives. 
The condition under the sum
in Eqs. (\ref{CLs_form-11}) and (\ref{CLs_form-12}) is replaced 
by
$Q(\vec{n}_{\gamma}, \vec{n}_{a \gamma}, \vec{n}_{b \gamma}, s) \leq Q(\vec{n}, \vec{n}_a, \vec{n}_b, s) \vee  \hat{s}(\vec{n}_{\gamma}, \vec{n}_{a \gamma}, \vec{n}_{b \gamma}) \leq \hat{s}(\vec{n}, \vec{n}_{a}, \vec{n}_{b})$ for the upper limit
and 
$Q(\vec{n}_{\gamma}, \vec{n}_{a \gamma}, \vec{n}_{b \gamma}, s) \leq Q(\vec{n}, \vec{n}_a, \vec{n}_b, s) \vee \hat{s}(\vec{n}_{\gamma}, \vec{n}_{a \gamma}, \vec{n}_{b \gamma}) \geq  \hat{s}(\vec{n}, \vec{n}_{a}, \vec{n}_{b})$ 
for the lower limit and for the $p$-value.
This method does not need the correction by $CL_b$,
because the check of $\hat{s}$ automatically provides the correct
behavior of the upper limit for the data microscopically dependent 
on the signal.
When the comparison of $\hat{s}$ is added to the 
likelihood-ratio methods with maximization over nuisance
parameters, it will be denoted by an additional suffix ``--FGML--''. 
For marginalization it will be denoted by ``--FMML--''.
This additional comparison is compatible with
any method of subgeneration.
For instance,
combining SSP--FMML and SSP--MM--CMR--NFLR
one obtains SSP--FMML--MM--CMR--FLR,
which is the most conservative method among these three ones
except for the 
upper limit, which is calculated without normalization
and can be slightly less than that for SSP--MM--CMR--NFLR.
The combined tests produce more reliable significance in many-channel experiments
and usually have negligible effect on intervals
if compared with the pure tests of $\hat{s}$.

One can perform the subgeneration with
varied assumed $\vec{a}$ and $\vec{b}$ 
and find the least possible lower limit and 
the highest possible upper limit
of the confidence internal.
These limits, if computed with random nuisance parameter measurements
and without the correction for $CL_b$,
are usually nontrivial (i.e. non-zero and finite) for
the FLR methods.
Since the FMML and FGML methods do not have this property
in the studied test cases,
the additional comparison of $\hat{s}$ values in the FLR methods,
as described in the previous paragraph, deprives FLR methods of this property. 
These limits should have frequentist coverage (conservative),
because the true limits calculated with the true unknown nuisance parameters
should not be wider than they.

The exact $p$-values $\rho_{\mathrm{ef}}$ and $\rho_{\mathrm{er}}$ 
for testing the FLR methods with fixed or random
nuisance parameter measurements
are obtained from Eq. (\ref{precise_p_form})
by replacements of  
$\hat{s}_{\gamma}(\vec{n}_{\gamma}, \vec{n}_{a}, \vec{n}_{b})$
by  
$Q(\vec{n}_{\gamma}, \vec{n}_{a}, \vec{n}_{b }, s)$
or
$Q(\vec{n}_{\gamma}, \vec{n}_{a \gamma}, \vec{n}_{b \gamma}, s)$, respectively,
and 
$\hat{s}_{\mathrm{obs}}(\vec{n}, \vec{n}_{a}, \vec{n}_{b})$
by $Q(\vec{n}, \vec{n}_a, \vec{n}_b, s)$.
The $p$-value $\rho_{\mathrm{er}}$
as a function of $\vec{a}$ and  $\vec{b}$
also has a nontrivial (usually non-zero) minimum.
As with the intervals, the subgeneration with fixed measurements of nuisance parameters 
(except for $\vec{n}_a$)  
as well as merging FLR with the FMML and FGML methods (testing
$Q$ and $\hat{s}$ simultaneously)
cancels this property.

The true significance calculated with the true unknown nuisance parameters
cannot be less than the minimal one.
So the minimal significance is very reliable,
although it may be conservative.
In particular, the best fitted values of the parameters
can be incompatible with observations $\vec{n}_{a}$ and $\vec{n}_{b}$.
Another problem is that it is very difficult to find these values,
because at slight variations of $a_i$ or  $b_i$
the corresponding $n_{i\gamma}$, $n_{ai\gamma}$ and $n_{bi\gamma}$ obtained by the subgeneration of
any given event switch to neighboring values
at different time. These discrete steps cause small fluctuations of
limits and significance and produce many false maxima and minima,
especially at the plateau which usually
exists at the dependence of significance and upper limits
on the background at large $b_i$.
However, low correlations between the channels for the problems considered 
herein
make it possible to fit the parameters one by one in
several rounds.
These methods will be denoted by the prefix ``Min/Max--RN''
with an appropriate continuation.

Let us consider the main features of all these methods in more detail.
The last equality in Eq. (\ref{CLs_form-11}) for the SARN--GM--CMR--NFLR 
method\footnote{The LHC-style $CL_s$ method \cite{CMS_Combined}.
As follows from the previous descriptions, the local abbreviation is decoded as 
Subgeneration with Adjusted Random Nuisance, Global Maximization,
Constrained-Maximum-Related Normalized Frequency of Likelihood Ratio.} 
is converted into
\begin{align}
\sum_{\substack{ \vec{n}_{\gamma}, \vec{n}_{a \gamma}, \vec{n}_{b \gamma}:  \\
\scriptsize{ \makebox[1.5cm][l]{\( 
Q(\vec{n}_{\gamma}, \vec{n}_{a \gamma}, \vec{n}_{b \gamma}, s) \leq
Q(\vec{n}, \vec{n}_a, \vec{n}_b, s) \) } } } }
& P( \vec{n}_{\gamma}, \vec{n}_{a \gamma}, \vec{n}_{b \gamma}  | 
s_u , \hat{ \hat {\vec{a}}}, \hat{ \hat {\vec{b}}} ) 
 = \alpha 
\label{CLs_form-12.5}
\end{align}
with $s_u = s$ in the case of $CL_{s+b}$.
In this equation it is assumed 
that the subgeneration is performed with the signal $s_u$ and with $\hat{ \hat {\vec{a}}}$ and
$\hat{ \hat {\vec{b}}}$ that maximize 
$P( \vec{n}, \vec{n}_{a}, \vec{n}_{b}  | s_u , \hat{ \hat {\vec{a}}}, \hat{ \hat {\vec{b}}})$.
The value $Q(\vec{n}, \vec{n}_a, \vec{n}_b, s)$ is used as the threshold.
It is compared with $Q(\vec{n}_\gamma, \vec{n}_{a\gamma}, \vec{n}_{b\gamma}, s)$
computed with the subgenerated data by Eq. (\ref{label_CLs_11}) 
with different
$\hat{\hat{\vec{a}}}_{\gamma}$ and $\hat{\hat{\vec{b}}}_{\gamma}$
that maximize $P( \vec{n_{\gamma}}, \vec{n}_{a \gamma}, \vec{n}_{b \gamma}  | s , \hat{ \hat {\vec{a}}}_{\gamma}, \hat{ \hat {\vec{b}}}_{\gamma})$,
as well as with different $s_{\mathrm{ref}}$, 
$\vec{a}_{\mathrm{ref}}$ and $\vec{b}_{\mathrm{ref}}$.
Of course, both $Q$-values are computed with the same definition of 
$s_{\mathrm{ref}}$.
The probabilities of
$Q(\vec{n}_\gamma, \vec{n}_{a\gamma}, \vec{n}_{b\gamma}, s) \leq Q(\vec{n}, \vec{n}_a, \vec{n}_b, s)$ 
are computed for both the upper and the lower limits
(with different $s_{\mathrm{ref}}$),
as well as for the $p$-values. 
If the definition of $s_{\mathrm{ref}}$ used corresponds to the upper limit
and the subgeneration is done with given $s$ ($s_u = s$), 
this probability is $CL_{s+b}$.
With the same definition of $s_{\mathrm{ref}}$
and with the subgeneration performed with a zero signal $s_u = 0$ ($s \neq s_u$)
this probability is $CL_{b}$.
The maximal value of $s$ at which $CL_{s+b}/CL_{b} = \alpha$
is the upper limit. 
The lower limit is found here as the minimal $s$ at which $CL_{s+b} = \alpha$
with the other definition of $s_{\mathrm{ref}}$ described in the 
previous section.
If $s_u = 0$,
$s=0$, and 
the definition of $s_{\mathrm{ref}}$ corresponds to the lower limit,
the left-hand side of this equation is the $p$-value.

The variation of given $s$ in this method leads to simultaneous variations of 
$\hat{\hat{\vec{a}}}$, $\hat{\hat{\vec{b}}}$,
$Q(\vec{n}, \vec{n}_a, \vec{n}_b, s)$,  $CL_{s+b}$,  $CL_{b}$, and $CL_{s}$.
The total result of this is very  difficult to trace.

The intervals computed by this method, as well as by all RN-methods,
do not provide a reasonable modeling interpretation.
The interpretation of $p$-values by this method assumes 
single-valued hypotheses
for nuisance parameters
without taking into account other possibilities.
The self-consistency of the $p$-value is not guaranteed by this method,
as well as by all other RN-methods.

\begin{figure}[t]
\begin{center}
\includegraphics[width=1.0\linewidth]
{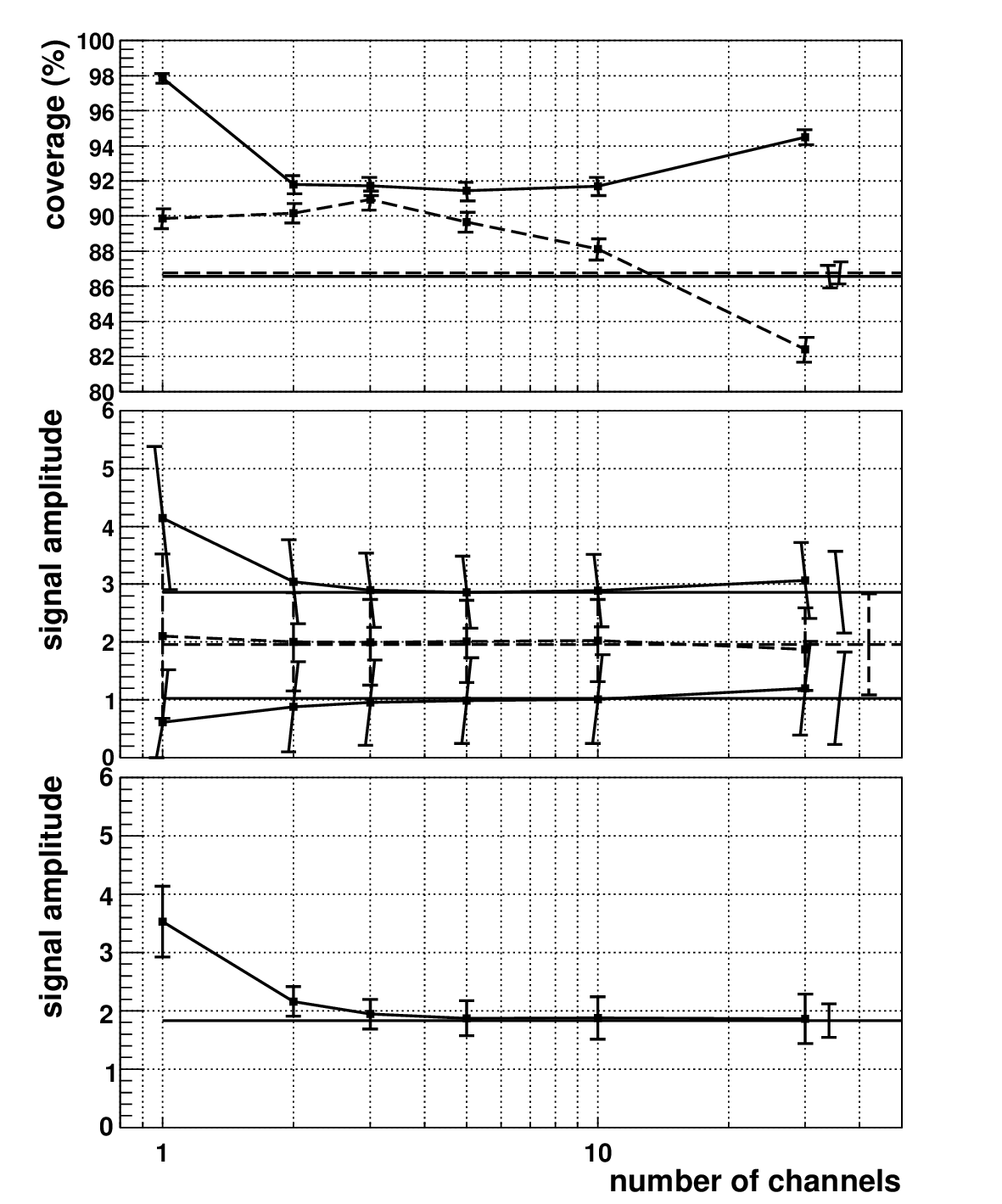}
\caption{
The SARN--GM--CMR--NFLR (LHC $CL_s$) method, unknown expected background,
confidence intervals for  90\% 
one-sided confidence ($\alpha=0.1$).
The horizontal lines show  
the optimization by the widths of the Bayesian modified central intervals.
Other details are described in Sections
\ref{Sect_SARN_CLs}, \ref{sec_Bayes_cover}, 
and in Fig. \ref{Bayes_center_mod_no_unc_0}.
}
\label{LHC_CLs_geny_anay_5}
\end{center}
\end{figure}

For the tested example
this method yields good frequentist coverage for the upper limit
at the ``standard'' parameters described in Section \ref{Parameters_sect}.
The coverage is worse at some other parameters,
but the insufficient coverage of the upper limit
has not been proved with statistical confidence. 
For the standard parameters
the average position of the upper limit is close to that of
SARN--FGML.
They are
the lowest among all other tested working (covering) methods,
see Fig. \ref{LHC_CLs_geny_anay_5}.
But the coverage of the lower limit, as well as the standard 
optimized coverage of both limits, is insufficient.
The most detailed division without zeros in the expected background 
provides good coverage (not shown in Fig. \ref{LHC_CLs_geny_anay_5}), but
it was explained in Sections \ref{Optimization} and \ref{SMRN-FGML_sec} that such
optimization is not applicable as a general method.
When only the uncertainty of the expected signal is present,
the coverage of the upper limit is not decreased with the increase
in the number of channels, but the standard optimized coverage is slightly
lower than necessary, about 88\%.
If the expected background contains many zeros,
the significance by this method
can be noticeably higher than $z_{\mathrm{er}}$ with the same test statistic,
see the results for Min--RN--GM--CMR--FLR in
the table in the next section.
The coverage of the lower limit can be partially corrected
at the standard parameters
if one obtains the least minimal limit by
this method and by the similar unconstrained method
SARN--GM--UMR--FLR (computed with $2\alpha$).
But since the latter method has undercoverage (if compared with $1-\alpha$) 
at low signal, this coverage does not always hold. 
The significance is identical for these two methods.
The combined method will be called SARN--GM--CMR--NFLR--UMR--FLR.
The upper limit is obtained by  SARN--GM--CMR--NFLR in it. 
A similar method with marginalization
will be called SARN--MM--CMR--NFLR--UMR--FLR.

There exists an asymptotic 
approximation of the significance estimate 
by SARN--GM--UMR--FLR, see  
Refs.  \cite{CMS_Combined,CMS_Paper_2012,Cowan_11}.
In general, 
the asymptotic values are reached at a large number of events.
The accuracy is good at a wide range of conditions
\cite{CMS_Combined}, but it is not clear how strongly it can vary
at a small or zero number of events in separate channels.
The asymptotic approximation gives the significance 
as 
\begin{eqnarray}
z = \sqrt{-2\ln{Q(0)}} \, .
\label{asympt}
\end{eqnarray}
This approximation allows one 
to avoid time-consuming  
subgeneration-based calculations,
which are unfeasible
for large significance.
The accuracy of this asymptotic estimate is
not clear in particular situations.
In our calculations this value was sometimes greater than
the exact significance calculated with the true parameters
by RN--GM--MR--CMR--FLR 
and usually greater than the minimal significance
by this method.
Note that the identical $Q(0)$ is obtained by
SARN--GM--UMR--FLR and SARN--GM--CMR--NFLR.
Removing unimportant characters 
we will denote this method by ``Asymptotic GM--MR--FLR''.

If marginalization is used in SARN--GM--CMR--NFLR
instead of maximization,
the coverage of the lower limit is correct at the standard parameters.
But this coverage is violated at some other parameters.
The significance obtained by
SARN--MM--CMR--NFLR usually exceeds the significance by Min--RN--MM--CMR--FLR
as well as SARN--GM--CMR--NFLR exceeds Min--RN--GM--CMR--FLR,
but the difference between the first pair is usually smaller,
than between the last pair.
However, this was observed only for the problems with expected-background uncertainty.
In the problems with expected-signal uncertainty
these methods behave differently in different conditions.
If we substitute the observed $Q$-value obtained 
in the SARN--MM--CMR--NFLR method  
into Eq. (\ref{asympt}),
the resulting ``marginalized'' (pseudo-) asymptotic significance
can slightly exceed the significance by Min--RN--MM--CMR--FLR too,
but the difference between them is usually smaller
than the difference between the regular asymptotic significance
and Min--RN--GM--CMR--FLR.
\linebreak
Therefore the ``marginalized'' asymptotic significance
is usually more reliable than the standard one.
We will symbolically denote this 
method 
by the notation ``Asymptotic MM--MR--FLR''.

The intervals obtained by the SMRN--GM--CMR--NFLR method
behave similarly to the intervals of SMRN--FGML, 
see Fig. \ref{SMRN_FGML_unc_5}.
The coverage of the lower limit is slightly better, but not perfect.
The significance is usually incorrect.
So this method cannot be recommended.
But its results can be combined with 
that of the SARN--GM--CMR--NFLR($CL_s$) method
by selecting minimal and maximal (for the upper limit) values.
This may be viewed as a very rough  approximation to
the Min/Max-RN methods which are extremely difficult to implement.
The same conclusion is valid for similar methods with
marginalization.

For the problem considered here,
the limits by SARN--FGML--GM--CMR--FLR (with additional comparison of $\hat{s}$, 
see p. \pageref{label_combined_page}) are close to 
the limits by SARN--FGML.
The significance is usually
more conservative (smaller) than that for SARN--GM--CMR--NFLR.
Similar features are exhibited by 
SARN--FMML--MM--CMR--FLR.
We can also combine SARN--FGML--GM--CMR--FLR
with 
SARN--GM--\linebreak
UMR--FLR,
taking the upper limit and significance
from SARN--FGML--GM--CMR--FLR
and the least lower limit from 
SARN--GM--UMR--FLR and 
SARN--FGML--GM--\linebreak
CMR--FLR.
This method will be referred as
SARN--FGML--GM--CMR--FLR--UMR--FLR.
The same method with 
\linebreak
marginalization is denoted as
SARN--FMML--MM--CMR--FLR--UMR--FLR.
The features of the two last methods are similar to the
features of 
SARN--FGML--GM--CMR--FLR and
SARN--FMML--MM--CMR--FLR,
respectively.

The intervals obtained by the SSPRN--GM--CMR--NLFR method
can be characterized by 
Fig. \ref{LHC_CLs_geny_anay_5} with the exception that
the last 3 points (for 5, 10, and 30 channels)
of the lower-limit coverage are on the level of
86--88\%. 
Another interesting observation is that
the upper limit can be calculated not only with the
inverse (safe) priors, but also with the uniform priors.
The result is shorter intervals with still sufficient coverage.
However, there is no way found to optimize the divisions
without the loss of coverage in this method.
Choosing the most detailed division without zeros in
the expected-background distribution
allows one to obtain the interval with coverage,
but this method cannot be recommended in general,
as explained in Section \ref{Optimization}.
The significance estimated by this method is
usually greater than that by the Min--RN--GM--CMR--FLR
and is therefore unreliable.
In conclusion, this method cannot be recommended.
The intervals obtained by the SSPRN--MM--MR--NLFR method
behave approximately as that by SSPRN--FMML in 
Fig. \ref{SSPRM_FMML_unc_5}. Its significance is large and unreliable. This method cannot be recommended either.

The features of intervals obtained by the SSP--BM--CMR--NLFR and
SSP--FBML--BM--CMR--FLR methods (see p. \pageref{label_combined_page}, 
here ``B'' stands for either ``M'' or ``G'' and 
is the same during the following comparisons)
are similar to those of SSP--FBML methods.
The modeling interpretation of intervals as well as 
the interpretation of significance
by SSP--FBML--BM--CMR--FLR
is similar to that of SSP--FBML.
The $p$-values are self-consistent. 
The significance by SSP--FMML--MM--CMR--FLR
is usually lower than
the significance by Min--RM--MM--CMR--FLR,
especially if the auxiliary parameters are constrained
at fitting, for example, as described in the next section.
These two significances are usually very close to each other.
Therefore, if the minimization is not feasible,
SSP--FMML--MM--CMR--FLR could be a good replacement
of minimization for the calculation of significance.
It has a good independent interpretation itself.
The SSP--FGML--GM--CMR--FLR method
could also be a replacement of Min--RM--GM--CMR--FLR,
but it is less reliable, since it exaggerates the significance
more frequently and for greater values in the general case.

\section{Examples of problems with
unknown $\vec{a}$ or $\vec{b}$ }
\label{on_off_Sec}

Table \ref{NewOnOff_table} illustrates features of different methods
for some examples.
\begin{table*}[t]
\centering

\caption{Comparison of significance obtained by different methods
for the simple one-channel examples. 
For each case either the nuisance parameter is known exactly or it is measured
as $n_a$ or $n_b$. 
The character ``B'' in method notations means either ``M'' or ``G'',
because the results are very similar for these cases.
Other details are described in the text.
}
\setlength{\tabcolsep}{0.35em} 
\begin{tabular}{|l|c|c|c|c|c|c|c|c|c|} \hline

$n$                                    &  67          &  60        & \multicolumn{2}{c|}{88}     & \multicolumn{2}{c|}{37, 51}         & standard      & standard     & 7         \\  
$t_{a}$                                &  1           &  1         & \multicolumn{2}{c|}{1}      & \multicolumn{2}{c|}{1}              & cond.,        & cond.,       & 1         \\  
$a$ / $n_{a}$                          &  1 / --      &  1 / --    & \multicolumn{2}{c|}{1 / --} & \multicolumn{2}{c|}{$a_1$, $a_2$ / --, --}  & 5             & 30           & -- / 3    \\  
$t_{b}$                                &  2           &  10        & \multicolumn{2}{c|}{10}     & \multicolumn{2}{c|}{10}             & channels      & channels     & 1         \\  
$b$ / $n_{b}$                          & -- / 15      &  -- / 0    & \multicolumn{2}{c|}{-- / 3} & \multicolumn{2}{c|}{--, -- / 3, 0 } &               &              & 2 / --    \\ \hline
                                       &  z           &  z         & z        & limits           & z              & limits             &  z            & z            & z         \\ \hline
     Bayesian central                  &   ---        &    ---     & ---      & 23.6--81.9       & ---            & 33.0--75.4         & ---           &  ---         &  ---      \\
     Bayesian central modified         &   ---        &    ---     & ---      & 23.6--81.9       & ---            & 33.1--75.4         & ---           &  ---         &  ---      \\
     Likelihood Ratio                  &   ---        &    ---     & ---      & 28.0--79.9       & ---            & 49.1--74.0         & ---           &  ---         &  ---      \\
     SSP--FGML                         &  2.89        &   2.72     & 1.89     & 18.8--81.8       & 2.46           & 32.0--74.7         & 2.62          &  0.65        &  2.61     \\
     SSP--FMML                         &  2.89        &   2.72     & 1.89     & 18.8--81.8       & 2.47           & 32.0--74.6         & 2.61          &  1.13        &  2.61     \\
     SSP--GM--CMR--NFLR                &  2.89        &   2.72     & 1.89     & 18.8--81.9       & 2.47           & 31.9--74.7         & 2.58          &  0.07        &  2.61     \\
     SSP--MM--CMR--NFLR                &  2.89        &   2.72     & 1.89     & 18.8--81.9       & 2.47           & 32.0--74.7         & 2.59          &  0.37        &  2.61     \\
     SSP--FGML--GM--CMR--FLR           &  2.89        &   2.72     & 1.89     & 18.8--81.8       & 2.40           & 30.8--74.7         & 2.52          &  0.07        &  2.61     \\
     SSP--FMML--MM--CMR--FLR           &  2.89        &   2.72     & 1.89     & 18.8--81.8       & 2.41           & 31.3--74.7         & 2.53          &  0.37        &  2.61     \\
     SEP--FBML                         &  3.09        & $\infty$   & 2.35     & 32.1--74.8       & $\approx$4.6   & 52.2--67.3         & $\approx3.4$  & $>$5.0       &  2.61     \\
     SEP--BM--CMR--NFLR                &  3.09        & $\infty$   & 2.35     & 32.1--75.0       & $\approx$4.5   &$\approx$52.3--67.3 & $\approx3.4$  & $>$5.0       &  2.61     \\
     SEP--FBML--BM--CMR--FLR           &  3.09        & $\infty$   & 2.35     & 32.1--74.8       & $\approx$4.5   &$\approx$52.2--67.3 & $\approx3.3$  & $>$5.0       &  2.61     \\
     SSPRN--FGML                       &  3.63        &   4.02     & 2.80     & 31.7--84.4       & 3.69           & 45.3--74.2         & 3.67          & 1.96         &  1.73     \\
     SSPRN--FMML                       &  3.70        &   3.93     & 2.85     & 33.1--83.2       & 3.61           & 44.2--73.9         & 3.70          & 2.51         &  2.08     \\
     SSPRN--GM--CMR--NFLR              &  3.01        &   4.30     & 2.01     & 23.7--79.6       & 3.90           & 47.6--74.2         & 3.07          & 1.24         &  2.70     \\
     SSPRN--MM--CMR--NFLR              &  3.03        &   4.29     & 2.20     & 27.5--79.8       & 3.90           & 45.8--73.8         & 3.35          & 1.65         &  2.71     \\
     SMRN--FGML                        &  4.07        & $\infty$   & 5.09     & 34.1--85.3       & $>$6.0         & 51.9--74.2         & $>$5.6        & 2.48         &  1.73     \\
     SARN--FGML                        &  2.94        &   2.93     & 2.08     & 25.4--79.7       & 2.74           & 44.7--74.2         & 2.74          & 2.51         &  1.73     \\
     SMRN--GM--CMR--NFLR               &  3.02        & $\infty$   & 1.76     & 24.4--79.7       & $>$6.0         & 51.9--74.2         & 4.06          & 2.10         &  2.65     \\
     SARN--GM--CMR--NFLR               &  3.01        &   3.07     & 2.09     & 25.4--80.0       & 2.85           & 45.6--74.2         & 2.76          & 2.42         &  2.65     \\
     SARN--GM--UMR--FLR                &  3.01        &   3.07     & 2.09     & 27.6--81.7       & 2.85           & 47.7--73.4         & 2.76          & 2.42         &  2.65     \\
Asymptotic GM--MR--FLR                 &  3.04        &   3.38     & 2.19     & ---              & 3.13           & ---                & 2.89          & 0.60         &  2.75     \\
\ \ \ \ \ Used nuisance parameters     &  \ \ 27.3    & \ \ 5.45   & \ \ 8.27 &                  & 3.6, 4.6       &                    &               &              &           \\
Min/Max RN--GM--CMR--FLR               &  2.83        &   3.02     & 1.74     & 13.9--79.6       & 2.77           & 42.1--75.3         & 2.60          & 0.24         &  2.64     \\
Min/Max RN--GM--UMR--FLR               & [2.87]       &  [3.02]    &[1.74]    & 21.6--81.7       &[2.78]          & 45.8--75.5         &[2.63]         &[0.41]        & [2.64]     \\
\ \ \ \ \ Best nuisance parameters     &  \ \ 4.5     &   \ \ 6.1  & \ \ 2.8  &                  & 29.2, 5.3      &                    &               &              &  \ \ 2.4  \\
     SARN--FGML--GM--CMR--FLR          &  2.91        &   2.93     & 2.04     & 24.3--80.7       & 2.68           & 43.4--74.3         & 2.65          & 2.41         &  1.71     \\
SARN--GM--CMR--NFLR--                  &              &            &          &                  &                &                    &               &              &           \\
\ \ \ \ \ \ \ \ \ \ \ \ \ \ \  
\ \ \ \ \ \ \ \ \ \ \ \     --UMR--FLR &  3.01        &   3.07     & 2.09     & 25.4--80.0       & 2.85           & 45.0--74.5         & 2.76          & 2.42         &  2.65     \\
SARN--FGML--GM--CMR--FLR--             &              &            &          &                  &                &                    &               &              &           \\
\ \ \ \ \ \ \ \ \ \ \ \ \ \ \ 
\ \ \ \ \ \ \ \ \ \ \ \     --UMR--FLR &  2.91  &   2.93     & 2.04     & 24.3--80.7       & 2.68           & 43.1--74.6         & 2.65          & 1.95         &  1.71     \\
     SMRN--MM--CMR--NFLR               &  3.02        & $\infty$   & 2.07     & 26.6--79.4       & $>$6.0         & 51.7--74.1         & 4.74          & 2.55         &  2.74     \\
     SARN--MM--CMR--NFLR               &  3.02        &   3.07     & 2.10     & 25.6--80.1       & 2.85           & 44.9--74.0         & 2.78          & 2.61         &  2.74     \\
     SARN--MM--UMR--FLR                &  3.02        &   3.07     & 2.10     & 27.1--86.5       & 2.85           & 40.4--74.8         & 2.78          & 2.61         &  2.74     \\
Asymptotic MM--MR--FLR                 &  3.02        &   3.11     & 2.16     & ---              & 2.85           & ---                & 2.80          & 0.53         &  2.62     \\
Min/Max--RN--MM--CMR--FLR              &  3.01        &   3.01     & 2.01     & 25.2--83.9       & 2.77           & 37.9--74.9         & 2.77          & 0.52         &  2.61     \\
Min/Max--RN--MM--UMR--FLR              & [3.01]       &  [3.01]    &[2.01]    & 20.1--87.3       &[2.77]          & 28.3--84.5         &[2.77]         &[0.52]        & [2.61]     \\
\ \ \ \ \ Best nuisance parameters     &  $\gtrsim15$ & \ \ 6.1    & \ \ 3.4  &                  &  $\gtrsim$4.2, $\gtrsim$5  &                    &               &              &  \ \ 12.8 \\
     SARN--FMML--MM--CMR--FLR          &  2.92        &   2.84     & 2.04     & 24.3--80.7       & 2.60           & 42.9--74.2         & 2.65          & 2.61         &  2.03     \\
SARN--MM--CMR--NFLR--                  &              &            &          &                  &                &                    &               &              &           \\
\ \ \ \ \ \ \ \ \ \ \ \ \ \ \
\ \ \ \ \ \ \ \ \ \ \ \     --UMR--FLR &  3.02        &   3.07     & 2.10     & 25.6--80.1       & 2.85           & 40.4--74.0         & 2.78          & 2.61         &  2.74     \\
SARN--FMML--GM--CMR--FLR--             &              &            &          &                  &                &                    &               &              &           \\
\ \ \ \ \ \ \ \ \ \ \ \ \ \ \ 
\ \ \ \ \ \ \ \ \ \ \ \     --UMR--FLR &  2.89  &   2.84     & 2.05     & 24.3--80.7       & 2.60           & 40.4--74.2         & 2.65          & 2.63         &  2.03     \\
Min/Max--E--RN--MM--CMR--FLR           &  0.30        &   3.01     & 0.0      & 0.0 -- 85.0      & 0.74           & 37.3--77.3         & 0.0           & 0.19         &  2.61    \\
\hline  
\end{tabular}
\label{NewOnOff_table}
\end{table*}
The first example is taken from table 1 of Ref.  \cite{Cousins_08}.
The other cases are artificial examples.
The first six examples (columns) have only the expected-background uncertainty.
The 5-th and 6-th columns represent one of the ``real'' events generated with 
standard parameters
from Section \ref{Parameters_sect} 
and divided in 5 and 
30 channels\footnote{The generated $\vec{n}_b$ for 30 channels from the first to the last is 
2, 1, 2x4, 1, 0, 3, 2x4, 3, 5, 4, 2, 1, 2, 1, 0, 1, 3x0, 1, 2, 2x1, 5x0.
$\vec{n}$ is
20, 17, 25, 28, 2x19, 17, 12, 9, 2x14, 10, 11, 8, 4, 7, 9, 7, 6, 4, 6, 7, 5, 4, 6, 7, 0, 9, 2x7.
They are summed up by 6 for 5 channels.}. 
The standard algorithm for division optimization from 
Section \ref{Optimization}
chooses the 5-channel division.
The last case is with only the expected-signal uncertainly.
The statistical uncertainty of significances
should not exceed $0.01$.

The significances and interval widths 
obtained by minimization over nuisance parameters
for many-channel cases 
are our estimates, which can in theory exaggerate the real values,
if local minima are found instead of the global ones,
although it is unlikely.
The significances and intervals by the methods 
with minimization are obtained without
significant constraints for the nuisance parameters, 
except for the numbers in square brackets,
whose constraints are described later.
Technical constraints are always present,
but they are believed to be very wide and unimportant.
The optimal nuisance parameters for significances are given
as values or as lower thresholds if the corresponding 
significance is not changed noticeably when the nuisance parameter is increased.

Note that significance is the same by definition for
CMR and UMR methods, whereas the lower limits are different.
Note also, that the Min/Max--RN--MM... methods include the analysis
with marginalization using safe priors.
The Min/Max--RN--MM--CMR--FLR method with inverse priors, 
denoted by Min/Max--E--RN--MM--CMR--FLR and
presented in the last line of this table,
yields usually smaller significances and they are frequently
equal to trivial zero. Therefore the inverse priors should not be used here.

One can see that the second example (column) is with zero result
of the auxiliary background experiment.
The fourth example has zero result
of the auxiliary background experiment in the last channel.
For this case $a_1 = 0.182426$ and $a_2 = 0.817574$,
which corresponds to the ``standard'' conditions of Section \ref{Parameters_sect}.
The previous third example is the same experiment
with unified channels.
The 30-channel example has many channels with zero expected background
and one channel (number 27) with zero expected background
and zero obtained signal.

The Min/Max--RN--BM--CMR--FLR and Min/Max--RN--BM--UMR--FLR methods
(``B'' is either ``M'' or ``G''), 
produce identical (by definition)
significance but different limits.
To avoid duplication of these numbers in
the columns with significances
and to give an idea of the effect of constraints
the significances given in the Min--RN--BM--UMR--FLR lines
are computed with the following simple
constraints.
\label{label_constraints}
Each $a_i$ or $b_i$ is limited by the Poisson frequentist central confidence
limits for measured $n_{ai}$ or $n_{bi}$ (used as the test statistic), 
respectively, 
and for one-sided $\alpha$ equal to
$( 1 - (1 - \rho)^{1/k})/2 \approx \rho / (2 k)$, where
$\rho = F( -\max(z, 1.0)$,
$F$ is defined in 
Eq. (\ref{IntGaussian}), and $z$ is the significance 
which is found
in the iterative fitting procedure. At each next step one can use the 
previously found value to set the new limits.
Thus, the total probability of limits violation
is equated to the $p$-value.
The left-hand side of the approximate equality above means that  
the violation of two limits in different channels
is considered as one violation, though this 
does not change the result for small $\rho$.  
In order to avoid zero-length intervals
the $z$-value is bounded from below by unity.
This  
constrained minimum is
the same 
by definition for 
...--CMR--FLR and ...--UMR--FLR methods as well.
To remind about constraints 
the constrained minimal significances are given in square brackets.
The best values of parameters given in next lines correspond to
unconstrained cases.
For GM-methods they turn out to be different
from the parameters that maximize the global likelihood and
used in the SARN--GM--... methods as well as in
the asymptotic GM--MR--FLR.
The latter parameters are given under the line of the latter method.

The minimal significances 
in the last column (as well as in the previous ones)
correspond to the subgeneration with the
{\it random} expected-signal measurement,
according to the notation.
It is useful to remind that
for the SSP methods and expected-signal uncertainty 
the significance $z_{\mathrm{ef}}$ calculated with any $a$
turns out to be identical to the approximate significance
calculated by the same method independently on $a$.

For the sake of briefness, the ``MM'' and ``GM'' variants
of  some less interesting methods
were merged in the table. If the results
differ, this is marked by the sign ``$\approx$''.
If the significance cannot be computed and it cannot be
proved that it is infinite, it is given as the lower limit (e.g. ``$>$5'').

It can be seen that significances calculated by SARN--MM--CMR--NFLR and 
asymptotic MM--MR--FLR
are very close to the significance by
Min--RN--MM--CMR--FLR everywhere except the 30-channel example.
This is not the case
for the corresponding GM methods:
SARN--GM--CMR--NFLR
and  Min--RN--GM--CMR--FLR.
In the first column SARN--GM--CMR--NFLR and
the asymptotic GM--MR--FLR indicate the evidence of the signal
according to the common agreement of ``$>3$''.
This is not confirmed by Min--RN--GM--CMR--FLR, which gives less than 3.
If one uses ``SARN--MM--...'' methods, one does not
miss the evidence, because all SARN--MM--CMR--NFLR,
asymptotic MM--MR--FLR, and Min--RN--MM--CMR--FLR
(with and without restrictions)
produce almost the equal significance greater than $3$.
Unfortunately, all SSP methods, including
SSP--MM--MR--NFLR and SSP--FMML--MM--MR--FLR, do not
confirm the evidence in this case.
But in the 30-channel example 
the significance by Min--RN--MM--CMR--FLR (0.52)
is much lower than 
that by SARN--MM--CMR--NFLR
(2.61, 5 times as much, because of the mentioned channel 27).
If we consider the former as the true significance
(which is extremely difficult to compute directly),
and if we want to be able to obtain a reasonable value of
significance for any division,
we have to reject SARN--MM--CMR--NFLR,
as well as SARN--GM--CMR--NFLR 
\linebreak
(usual LHC $CL_s$)
and many other methods in this table.
The first and the third columns indicate that
we might also reject SSP--GM--CMR--NFLR and SSP--FGML--GM--CMR--FLR,
because their significance is slightly greater than
that by Min--RN--GM--CMR--FLR.
The SSP--MM--CMR--NFLR and SSP--FMML--MM--CMR--FLR
methods produce $z$ which is lower than that by
Min--RN--MM--CMR--FLR
for all columns with uncertain expected background.
Unfortunately, this does not hold in the general case.
However, the SSP--FMML--MM--CMR--FLR method is currently
the only known method for which the case with its significance
greater than the significance by the corresponding
method with {\it constrained} minimization over the nuisance parameters
(Min--RN--MM--CMR--FLR in given case, 
and for GM methods it is Min--RN--GM--CMR--FLR)
in the presence of
{\it only the expected-background uncertainty} 
has not been found so far 
(in the analysis of hundreds of fictional experiments mostly with one or two channels
with constraints for minimization described earlier in this section).

In the one-channel problems with only the expected-signal uncertainty 
GM-methods usually compete very well with MM-methods, as it is
seen in the table.
This is not the case for many-channel problems with 
only the expected-signal uncertainty,
in which the results fluctuate depending on the details
of the problem.

Looking at the last column
with the expected-signal uncertainty
one can find that the asymptotic 
GM--MR--FLR gives greater significance than
Min--RN--GM--CMR--FLR,
whereas the asymptotic MM--MR--FLR is almost equal to 
Min--RN--MM--CMR--FLR.
These relations were observed in many other
examples with only the expected-signal uncertainty.
In general, the asymptotic MM--MR--FLR is more reliable than 
the asymptotic GM--MR--FLR.

If there are uncertainties of both expected background and expected signal,
we can calculate the minimum of the significance 
with random expected background
and with either random or fixed expected signal.
In several tested examples the minimal $z$ with fixed expected signal
was slightly greater than the minimal $z$ with random expected signal,
and the latter was slightly greater than $z$ by
SSP--BM--CMR--NFLR (``B'' is either G or M and 
is the same for all compared methods).

If the significance can be minimized,
one can choose the division that
provides the greatest value of this minimum.
Otherwise, 
random differences between the
minimum and $z$ by the other methods could, in theory, 
lead to too optimistic results from the frequentist viewpoint, 
if too many choices are available and the look-elsewhere effect is involved.
Then it can be recommended to use the standard optimization method
based on the minimum of interval width and rejection of zero
expected signals, but the accuracy of this, as well as of any other methods
of optimization of division for calculation of significance,
cannot be currently verified.

Thus, if the minima cannot be calculated
because of technical difficulties,
it can be recommended to calculate significance by 
SSP--FMML--MM--MR--FLR, 
which, in the general case, according to the available calculations,
is the most reliable method among the methods without minimization,
or by SSP--MM--CMR--NFLR, SSP--FGML--GM--CMR--FLR, or SSP--GM--CMR--NFLR,
which are less reliable.
All these methods have own interpretations and attractive features, 
and they can be used 
even if the minima are available.
If these subgeneration-based methods are not
feasible either (for instance, because of too high $z$),
the asymptotic MM--MR--FLR is preferable.
Note that the confidence intervals can be calculated
also by SSP--FBML, whose limits
are usually very close to SSP--FBML--BM--CMR--FLR.

\section{Conclusion}

The Bayesian credible intervals provide frequentist coverage
(sometimes conservative) for the tested examples, 
if the upper limit is computed
with inverse priors for nuisance parameters
and the lower limit is computed with uniform priors
for nuisance parameters. This combination
of priors is called ``safe priors'' in this work.
The prior for the main parameter should be uniform in both cases.
The modified central intervals should be used in order to 
provide frequentist coverage. 
There is a way to choose the optimal number of channels,
or the ``optimal division'', and to retain coverage.
This Bayesian method is applicable for a wide class of problems,
but does not allow the user to calculate the classical statistical significance.

The likelihood ratio (profile) method is technically simple.
It has an asymptotic frequentist interpretation for large statistics
and does not have any meaning for non-Gaussian problems with small statistics. 
The intervals obtained with fixed divisions have slightly insufficient
coverage for the problems studied here. 
All reasonable methods of division (or binning) 
optimization lead to significant undercoverage.
The significance is not calculated.

The frequentist approach provides both the confidence intervals and the statistical significance.
There are many frequentist methods which yield different results
and there is no strict rule for the selection of the best method.
All frequentist methods (except asymptotic approximations) 
are complex and computationally expensive.
The direct calculations of very large significance are unfeasible.
The test statistic can be either the maximum likelihood
estimate or the likelihood ratio in different forms, or both.
For generation of pseudo-experiments 
(subgeneration, according to our terminology)
nuisance parameters can be obtained by the Bayesian approach
with safe priors or by the maximization of likelihoods.
Nuisance measurements can be subgenerated or
not. 
During the analysis, 
the nuisance parameters can be eliminated either by maximization
of the likelihood 
or by marginalization. 
Nontrivial minimal lower limits and the minimal significance,
as well as nontrivial maximal upper limits 
with respect to nuisance parameters 
(i.e. by fitting nuisance parameters), 
can usually be obtained
for some likelihood ratio-based methods with subgeneration 
of nuisance parameter measurements,
though it is very difficult to obtain them.
The optimal values of the nuisance parameters
can be inconsistent with their measurements.
There are a number of more or less adequate methods
that do not include fitting nuisance parameters,
with different features, advantages, and disadvantages.
Among these methods, 
the method denoted here by SSP--FMML--MM--MR--FLR
seems preferable for the calculation of significance.
It has convincing independent interpretation and provides
self-consistent signifcance.
The significance by it
does not usually exceed or exceeds only slightly the minimal significance,
which allows one to interpret it in a purely frequentist way too.
For frequentist confidence intervals the preferable methods
are this method too or the simpler SSP--FMML or SSP--FGML methods.
All tested methods in which only the likelihood ratio 
is tested and the nuisance parameters are not fitted
are less reliable in the general case
than SSP--FMML--MM--MR--FLR for the calculation of significance.
Some of them (including the popular method usually denoted by $CL_s$)
can greatly overestimate the significance
compared to the minimal significance.
This happens rarely, but it is undesirable for significance at all.
In the numerical tests,
the asymptotic approximations to these methods
turn out to be more reliable than the methods themselves,
and the approximation obtained with marginalization
over the nuisance parameters is more reliable than
the regular approximation obtained with maximization.

\clearpage

\end{document}